\documentclass[a4paper,12pt,oneside]{article}

\usepackage{pdfpages}
\usepackage{rotating}
\usepackage{caption}
\usepackage{subfig}

\usepackage{amsmath, amssymb, graphics}
\usepackage{float}

\usepackage{datetime}
\usepackage{changepage}
\usepackage{refstyle}
\usepackage{bookmark}
\usepackage[framemethod=TikZ]{mdframed}
\usepackage{showexpl}
\usepackage{kvsetkeys}

\usepackage{yfonts}
\usepackage[utf8]{inputenc}

\graphicspath{ {figures/} }
\usepackage{array}
\usepackage[english]{babel}
\usepackage{multicol}

\usepackage{tikz-cd}
\usepackage[utf8]{inputenc}
\usepackage[super,sort&compress,comma]{natbib}
\usepackage{braket}
\usepackage{adjustbox}
\usepackage{epstopdf}
\usepackage{soul}
\usepackage{setspace}
\hypersetup{
    colorlinks=true,         
    linkcolor=black,         
    citecolor=blue,          
    }

\newcommand*\rfrac[2]{{}^{#1}\!/_{#2}}
\begin{document}

\setcounter{page}{1}

\begin{center}
{\large \textbf{Effective Floquet Hamiltonian theory of multiple-quantum NMR in anisotropic solids involving quadrupolar spins: Challenges and Perspectives}}
\end{center}
\begin{center}
Vinay Ganapathy and Ramesh Ramachandran$^{*}$ \\
Department of Chemical Sciences \\
Indian Institute of Science Education and Research (IISER) Mohali \\
Sector 81, Mohali-140306, Punjab, INDIA
\end{center}

\begin{center}
$*$ Author to whom correspondence should be addressed: 
E-mail:rramesh@iisermohali.ac.in
\end{center}

\vspace{2cm}

{\textbf{Abstract}} \\
\\
The response of a quadrupolar nucleus (nuclear spin with I>$\rfrac{1}{2}$) to an oscillating radio-frequency (RF) pulse/field is delicately dependent on the ratio of the quadrupolar coupling constant to the amplitude of the pulse in addition to its duration and oscillating frequency. Consequently, analytic description of the excitation process in the density operator formalism has remained less transparent within existing theoretical frameworks. As an alternative, the utility of the "concept of effective Floquet Hamiltonians" is explored in the present study to explicate the nuances of the excitation process in multilevel systems. Employing spin $I=\rfrac{3}{2}$ as a case study, a unified theoretical framework for describing the excitation of multiple-quantum (MQ) transitions in static isotropic and anisotropic solids is proposed within the framework of perturbation theory. The challenges resulting from the anisotropic nature of the quadrupolar interactions are addressed within the effective Hamiltonian framework. The possible role of the various interaction frames on the convergence of the perturbation corrections is discussed along with a proposal for a "hybrid method" for describing the excitation process in anisotropic solids. Employing suitable model systems, the validity of the proposed hybrid method is substantiated through a rigorous comparison between simulations emerging from exact numerical and analytic methods.
\clearpage
\section{Introduction}

Theoretical descriptions of NMR experiments involving quadrupolar nuclei have always been fraught with difficulty owing to the domineering presence of the quadrupolar interactions. In contrast to the internal spin interactions commonly prevalent in spin I=$\rfrac{1}{2}$ systems (such as chemical shift, dipolar interactions (both isotropic and anisotropic)), the quadrupolar interactions are four to six orders of magnitude larger and are solely responsible for compromising the resolution of the NMR spectra, both in the solution and solid state\cite{Jerschow200563,manqi,freude2000,wasylishen2012nmr}. Nevertheless, the importance of quadrupolar nuclei as molecular probes for identifying the distinct sites of a particular nucleus in material science and inorganic clusters\cite{Hanna20101,wasylishen2012nmr} is known and well documented. Since the relevance of NMR spectroscopy as an analytical tool largely depends on both the availability and reliable estimate of the available molecular constraints, parallel development of theoretical methods along with experiments remain indispensable.

Although, the introduction of magic angle spinning (MAS) \cite{andrew} has enhanced the spectral resolution in the study of spin I=$\rfrac{1}{2}$ systems, the line broadening effects of the quadrupolar interactions do persist in the solid state. From a practical viewpoint, the advent of multiple quantum MAS (or MQMAS) \cite{frydman1995, doi:10.1021/ja00156a015} NMR experiments has largely been instrumental in reviving the NMR spectroscopy of quadrupolar nuclei. Since then, several modifications to the original scheme have emerged in recent literature \cite{Rocha2005,goldbourt2002multiple} and are beyond the scope of the present article. Nevertheless, the extent of development of NMR methodology in the study of quadrupolar nuclei is only modest when compared to their spin I=$\rfrac{1}{2}$ counterparts. This could be largely due to the lack of a formal less cumbersome theoretical approach for describing the time-evolution of a quadrupolar nucleus in solid-state MAS experiments \cite{SANCTUARY199079,mangen,Eden20091}.

From a theoretical standpoint, the major contributing factor to the complexity arises from the time-dependent nature of the spin interactions \cite{mmehring,slichter2013principles,Jerschow200563,manqi}. Although, the spin interactions in the solid-state are anisotropic and time-independent to begin with, the introduction of sample rotation along with multiple pulses renders the spatial and spin parts of the interaction Hamiltonians time-dependent, respectively\cite{mmehring,haeberlen2012high}. Consequently, simulations of NMR experiments based on numerical methods have become indispensable in the solid state. With the continued increase in the complexity of NMR techniques and its extended applications, numerical simulations have become an integral part of modern NMR research methodology \cite{BAK2000296,Bak2011366,Veshtort2006248}. Employing numerical simulations, optimal parameters for a given experiment are deduced by trial and error, resulting in the development of sophisticated experiments both in the solution and solid state \cite{BAK2000296,Bak2011366,Veshtort2006248}. Nevertheless, understanding the nuances of the underlying spin physics is quintessential to the design of new pulse sequences besides extending the range of applications of NMR spectroscopy. Since extraction of molecular constraints in NMR experiments involves iterative fitting of the experimental data, simplified analytic expressions that are computationally efficient are essential. Additionally, the accuracy of the analytic expressions needs to be constantly validated through simulations from exact numerical methods. 

Here in this article, we confine our discussion towards the development of analytic methods for describing NMR experiments involving quadrupolar nuclei. Specifically, we focus our attention to the single-pulse based excitation of multiple-quantum (MQ) transitions in spin I=$\rfrac{3}{2}$ nucleus in static solids. Although, the effects of an RF pulse on a spin I=$\rfrac{1}{2}$ system is well understood \cite{mmehring,slichter2013principles}, the same does not hold true with regard to a quadrupolar spin. From an operational point of view, the main complexity arises in the description of the time-evolution of the spin system during an RF pulse. In the description of spin I=$\rfrac{1}{2}$ systems, the amplitude of the RF pulse often exceeds the magnitude of the internal spin interactions. Hence, the time-evolution of the spin system during an RF pulse is approximately governed by the RF interaction and is conveniently described through rotation operators \cite{mmehring,slichter2013principles}. By contrast, in the case of quadrupolar spins, the magnitude of the quadrupolar interaction (described in terms of the quadrupolar coupling constant) often exceeds the available RF amplitudes besides other internal spin interactions. Hence, during an RF pulse, a quadrupolar nucleus evolves under both the RF and the quadrupolar interaction Hamiltonians. Consequently, the rotation operators employed in the description of spin I=$\rfrac{1}{2}$ nuclei are redundant in the description of quadrupolar nuclei.

\begin{center}
\begin{figure}[H]
    \centering
     \includegraphics[width=0.6\textwidth]{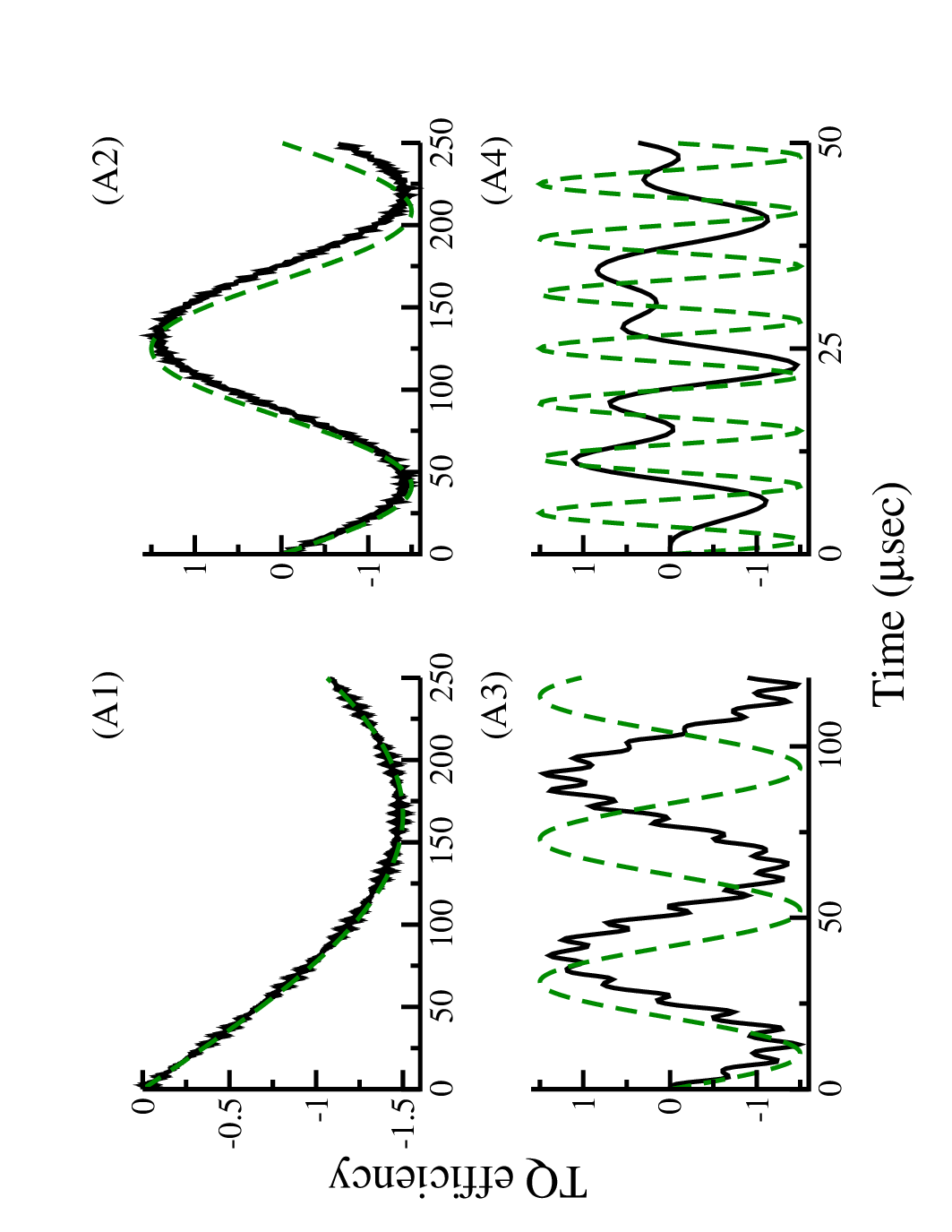}  
\caption{Simulations depicting the efficiency of triple quantum (TQ) excitation in static I=$\rfrac{3}{2}$ system (single crystal) derived from analytic \cite{doi:10.1063/1.441857}  (green dotted lines) and numerical (black thick lines) methods. In the simulations depicted, the quadrupole coupling constant $\left(C_{Q}=\rfrac{\omega_{Q}}{\pi}\right)$ is varied A1) $C_{Q}=2$ MHz, A2) $C_{Q}=1$ MHz, A3) $C_{Q}=500$ kHz, A4) $C_{Q}=200$ kHz, employing an excitation pulse of constant RF amplitude, $(\rfrac{\omega_{1}}{2\pi})=100$ kHz.}       
\label{fig:vegahr}
    \end{figure} 
\end{center}

To this end, Vega and Naor \cite{doi:10.1063/1.441857} developed a theoretical framework for describing MQ transitions in spin I=$\rfrac{3}{2}$ system. Employing the fictitious spin operator algebra \cite{doi:10.1063/1.435679,abragam1961principles}, an analytic expression describing the excitation of triple-quantum (TQ) transitions in static single crystals was proposed in 1980. Below, in Figure.~\ref{fig:vegahr}, numerical simulations (solid lines, emerging from SIMPSON \cite{BAK2000296}) depicting the dependence of the triple-quantum (TQ) excitation efficiency on the quadrupolar coupling constants ($C_{Q}$) is presented along with a comparison of the simulations emerging from their analytic expressions.  As depicted, the simulations emerging from their analytic results agree only in the strong coupling regime ($C_{Q}>(\rfrac{\omega_{1}}{2\pi})$) and deviate when the magnitude of the quadrupolar constant approaches to that of the amplitude of the RF pulse.
Since the magnitude of the quadrupolar frequency ($C_{Q}$) is always greater than both the RF amplitude and the internal spin interactions, the deviations reported above have often been ignored and remain unexplained. Additionally, the extension of their approach for describing the excitation of MQ transitions in a powder (anisotropic) sample is less straightforward and has remained elusive for almost four decades\cite{KWAK200171}. In an alternate formulation, Nielsen and coworkers\cite{NIELSEN1992205} proposed an approach based on numerical methods for understanding the excitation process in isotropic and anisotropic solids. In their approach, the evolution of the spin system during the pulse was evaluated numerically using time-ordered integration of the evolution operator comprising of the quadrupolar and the RF Hamiltonians. Although, their approach yields results in agreement with exact numerical methods, the methodology employed involves diagonalization of matrices and is computationally less efficient when integrated with iterative fitting routines for extracting molecular constraints from experimental data. Additionally, lack of physical insights into the excitation process limits the utility of such methods.

As an alternative to these existing frameworks, an analytic method based on the concept of effective Floquet Hamiltonians \cite{doi:10.1063/1.1354147,C3RA45195J,C2CP43103C,C3CP44296A} is proposed to explain the nuances of the excitation of TQ transitions in both isotropic and anisotropic solids. Although, understanding the excitation process in spinning samples is experimentally more relevant, from a theoretical perspective, addressing the deviations observed in the static case is a prerequesite to developing sophisticated models for describing the excitation process in solid state MAS experiments involving quadrupolar nuclei. For illustrative purposes, the calculations presented in this article are confined only to static cases. The proposed effective Floquet Hamiltonians are derived systematically from the contact transformation procedure \cite{aliev1,aliev2,papousek1982molecular,PhysRev.33.467}. Although, effective Floquet Hamiltonians have found their importance in the description of solid-state NMR experiments involving spin I=$\rfrac{1}{2}$ nuclei, \cite{doi:10.1063/1.1875092, doi:10.1063/1.2216711, doi:10.1063/1.4794856, doi:10.1080/00268976.2011.572088} their utility in the description of the excitation profile in quadrupolar systems is less realized \cite{C3RA45195J, C2CP43103C, C3CP44296A}. To this end, a unified approach that is suitable for describing both isotropic and anisotropic systems is presented in this article. The deviations observed in Figure.~\ref{fig:vegahr} forms the major emphasize of the present article and are addressed in detail. The importance of the interaction frames and their role in the convergence of the perturbation corrections employed in the derivation of effective Floquet Hamiltonians is discussed extensively through comparisons with analytic and numerical simulations. Additionally, need for the "hybrid method" based on the concept of effective Floquet Hamiltonians derived from different interaction frames is discussed to quantify the excitation profiles observed in anisotropic solids. To substantiate the validity of the proposed analytic approach, simulations emerging from the proposed effective Hamiltonians are compared with simulation results emerging from exact numerical methods in all-possible regimes. To present a pedagogical description of the results obtained, the article is organized as illustrated below.

In section II, a detailed description of the basic theory along with simulations is presented both for static isotropic (single crystal) and anisotropic (powder) solid samples. The convergence of the perturbation corrections are discussed extensively and explained through analytic simulations of the excitation profiles. The importance of the "hybrid method" in the description of the excitation process in static anisotropic solids is discussed along with a comparison of the simulations emerging from the proposed analytic hybrid method and exact numerical methods. A brief summary of the present study along with possible extensions is discussed in the final section.
\clearpage
\section{Theory and Simulations}
To understand the response of a quadrupolar nuclei under RF pulses, the truncated Hamiltonian (non-commuting terms with respect to the Zeeman interaction are ignored under secular approximation) described in the laboratory frame (see Eq.~\ref{eq:geq}) is transformed into a frame of reference, wherein, the dominant contributions (arising) due to the Zeeman ($H_{z}$) and quadrupolar ($H_{Q}$) interactions are absent.
\begin{align}
H_{lab}(t)&=\underbrace{-\hbar  \omega_{0}   I_{z}}_{H_{z}}\underbrace{-2 \hbar \omega_{1} \cos(\omega t - \phi_{1}) \ I_{x}}_{H_{RF}}\underbrace{-\hbar \Omega_{Q} T^{(2)0}}_{H_{Q}}
\label{eq:geq}
\end{align}
In Eq.~\ref{eq:geq}, `$\omega_{0}$' represents the Larmor frequency, `$\Omega_{Q}$' the quadrupolar frequency (in angular frequency units). The effect of the RF pulse is  represented by the Hamiltonian `$H_{RF}$' and is often characterized in terms of it's amplitude `$\omega_{1}$'(rad/s), phase `$\phi_{1}$' and oscillating frequency `$\omega$'(rad/s).\\
The initial step involves the transformation into the Zeeman interaction frame ($U_{1}=e^{-\left(i \omega_{0}t\right)  I_{z}}$ )
\begin{align}
\tilde{H}(t)&=U_{1}\ H_{lab}(t)\ U_{1}^{-1} = e^{-\left(i \omega_{0}t \right) I_{z}}\ {H}_{lab}(t)\ e^{\left(i \omega_{0}t\right) I_{z}} \nonumber \\ 
&=-\hbar  \omega_{1} \left(i\sqrt{\dfrac{5}{2}} \right) \left\lbrace \left(e^{i \left(\omega-\omega_{0}\right)t}+e^{-i \left(\omega+\omega_{0}\right)t} \right) \Phi_{1} T^{(1)1} \right. \nonumber \\
& \hspace{3cm} \left. -\left(e^{i \left(\omega+\omega_{0}\right)t}+e^{-i \left(\omega-\omega_{0}\right)t} \right) \Phi_{1}^{-1} T^{(1)-1} \right\rbrace  -\hbar \Omega_{Q} T^{(2)0} 
\label{eq:zeeint}
\end{align}
where, $\Phi_{1}^{n}=e^{-in\phi_{1}} $ denotes the phase factor of the pulse.\\
In the Zeeman interaction frame, the quadrupolar interaction (to first order) is invariant and the RF Hamiltonian acquires an additional time-dependent phase factor due to the Larmor frequency `$\omega_{0}$'. Depending on the nature of the sample, the form of `$\Omega_{Q}$' varies. For example, in the case of a single crystal, $\Omega_{Q}=\omega_{Q}$, (`$\omega_{Q}$'(rad/s) represents the quadrupolar frequency and is related to the quadrupolar coupling constant `$C_{Q}$'(Hz), (i.e., $\omega_{Q}=\rfrac{3 (2\pi) C_{Q}}{2I(2I-1)} \quad ; \quad C_{Q}=\rfrac{e^{2}Qq}{h}$); while in a powder sample, the quadrupolar interaction is anisotropic and is represented by, $\Omega_{Q}=\omega_{Q}^{(\alpha \beta \gamma)}$.
The orientation dependence of the quadrupolar interaction is represented by, `$\omega_{Q}^{(\alpha \beta \gamma)}$' 
\begin{align}
\omega_{Q}^{(\alpha \beta \gamma)}=\omega_{Q} \left\lbrace D_{0,0} (\Omega_{PL}) +\dfrac{\eta}{\sqrt{6}} \left( D_{-2,0} (\Omega_{PL})+D_{2,0} (\Omega_{PL})  \right) \right\rbrace
\end{align}
where, `$D (\Omega_{PL})$' represents the Wigner rotation matrix\cite{mmehring} that essentially describes the transformation from the Principal axis (PAS) to the laboratory axis. In the case of a static powder sample, the transformation from the PAS to lab frame is derived through two sets of Euler angles $\Omega_{PM}=\left(\alpha_{PM}, \beta_{PM}, \gamma_{PM}\right)$ and $\Omega_{ML}=\left(\alpha_{ML}, \beta_{ML}, \gamma_{ML}\right)$.
\begin{align}
D_{q,0} (\Omega_{PL})=\sum_{q_{1}=-2}^{2} D_{q,q_{1}} (\Omega_{PM})\  D_{q_{1},0} (\Omega_{ML})
\end{align}
The Euler angles $(\Omega_{PM})$ relating the PAS to the molecular axis are unique and identical for all the crystallites present in a powder sample. The transformation from the molecular axis to the laboratory axis is orientation dependent (varies for each crystallite and is represented by $\Omega_{ML}$). In the case of a spinning sample, an additional transformation defining the orientation of the rotor axis with respect to lab axis is defined and the Hamiltonian becomes periodically time-dependent.
\begin{align}
D_{q,0} (\Omega_{PL})=\sum_{q_{1},q_{2}=-2}^{2} D_{q,q_{1}} (\Omega_{PM})\ D_{q_{1},q_{2}} (\Omega_{MR})\ D_{q_{2},0} (\Omega_{RL})
\end{align}
where, `$D (\Omega_{RL})=\left(\omega_{r}t, \beta_{m}, 0 \right)$' represents the time-dependent transformation from the rotor axis to the lab-frame (where $\omega_{r}$ denotes the sample spinning frequency and $\beta_{m}$ the magic angle).\\
Depending on the magnitude of the quadrupolar interaction ($\omega_{Q}$) relative to the amplitude of the RF pulse ($\omega_{1}$), the Hamiltonian in the Zeeman interaction frame is further transformed. Based on the above criteria, the following regimes are identified and discussed for both single crystal and powder samples under static conditions.
\subsection{Single Crystal ($\Omega_{Q}=\omega_{Q}$)}
\subsubsection{Strong coupling ($\omega_{Q} >> \omega_{1}$)}
\begin{center}
\begin{figure}[H] 
    \centering
     \includegraphics[width=1.0\textwidth]{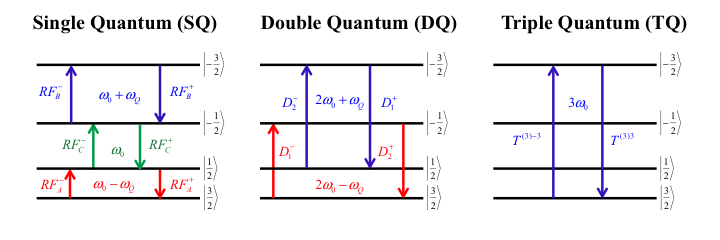}    
\caption{Schematic depiction of transitions along with operators for a spin I=$\rfrac{3}{2}$ system}
\label{fig:tenopt}
    \end{figure} 
\end{center}

When the magnitude of the quadrupolar coupling constant exceeds the amplitude of the pulse, `$\omega_{1}$', the Hamiltonian in the Zeeman interaction frame\cite{C2CP43103C} (Eq.~\ref{eq:zeeint}) is further transformed into the quadrupolar interaction frame, defined below.
\begin{align}
\tilde{\tilde{H}}(t)&=U_{2}\  \tilde{H}(t) \  U_{2}^{-1} =e^{-\left(i \omega_{Q}t\right) T^{(2)0}}\ \tilde{H}(t)\ e^{\left(i \omega_{Q}t\right) T^{(2)0}}\\
&=\tilde{\tilde{H}}_{Q}+\tilde{\tilde{H}}_{RF}(t)
\end{align}
In the combined Zeeman-Quadrupolar interaction frame, the quadrupolar interaction acquires an offset term defined by `$\Delta$', (i.e., $\Delta =\omega_{Q}-\Omega_{Q}$) 
\begin{align}
\tilde{\tilde{H}}_{Q}&=\hbar \Delta T^{(2)0}
\end{align}
Consequently, the offset is zero ($\Delta=\omega_{Q}-\omega_{Q}=0$) for a single crystal and is orientation dependent for a powder sample ($\Delta=\omega_{Q}-\omega_{Q}^{(\alpha \beta  \gamma)}$). \\
To understand the effect of an RF pulse, the RF Hamiltonian ($\tilde{\tilde{H}}_{RF}$) in the Zeeman-Quadrupolar interaction frame is grouped in terms of operators depicting the possible transitions specific to a given spin system\cite{Vinay2016123}. A schematic depiction of the possible transitions along with their frequencies and operators is illustrated in Figure.~\ref{fig:tenopt}. 
Employing secular approximation, a truncated form of the RF Hamiltonian is derived as represented below.
\begin{align}
\tilde{\tilde{H}}_{RF}(t)=&-\hbar \omega_{1} \left(\Phi_{1}RF_{C}^{+} \ e^{i \left(\omega- \omega_{0}\right)t}+\Phi_{1}^{-1}RF_{C}^{-}\ e^{-i \left(\omega- \omega_{0}\right)t} \right)  \nonumber \\
& -\dfrac{\hbar \omega_{1}}{2}\left\lbrace \left(\Phi_{1}RF_{A}^{+} \ e^{i \left(\omega- \left(\omega_{0}-\omega_{Q}\right)\right)t} +\Phi_{1}^{-1}RF_{B}^{-}\ e^{-i \left(\omega- \left(\omega_{0}+\omega_{Q}\right)\right)t} \right)  \right. \nonumber \\
& \left. +\left(\Phi_{1}^{-1}RF_{A}^{-} \ e^{-i \left(\omega- \left(\omega_{0}-\omega_{Q}\right)\right)t} +\Phi_{1}RF_{B}^{+}\ e^{i \left(\omega- \left(\omega_{0}+\omega_{Q}\right)\right)t} \right)  \right\rbrace 
\label{eq:hinqi}
\end{align}
A detailed description of the operators along with their relationship to the spherical tensor operators \cite{Vinay2016123} is tabulated in Table.~\ref{tab:def} 
\begin{table}[H]
\begin{center}
\caption{ Definition of the spin operators corresponding to the possible transitions in a spin $I=\rfrac{3}{2}$ system} \label{tab:def}
\resizebox{10.00cm}{!} {
\begin{tabular}{||c|c|c|}
\hline 
Operator & Tensoral Operators & Frequency\\ 
\hline \hline
 \multicolumn{3}{||c|}{{\textbf{Zero Coherence Operators}}}  \\ \hline 
 & & \\
$ZQ_{A}$  &$  \dfrac{i}{\sqrt{5}}  \left( T^{(1)0}-2T^{(3)0} \right) $ & \\ [.5cm]
$ZQ_{B}$  &$  \dfrac{i}{\sqrt{5}}  \left( 2T^{(1)0}+T^{(3)0} \right) $ & \\ [.5cm]
$ZQ_{C}$ &$\dfrac{i}{\sqrt{5}}  \left( T^{(1)0}+3T^{(3)0} \right)$& \\ [.5cm]
$ZQ_{D}$ &$\dfrac{i}{\sqrt{5}}  \left( 3T^{(1)0}-T^{(3)0} \right)$ & \\ [.5cm]
 \hline
 \multicolumn{3}{||c|}{{\textbf{+1 Coherence Operators}}}  \\ \hline
 & & \\
$RF^{+}_{A}$ & $i \dfrac{3}{\sqrt{10}}\ T^{(1)1}+\sqrt{\dfrac{3}{2}}\ T^{(2)1}-i \sqrt{\dfrac{3}{5}}\ T^{(3)1}$ & $\omega=\omega_{0}-\omega_{Q}$ \\  [.5cm]
$RF^{+}_{B}$ & $i \dfrac{3}{\sqrt{10}}\ T^{(1)1}-\sqrt{\dfrac{3}{2}}\ T^{(2)1}-i \sqrt{\dfrac{3}{5}}\ T^{(3)1} $ & $\omega=\omega_{0}+\omega_{Q}$\\ [.5cm]
$RF^{+}_{C}$ & $ i \sqrt{\dfrac{2}{5}}\ T^{(1)1}+i \sqrt{\dfrac{3}{5}}\ T^{(3)1} $  & $\omega=\omega_{0}$ \\ [.5cm]
\hline  \multicolumn{3}{||c|}{{\textbf{-1 Coherence Operators}}}  \\ \hline
& & \\
$RF^{-}_{A}$ & $-i \dfrac{3}{\sqrt{10}}\ T^{(1)-1}-\sqrt{\dfrac{3}{2}}\ T^{(2)-1}+i \sqrt{\dfrac{3}{5}}\ T^{(3)-1}$ & $\omega=\omega_{0}-\omega_{Q}$\\ [.5cm]
$RF^{-}_{B}$ & $-i \dfrac{3}{\sqrt{10}}\ T^{(1)-1}+\sqrt{\dfrac{3}{2}}\ T^{(2)-1}+i \sqrt{\dfrac{3}{5}}\ T^{(3)-1} $ & $\omega=\omega_{0}+\omega_{Q}$\\ [.5cm]
$RF^{-}_{C}$ & $ -i \sqrt{\dfrac{2}{5}}\ T^{(1)-1}-i \sqrt{\dfrac{3}{5}}\ T^{(3)-1} $  &$\omega=\omega_{0}$ \\  [.5cm] 
\hline  \multicolumn{3}{||c|}{{\textbf{+2 Coherence Operators}}}  \\ \hline
& & \\
$D^{+}_{1}$ & $ T^{(2)2}+i\ T^{(3)2}$ & $\omega=2\omega_{0}+\omega_{Q}$ \\ [.5cm]
$D^{+}_{2}$ & $ T^{(2)2}-i\ T^{(3)2} $ & $\omega=2\omega_{0}-\omega_{Q}$\\ [.5cm]
\hline  \multicolumn{3}{||c|}{{\textbf{-2 Coherence Operators}}}  \\ \hline 
& & \\
$D^{-}_{1}$ & $ T^{(2)-2}-i \ T^{(3)-2}$  & $\omega=2\omega_{0}-\omega_{Q}$\\ [.5cm]
$D^{-}_{2}$ & $ T^{(2)-2}+i \ T^{(3)-2} $ & $\omega=2\omega_{0}+\omega_{Q}$ \\  [.5cm]
\hline  \multicolumn{3}{||c|}{{\textbf{+3 Coherence Operators}}}  \\ \hline
& & \\
$T^{+} $ & $T^{(3)3}$ & $\omega=3\omega_{0}$ \\ [.5cm]
\hline  \multicolumn{3}{||c|}{{\textbf{-3 Coherence Operators}}}  \\ \hline 
& & \\
$T^{-} $ & $T^{(3)-3}$ & $\omega=3\omega_{0}$   \\ [.5cm]
\hline
\end{tabular}
}
\end{center}
\end{table}
When the oscillating (or carrier) frequency `$\omega$' of the RF pulse is set to `$\omega_{0}$', the RF Hamiltonian (Eq.~\ref{eq:hinqi}) reduces to a set of operators comprising of time-dependent and time-independent terms. The operators corresponding to the central transition (see Figure.~\ref{fig:tenopt}) are associated with time-independent coefficients, while the operators corresponding to the satellite transitions are identified through time-dependent terms. The combined Hamiltonian in the Zeeman-Quadrupolar interaction frame is represented by,
\begin{align}
 \tilde{\tilde{H}}&(t)= \hbar \Delta T^{(2)0} -\hbar \omega_{1} \left(\Phi_{1}RF_{C}^{+}+\Phi_{1}^{-1}RF_{C}^{-}\right) \nonumber \\
& -\dfrac{\hbar \omega_{1}}{2}\left\lbrace \left(\Phi_{1}RF_{A}^{+}+\Phi_{1}^{-1}RF_{B}^{-}\right) e^{i \omega_{Q}t}
 +\left(\Phi_{1}^{-1}RF_{A}^{-}+\Phi_{1}RF_{B}^{+}\right) e^{-i \omega_{Q}t} \right\rbrace
\label{eq:hst} 
\end{align}
\begin{table}[H]
\begin{center}
\caption{ Symmetric and Anti-symmetric combination of spin operators employed in spin $I=\rfrac{3}{2}$ system} \label{tab:symasym}
\resizebox{12cm}{!} {
\begin{tabular}{||c|c||c|c|}
\hline 
Operator & Combination  &Operator & Combination \\
\hline  \hline
&  & & \\ 
$\hat{CT}_{S}$ & $ \left(\Phi_{1}RF_{C}^{+} + \Phi_{1}^{-1}RF_{C}^{-}\right) $  &
$\hat{CT}_{AS}$ & $ \left(\Phi_{1}RF_{C}^{+} - \Phi_{1}^{-1}RF_{C}^{-}\right) $\\ [.3cm]
$\hat{ST}_{S}^{(r)}$ & $ \left(\Phi_{1}^{-1}RF_{A}^{-} + \Phi_{1}RF_{B}^{+}\right) $  &
$\hat{ST}_{AS}^{(r)}$ & $ \left(\Phi_{1}^{-1}RF_{A}^{-} - \Phi_{1}RF_{B}^{+}\right) $ \\ [.3cm]
$\hat{ST}_{S}^{(cr)}$ & $ \left(\Phi_{1}RF_{A}^{+} + \Phi_{1}^{-1}RF_{B}^{-}\right) $  &
$\hat{ST}_{AS}^{(cr)}$ & $ \left(\Phi_{1}RF_{A}^{+} - \Phi_{1}^{-1}RF_{B}^{-}\right) $ \\ [.3cm]
$\hat{D}_{S}^{(r)}$ & $ \left(\Phi_{1}^{2} D_{1}^{+} + \Phi_{1}^{-2} D_{1}^{-}\right) $  &
$\hat{D}_{AS}^{(r)}$ & $ \left(\Phi_{1}^{2} D_{1}^{+} - \Phi_{1}^{-2} D_{1}^{-}\right) $\\ [.3cm]
$\hat{D}_{S}^{(cr)}$ & $ \left(\Phi_{1}^{2} D_{2}^{+} + \Phi_{1}^{-2} D_{2}^{-}\right) $  &
$\hat{D}_{AS}^{(cr)}$ & $ \left(\Phi_{1}^{2} D_{2}^{+} - \Phi_{1}^{-2} D_{2}^{-}\right) $  \\ [.3cm]
$\hat{T}_{S}$ & $ \left(\Phi_{1}^{3} T^{+} + \Phi_{1}^{-3} T^{-}\right) $  &
$\hat{T}_{AS}$ & $ \left(\Phi_{1}^{3} T^{+} - \Phi_{1}^{-3} T^{-}\right) $  \\ [.3cm]
\hline
\end{tabular}
}
\end{center}
\end{table}
The time-dependent phase factor due to `$e^{\pm i \omega_{Q} t}$' is further classified into rotating ($e^{-i\omega_{Q}t}$) and counter rotating terms ($e^{+i\omega_{Q}t}$) .
To elucidate the mechanisms of MQ excitation, the Hamiltonian (Eq.~\ref{eq:hst}) is re-expressed in terms of the symmetric and anti-symmetric combination of spin operators (see Table. ~\ref{tab:symasym}). Along with the quadrupolar offset term, the Hamiltonian describing an RF pulse is represented by,
\begin{align}
\label{eq:tdh}
&H_{pulse} (t)=\tilde{\tilde{H}}_{Q,off}+\tilde{\tilde{H}}_{CT}+\tilde{\tilde{H}}_{ST}(t) 
\intertext{where,}
&\tilde{\tilde{H}}_{Q,off}= \hbar \Delta T^{(2)0} \\
&\tilde{\tilde{H}}_{CT}= -\hbar \omega_{1} \left(\hat{CT}_{S}\right) \\
&\tilde{\tilde{H}}_{ST}(t)=-\dfrac{\hbar \omega_{1}}{2}\left\lbrace \left(\hat{ST}_{S}^{(cr)} \right) e^{i \omega_{Q}t}
 +\left(\hat{ST}_{S}^{(r)}\right) e^{-i \omega_{Q}t} \right\rbrace \\
 &\hspace{1.3cm}=\tilde{\tilde{H}}_{ST}^{(cr)}(t)+\tilde{\tilde{H}}_{ST}^{(r)}(t)
\end{align}
In the above equation, `$\tilde{\tilde{H}}_{CT}$' represents the Hamiltonian depicting the central transition and is expressed through the symmetric combination of tensor operators (see Table.~\ref{tab:symasym}). The time-dependent Hamiltonian representing the satellite transition ($\tilde{\tilde{H}}_{ST}(t)$) is expressed through a symmetric combination of tensor operators corresponding to both rotating and counter rotating components of the satellite transitions. Since analytic descriptions of spectroscopic phenomena under time-dependent Hamiltonians is less straightfarward, an alternate framework based on Floquet theory \cite{PhysRev.138.B979} is employed in the present study to explicate the nuances of the excitation of MQ transitions. Employing Floquet theorem, the time-dependent Hamiltonian (Eq.~\ref{eq:tdh}) is transformed into a time-independent Floquet Hamiltonian and is represented through a set of operators (Floquet operators) defined in an infinite-dimensional vector space. Accordingly, in the Floquet framework, the Hamiltonian depicting an RF pulse is represented by,
\begin{align}
 H_{F}=\ \omega_{Q}I_{F} + \hbar  \Delta  \left(T^{(2)0}\right)_{0} - \hbar \omega_{1} \left(\hat{CT}_{S}\right)_{0} 
 -\dfrac{\hbar \omega_{1}}{2}\left\lbrace \left(\hat{ST}_{S}^{(cr)}\right)_{-1} +\left(\hat{ST}_{S}^{(r)}\right)_{+1} \right\rbrace 
 \label{Floquetham}
\end{align}
In the above equation `$I_{F}$' represents the identity operator defined in the Floquet space, 
\begin{gather}
I_{F}= N \otimes I \quad ; \quad 
N = \sum_{n=-\infty}^{\infty} n  \ket{n} \bra{n} 
\label{Floqop1}
\end{gather}
where, `I' is the identity operator. The Floquet operators `$(I_{\alpha})_{m}$' are constructed from a direct product between the Fourier `$F_{m}$' and spin `$I_{\alpha}$' operators
\begin{gather}
(I_{\alpha})_{m}= F_{m} \otimes I_{\alpha} \quad  ; \quad 
F_{m} = \sum_{n=-\infty}^{\infty}   \ket{n} \bra{n+m} 
\label{Floqop2}
\end{gather}
A detailed description and derivation of the Floquet operators are well-documented and have been omitted here to avoid repetition\cite{doi:10.1063/1.1875092,rajat}. It is important to realize here that the Floquet Hamiltonian derived above (Eq.~\ref{Floquetham}) is defined in an infinite-dimensional vector space and only the non-zero terms associated with the Floquet operators are illustrated in Eq.~\ref{Floquetham} . Consequently, analytic descriptions have always remained elusive. To alleviate this problem, the concept of `Effective Floquet Hamiltonians' based on the Contact tranformation method \cite{aliev1,aliev2,papousek1982molecular,PhysRev.33.467} is employed in the present study. Although,the utility of the effective Floquet Hamiltonians in the description of spin $I=\rfrac{1}{2}$ systems is known, the application of this approach has not been extended to the description of quadrupolar nuclei. This could largely be due to the presence of dominant contributions arising from the quadrupolar Hamiltonian. Since the proposed effective Floquet Hamiltonians are derived from the contact transformation procedure (an operator equivalent of perturbation theory), the definition of the zero order and perturbing Hamiltonian, is crucial in the overall convergence of the perturbation corrections. Depending on the magnitudes of the interaction parameters (such as quadrupolar coupling constant, RF amplitude, offsets etc.), the definition of the zero order and perturbing Hamiltonian is problem specific and forms the basis for this article.\\
When the magnitude of the quadrupolar interaction (expressed here in terms of the quadrupolar frequency) exceeds the amplitude of the excitation pulse ($\omega _{1}$), the choice of the zero order and perturbing Hamiltonian plays an important role in the description of the excitation process.
To begin with, the Floquet operators that are diagonal `$(I_{\alpha})_{0}$' are retained along the zero order Hamiltonian, while, the off-diagonal operators are included along `$H_{1}$'.
\begin{gather}
H_{F}=H_{0}+H_{1}  \\
H_{0}=\omega_{Q}I_{F} + \hbar  \Delta  \left(T^{(2)0}\right)_{0} 
\end{gather}
The perturbing Hamiltonian in the present study is chosen to contain both diagonal and off-diagonal terms.
\begin{gather}
H_{1}=H_{1,d}+H_{1,od} \nonumber\\
H_{1,d}=- \omega_{1} \left( \hat{CT}_{S} \right)_{0} ;\quad H_{1,od}= -\dfrac{\omega_{1}}{2}\left\lbrace \left( \hat{ST}_{S}^{(r)} \right)_{+1} +
 \left( \hat{ST}_{S}^{(cr)} \right)_{-1}  \right\rbrace 
\end{gather}
Such a choice of the grouping of the spin Hamiltonians is problem specific and its validity could only be verified through a comparison with exact numerical simulations. In the contact transformation procedure, the original Floquet Hamiltonian is transformed through a single or a series of unitary transformations. The choice of this procedure is again problem specific and would be described in detail in this section. Employing the transformation function `$S_{1}$' the original untransformed Floquet Hamiltonian (Eq.~\ref{Floquetham}) is transformed through a unitary transformation illustrated below.
\begin{gather}
H_{eff}= e^{i \lambda S_{1}} \ H_{F}\  e^{-i \lambda S_{1}} \\
S_{1}=\ C_{ST}^{(1)} \left\lbrace  \left( \hat{ST}_{S}^{(r)} \right)_{+1} - \left( \hat{ST}_{S}^{(cr)} \right)_{-1}  \right\rbrace 
\label{eq:s1tr}
\intertext{where,}
C_{ST}^{(1)}=-i\left(\dfrac{\omega_{1}} {2\Omega_{Q}}\right)
\end{gather}
The transformation function `$S_{1}$' defined in Eq.~\ref{eq:s1tr} is carefully chosen to compensate the off-diagonal terms in `$H_{1}$'(i.e. $H_{1,od}$) and is derived through the procedure \cite{doi:10.1063/1.1354147} given below,
\begin{gather}
H^{(1)}_{1} = H_{1,d}+ H_{1,od}+i [S_{1}, H_{0}] = H_{1,d} \nonumber \\
i [S_{1}, H_{0}] =- H_{1,od}
\end{gather}
Consequently, to first-order, the effective Hamiltonian comprises of only `$H_{1,d}$' 
\begin{align}
H_{1}^{(1)}&=-\omega_{1} \left( \hat{CT}_{S} \right)_{0}
\end{align}
Subsequently, through Baker-Campbell-Hausdorff (BCH) expansion \cite{slichter2013principles}, the higher-order corrections to the effective Hamiltonian are derived and described in detail in Table.~\ref{tab:coneffham}. In the description that  follows, `$H_{n}^{(1)}$' represents the $n^{th}$ order corrections obtained from the first transformation, `$S_{1}$'.
\begin{table}[H]
\caption{Description of higher-order corrections to the effective Hamiltonian derived from BCH expansion}
\centering
\resizebox{13cm}{!} {
\begin{tabular}{||c|l|}
\hline 
\textbf{$n^{th}$ order} &\multicolumn{1}{c|}{}  \\ 
\textbf{Correction} & \multicolumn{1}{c|}{\textbf{Expression for the $n^{th}$ order Correction}}\\ [0.5ex] \hline \hline
& \multicolumn{1}{c|}{} \\
Zero order $(\lambda^{0})$ & $ H^{(1)}_{0} = H_{0} $ \\ [2.0ex]
I order $(\lambda^{1})$ & $ H^{(1)}_{1} = i [S_{1}, H_{0}]+H_{1} $ \\ [2.0ex]
II order $(\lambda^{2})$ & $ H^{(1)}_{2} = -\dfrac{1}{2!} [S_{1},[S_{1}, H_{0}]]+ i [S_{1}, H_{1}] $ \\ [2.0ex]
III order $(\lambda^{3})$ & $ H^{(1)}_{3} = -\dfrac{i}{3!} [S_{1},[S_{1},[S_{1}, H_{0}]]] - \dfrac{1}{2!} [S_{1},[S_{1}, H_{1}]] $ \\ [2.0ex]
IV order $(\lambda^{4})$ & $ H^{(1)}_{4} = \dfrac{1}{4!} [S_{1},[S_{1},[S_{1},[S_{1}, H_{0}]]]]  -\dfrac{i}{3!} [S_{1},[S_{1},[S_{1}, H_{1}]]] $ \\ [2.0ex]
V order $(\lambda^{5})$ & $ H^{(1)}_{5} = \dfrac{i}{5!} [S_{1},[S_{1},[S_{1},[S_{1},[S_{1}, H_{0}]]]]] +\dfrac{1}{4!} [S_{1},[S_{1},[S_{1},[S_{1}, H_{1}]]]] $ \\ [2.0ex]
VI order $(\lambda^{6})$ & $ H^{(1)}_{6} = -\dfrac{1}{6!} [S_{1},[S_{1},[S_{1},[S_{1},[S_{1},[S_{1}, H_{0}]]]]]]  $ \\ [2.0ex]
 & \multicolumn{1}{c|}{$ + \dfrac{i}{5!} [S_{1},[S_{1},[S_{1},[S_{1},[S_{1}, H_{1}]]]]] $} \\ [2.0ex]
VII order $(\lambda^{7})$ & $ H^{(1)}_{7} = -\dfrac{i}{7!} [S_{1},[S_{1},[S_{1},[S_{1},[S_{1},[S_{1},[S_{1}, H_{0}]]]]]]]  $ \\ [2.0ex]
 & \multicolumn{1}{c|}{$ - \dfrac{1}{6!} [S_{1},[S_{1},[S_{1},[S_{1},[S_{1},[S_{1}, H_{1}]]]]]] $} \\ [2.0ex]
VIII order $(\lambda^{8})$ & $ H^{(1)}_{8} = \dfrac{1}{8!} [S_{1},[S_{1},[S_{1},[S_{1},[S_{1},[S_{1},[S_{1},[S_{1}, H_{0}]]]]]]]]  $ \\ [2.0ex]
 & \multicolumn{1}{c|}{$  -\dfrac{i}{7!} [S_{1},[S_{1},[S_{1},[S_{1},[S_{1},[S_{1},[S_{1}, H_{1}]]]]]]] $} \\ [2.0ex]
IX order $(\lambda^{9})$ & $ H^{(1)}_{9} = \dfrac{i}{9!} [S_{1},[S_{1},[S_{1},[S_{1},[S_{1},[S_{1},[S_{1},[S_{1},[S_{1}, H_{0}]]]]]]]]]  $ \\ [2.0ex]
 & \multicolumn{1}{c|}{$ +\dfrac{1}{8!} [S_{1},[S_{1},[S_{1},[S_{1},[S_{1},[S_{1},[S_{1},[S_{1}, H_{1}]]]]]]]] $} \\ [2.0ex]
X order $(\lambda^{10})$ & $ H^{(1)}_{10} = -\dfrac{1}{10!} [S_{1},[S_{1},[S_{1},[S_{1},[S_{1},[S_{1},[S_{1},[S_{1},[S_{1},[S_{1}, H_{0}]]]]]]]]]]  $ \\ [2.0ex] 
& \multicolumn{1}{c|}{$ +\dfrac{i}{9!} [S_{1},[S_{1},[S_{1},[S_{1},[S_{1},[S_{1},[S_{1},[S_{1},[S_{1}, H_{1}]]]]]]]]] $} \\ [2.0ex] \hline
\end{tabular}
}
\label{tab:coneffham} 
\end{table}
In general, the higher order corrections to the effective Hamiltonian comprises of both diagonal and off-diagonal contributions. As a standard procedure, the diagonal corrections are often retained, while neglecting the off-diagonal terms. Nevertheless, the validity of such approximations could only be verified through a rigorous comparison of simulations emerging from analytic (based on effective Hamiltonians) and numerical based exact methods. \\
In the present problem, the higher order corrections mainly arise from commutator expressions involving the transformation function `$S_{1}$' and the perturbing Hamiltonian (`$H_{1,d}$' and `$H_{1,od}$'). The commutator of the transformation function `$S_{1}$' with `$H_{1,d}$' to various orders could be derived through the expression.
\begin{align}
H_{n,d}^{(1)}=\sum_{n=1}^{\infty} \dfrac{(i)^{n-1}}{(n-1)!} \left[ \underbrace{\left[S_{1},...............\left[S_{1} \right. \right.}_{n-1}  ,H_{1,d} {\left. \left. \right].............. .\right]}  \right]
\end{align}
Based on the expressions illustrated in Table.~\ref{tab:coneffham}, the diagonal corrections are represented by $\left(\text{central transition (CT) $\left( \hat{CT}_{S} \right)_{0}$ and triple-quantum (TQ) $\left( \hat{T}_{AS} \right)_{0}$} \right)$ operators, while the off-diagonal contributions are represented through \\
$\left( \text{double-quantum (DQ) $\left( \hat{D}_{S}^{(r,cr)} \right)_{\pm1}$} \right) $ operators. \\
In a similar vein, the commutator of `$S_{1}$' with `$H_{1,od}$' is derived (through the general expression Eq.~\ref{eq:h1od}) and comprises of both diagonal $\left(\text{ zero-quantum (ZQ) $(T^{(2)0})_{0}$} \right)$ and off-diagonal $\left( \text{ single-quantum (SQ) satellites $\left( \hat{ST}_{S}^{(r,cr)} \right)_{\pm1}$} \right) $ operators. 
\begin{align}
H_{n,od}^{(1)}=\sum_{n=2}^{\infty} \dfrac{(i)^{n-1}}{n \times (n-2)!} \left[ \underbrace{\left[S_{1},...............\left[S_{1} \right. \right.}_{n-1} ,H_{1,od} {\left. \left. \right].............. .\right]}  \right]
\label{eq:h1od}
\end{align}
A detailed description of the commutator relations involving the transformation function `$S_{1}$' and `$H_{1}$' to various orders of `$\lambda$' are given in supplementary information. 

To illustrate the importance of the various contributions to the excitation process, a systematic study that includes perturbation correction to the desired order are included in the effective Hamiltonian. To substantiate the proposed method, a comparison between simulations emerging from the effective Hamiltonians and exact numerical simulations based on SIMPSON (a numerical based software for simulating NMR experiments in solid state) is discussed in the following sections.\\
As illustrated in Table. S.1 (see supplementary information), the higher order contributions mainly arise from the diagonal $\left(\text{arising from ZQ} (T^{(2)0})_{0}, \right. \\ \left. \text{CT} \left( \hat{CT}_{S} \right)_{0}  \text{and TQ} \left( \hat{T}_{AS} \right)_{0} \right)$ and off-diagonal $\left( \text{from SQ} \left( \hat{ST}_{S}^{(r,cr)} \right)_{\pm1} \right.  \\ \left. \text{and DQ} \left( \hat{D}_{S}^{(r,cr)} \right)_{\pm1} \right) $ operators. Depending on the magnitude of the off-diagonal contributions, the convergence of the perturbation corrections could in principle, be accomplished through a single or a series of transformations (often termed as contact transformations) as illustrated below.
\begin{align}
H_{eff}=e^{i \lambda^{n} S_{n}}\ ........\ e^{i \lambda^{2} S_{2}}\ e^{i \lambda S_{1}}\ H_{F}\ e^{-i \lambda S_{1}}\ e^{-i \lambda^{2} S_{2}}\ ........ \  e^{-i \lambda^{n} S_{n}}
\end{align}
The first transformation function `$S_{1}$' folds off-diagonal contributions to order `$\lambda$,' while the off-diagonal contributions to order`$\lambda^{2}$' (arising from the residual terms from the first transformation) are folded by a second transformation function `$S_{2}$'. In a similar vein, the third transformation folds off-diagonal corrections to order `$\lambda^{3}$' present in the perturbing Hamiltonian, besides the residual off-diagonal contributions (to order $\lambda^{3}$) resulting from the first and second transformations. Nevertheless, it is important to realize here that the corrections obtained from successive transformations do not alter the results (coefficients) obtained from the previous transformations. A pedagogical description illustrating the role of higher order corrections in the excitation of TQ transitions in spin $I=\rfrac{3}{2}$ is discussed below along with simulations.\\
\textbf{I. Effective Hamiltonians from single transformation, `$S_{1}$'}\\
To begin with, the general form of the effective Hamiltonian (comprising of diagonal corrections only) describing the excitation from a single pulse is represented by, 
\begin{align}
H_{eff}&= e^{i \lambda S_{1}} \ H_{F} \  e^{-i \lambda S_{1}} \nonumber \\ 
&=\omega_{Q}I_{F}
            +G_{CT}^{(1)}  \left( \hat{CT}_{S} \right)_{0} 
            +i\ G_{TQ}^{(1)} \left( \hat{T}_{AS} \right)_{0} 
            +G_{ZQ}^{(1)}  \left(T^{(2)0}\right)_{0} 
            \label{eq:effham1} 
\intertext{where,}
G_{CT}^{(1)} &=\sum_{i=0}^{N_{1}} G_{CT,i}^{(1)}   \quad ; \quad 
G_{TQ}^{(1)} =\sum_{i=0}^{N_{1}} G_{TQ,i}^{(1)}    \quad ; \quad    
G_{ZQ}^{(1)} =\sum_{i=0}^{N_{1}} G_{ZQ,i}^{(1)}  
\end{align}
In Eq.~\ref{eq:effham1}, `$G_{CT}^{(1)} $' denotes the coefficients obtained from the first transformation (denoted by the superscript) corresponding to the central transition operator. The contributions from the various higher orders ($N_{1}$ denotes the desired order, power of $\lambda$) are included in $G_{CT}^{(1)} $. A detailed description of the coefficients illustrating the contributions from various orders is listed in Table. S.2 (see supplementary information).

To have a consistent description, the initial density operator ($\rho_{F}(0)=(I_{z})_{0}$) along with the detection operator `$T^{(3)-3}$,' (corresponding to TQ transition) is transformed by the transformation function `$S_{1}$' (refer Eq. S.1 and Eq. S.2 in the supplementary information). 

Although, from an experimental perspective, TQ transitions cannot be detected through direct means (they need to be reconverted back to detectable SQ(-1) transitions), the TQ excitation efficiency is evaluated through the standard procedure illustrated below.
\begin{align}  
 \left\langle T^{(3)-3} (t_{p1}) \right\rangle  &=Tr \left[\tilde{\rho}_{F}(t_{p1}).\tilde{T}^{(3)-3}_{F} \right]  \nonumber \\
 &=R_{CT}^{(1)}. P_{CT}^{(1)}\ S_{(\theta_{CT})}+2 \ R_{DQ}^{(1)}. P_{DQ}^{(1)} \ S_{(\theta_{RF})} \left( e^{i(\theta_{ZQ})} +e^{-i(\theta_{ZQ})} \right) \nonumber \\
 &+R_{TQ}^{(1)}. P_{TQ}^{(1)}\ S_{(\theta_{TQ})} 
\end{align}
A detailed description of the procedure along with the coefficients is illustrated in the supplementary information.
On further simplification, the `TQ' efficiency observed in experiments is calculated using the expression given below
\begin{mdframed}[innertopmargin=0pt,userdefinedwidth=14.3cm,align=center,middlelinewidth=1pt,roundcorner=10pt] 
\begin{align}
  \left\langle T^{(3)-3} (t_{p1}) \right\rangle=  \left( \Phi_{1}^{3} \right) & 
\left\lbrace -\dfrac{1}{4} \ J_{1}^{\left( \theta \right)}\  J_{2}^{\left( \theta \right)}\  S_{(\theta_{TQ})}   
 +\dfrac{1}{4} \ J_{-1}^{\left( \theta \right)}\  J_{-2}^{\left( \theta \right)}\  S_{(\theta_{CT})} \right. \nonumber \\  
 &\left. -\dfrac{1}{2} \left( S_{(\theta)}\right)^{2} S_{(\theta_{RF})} \ C_{(\theta_{ZQ})} \right\rbrace
\label{eq:lambda3}
\end{align}
\end{mdframed} 
In Eq.~\ref{eq:lambda3}, the notation $J_{\pm n}^{(\theta)}=\cos(\theta)\pm n$ with $ \theta = \left(\dfrac{\sqrt{3}\omega_{1}}{\Omega_{Q}}\right)$ (where $(\Omega_{Q}=\omega_{Q})$) has been employed. $\left(\theta_{CT}= 2 G_{CT}^{(1)} t_{p1} \quad ; \quad \theta_{ZQ}=\left(G_{ZQ}^{(1)}-\omega_{Q} \right)  t_{p1}  \quad ; \quad \theta_{RF}= \omega_{1} t_{p1}  \right. $ \\
$ \left. \theta_{TQ}= 2 G_{TQ}^{(1)} t_{p1} \quad ; \quad C_{(\theta)}=\cos\left(\theta \right) \quad ; \quad S_{(\theta)}=\sin\left(\theta \right)  \quad ; \quad \Phi_{1}^{n}=e^{-in\phi}\right)$ \\ 

When the initial density matrix and the detection operators are untransformed (i.e., ${\rho}_{F}(0)= \left(I_{z} \right)_{0} \quad ; \quad {T}^{(3)-3}_{F}=e^{3i \omega_{0}t_{2}} \ \Phi_{R}  \left\lbrace  \left( T^{(3)-3} \right)_{0} \right\rbrace $), the expression for TQ signal reduces to a much simpler form.
\begin{mdframed}[innertopmargin=0pt,userdefinedwidth=14.3cm,align=center,middlelinewidth=1pt,roundcorner=10pt] 
\begin{align}
  &\left\langle T^{(3)-3} (t_{p1}) \right\rangle=Tr \left[{\rho}_{F}(t_{p1}).{T}^{(3)-3}_{F} \right] = \left( \Phi_{1}^{3} \  \right) \left\lbrace -\dfrac{3}{2} \sin\left(\dfrac{3 \omega_{1}^{3} t_{p1}}{2 \Omega_{Q}^{2}} \right)   \right\rbrace
  \label{eq:t3gen}
\end{align}
\end{mdframed} 
The above equation partially resembles to those proposed by Vega and Naor \cite{doi:10.1063/1.441857} (it is opposite in sign and twice in magnitude to their TQ expression).\\
Although, to first order only the SQ transition operators (corresponding to the central transition) are present in the pulse Hamiltonian, the cross terms between the central transition and satellite transition operators results in the emergence of MQ operators.
As represented in Table. S.1 (refer supplementary information), the diagonal corrections to odd orders results from cross-terms between the SQ satellite and DQ transitions operators and are expressed in terms of the central (CT) and triple quantum (TQ) operators. In a similar vein, the diagonal corrections to even order results from cross terms between the SQ satellite transition operators and are represented through the `$T^{(2)0}$' operator. To illustrate the role of the higher order corrections in the exactness of the proposed approach, a systematic study incorporating their contributions is discussed below. To begin with, the following two cases are considered.\\
In Case-I, diagonal corrections to order $\lambda^{3}$ are included  \\
(i.e. $G_{CT}^{(1)} =\sum\limits_{i=0}^{3} G_{CT,i}^{(1)}   \quad ; \quad G_{TQ}^{(1)} =\sum\limits_{i=0}^{3}G_{TQ,1}^{(1)}    \quad ; \quad    
G_{ZQ}^{(1)} =\sum\limits_{i=0}^{3} G_{ZQ,i}^{(1)} $),\\
while, diagonal corrections to $n^{th}$ order are included in Case-II \\
(i.e. $G_{CT}^{(1)} =\sum\limits_{i=0}^{N_{1}} G_{CT,i}^{(1)}   \quad ; \quad 
G_{TQ}^{(1)} =\sum\limits_{i=0}^{N_{1}} G_{TQ,i}^{(1)}    \quad ; \quad    G_{ZQ}^{(1)} =\sum\limits_{i=0}^{N_{1}} G_{ZQ,i}^{(1)} $).\\
A detailed description of coefficients employed in the effective Hamiltonians for the above two cases are given in Table.~\ref{tab:coeffham1}
\begin{center}
\begin{table}[H]
\caption{Coefficients employed in the derivation of Effective Hamiltonians for Case-I and Case-II}
\centering
\resizebox{13cm}{!} {
\begin{tabular}{||l|c|c|c|}
\hline 
 &$G_{CT}^{(1)}$ &$G_{TQ}^{(1)}$ &$G_{ZQ}^{(1)}$ \\ \hline \hline
 & & & \\
\shortstack{\textbf{Case-I} \\[0.2cm] $\left(N_{1} =3\right)$}&$-\omega_{1}+\dfrac{1}{2!} \left(\dfrac{\omega_{1}}{2} \right)  \left(\theta\right)^{2} $ & $+\dfrac{1}{2!} \left(\dfrac{\omega_{1}}{2} \right)  \left(\theta \right)^{2}$ & $\Delta -\dfrac{\Omega_{Q}}{2}  \left( \theta \right)^{2} $ \\  [.5cm]
\shortstack{\textbf{Case-II} \\[0.2cm] $\left(N_{1} >3\right)$}&$-\dfrac{\omega_{1}}{2}\left\lbrace C_{(\theta)} +1  \right\rbrace$ &$-\dfrac{\omega_{1}}{2}\left\lbrace C_{(\theta)} -1  \right\rbrace$ &$ \Delta+\left(\Omega_{Q} \right) \left\lbrace -\dfrac{1}{2\times 0!} \left(\theta \right)^{2} \right. $  \\ [.5cm]
 & & &$\left. +\dfrac{1}{4\times 2!} \left(\theta \right)^{4} -\dfrac{1}{6\times 4!} \left(\theta \right)^{6} +....\right\rbrace  $ \\ [.5cm]
   & & & \\ \hline
\multicolumn{4}{||c|}{\textbf{}} \\
\multicolumn{4}{||c|}{\textbf{$\theta = \left(\dfrac{\sqrt{3}\omega_{1}}{\Omega_{Q}}\right) \quad ; \quad C_{(\theta)}=\cos\left(\theta \right)$}} \\ 
\multicolumn{4}{||c|}{\textbf{}} \\     \hline 
\end{tabular}
}
\label{tab:coeffham1} 
\end{table}
\end{center}
\begin{center}
\begin{figure}[H]
    \centering
     \includegraphics[width=0.6\textwidth, angle=0]{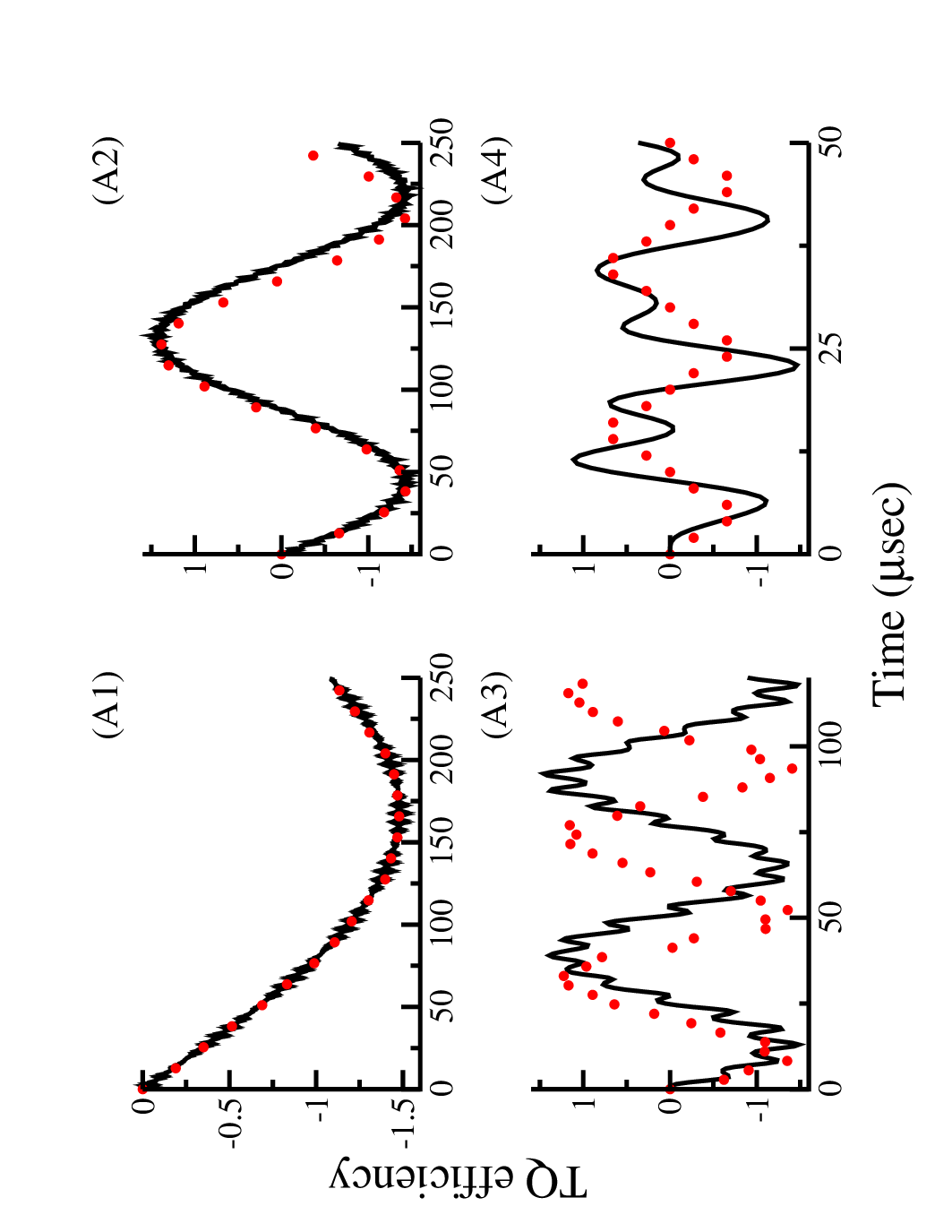}    
\caption{Case-I: Comparison of numerical (black thick line) and analytic simulations (red dots) based on effective Hamiltonians derived from a single transformation comprising of diagonal corrections to order $\lambda^{3} (N_{1}=3)$.  In the simulations depicted, the quadrupole coupling constant $\left(C_{Q}=\rfrac{\omega_{Q}}{\pi}\right)$ is varied A1) $C_{Q}=2$ MHz, A2) $C_{Q}=1$ MHz, A3) $C_{Q}=500$ kHz, A4) $C_{Q}=200$ kHz, employing an excitation pulse of constant RF amplitude, $(\rfrac{\omega_{1}}{2\pi})=100$ kHz. The simulations correspond to a single crystal.}
\label{fig:lambda3}
    \end{figure} 
\end{center}   
In Figure.~\ref{fig:lambda3}, simulations depicting the efficiency of TQ excitation are plotted as a function of the pulse duration ($t_{p1}$). In these simulations, the diagonal corrections to order  `$\lambda^{3}$' are only incorporated (representing Case:I). As depicted, when the magnitude of the quadrupolar frequecy ($\omega_{Q}$) largely exceeds the amplitude of the RF pulse, the analytic simulations are in excellent agreement with those obtained from SIMPSON. In the extreme strong coupling limit, the TQ signal in Eq.~\ref{eq:lambda3} reduces to the familiar form proposed by Vega and Naor \cite{doi:10.1063/1.441857}.
\begin{align}
 \left\langle T^{(3)-3} (t_{p1}) \right\rangle=-\dfrac{3}{2} \sin\left(\dfrac{3 \omega_{1}^{3} t_{p1}}{2 \Omega_{Q}^{2}} \right)
 \label{eq:t3vn}
\end{align}
\begin{center}
\begin{figure}[H]
    \centering
     \includegraphics[width=0.5\textwidth, angle=0]{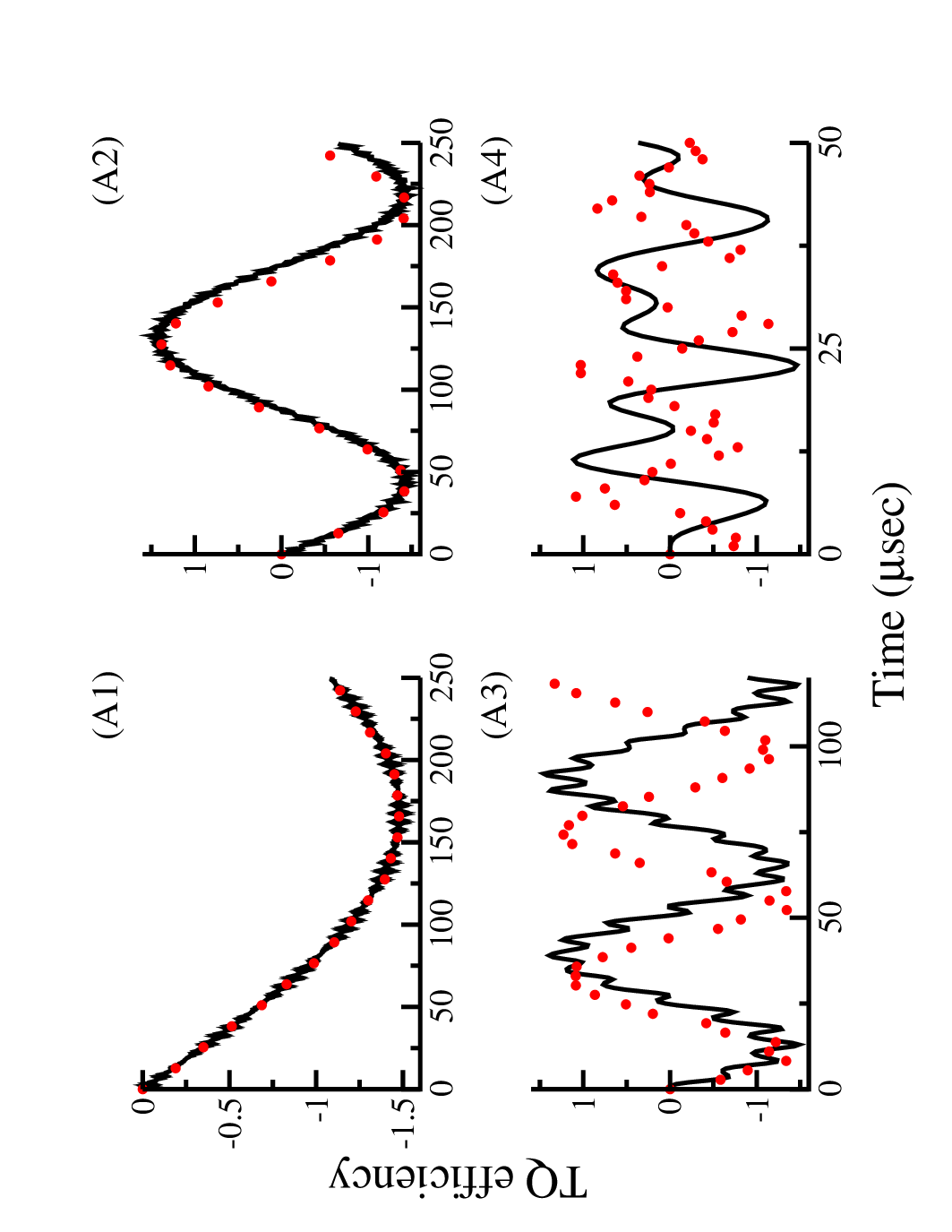}    
\caption{Case-II: Comparison of numerical (black thick line) and analytic simulations (red dots) based on effective Hamiltonians derived from a single transformation comprising of diagonal corrections to $n^{th}$ - order $\lambda^{3} (N_{1}>3)$.  In the simulations depicted, the quadrupole coupling constant $\left(C_{Q}=\rfrac{\omega_{Q}}{\pi}\right)$ is varied A1) $C_{Q}=2$ MHz, A2) $C_{Q}=1$ MHz, A3) $C_{Q}=500$ kHz, A4) $C_{Q}=200$ kHz, employing an excitation pulse of constant RF amplitude, $(\rfrac{\omega_{1}}{2\pi})=100$ kHz. The simulations correspond to a single crystal.}
\label{fig:lambdan}
    \end{figure} 
\end{center}
However, with decreasing magnitudes of the quadrupolar frequency, the discrepancy between the analytic and numerical simulations increases and is maximum when the magnitude of the quadrupolar coupling frequency is equal to the RF amplitude. To address this aspect, effective Hamiltonians comprising of diagonal contributions to ($n^{th}$ order, $N_{1}>3$) were employed in the simulations depicted in Figure.~\ref{fig:lambdan}. As depicted, the discrepancy still prevails in panels A3 and A4, inspite of the inclusion of higher order diagonal corrections. Hence, the residual off-diagonal terms ignored from the first transformation might play an important role in the excitation process. \\
\textbf{II. Effective Hamiltonians from second transformation, `$S_{2}$'} \\
To resolve the discrepancy observed in the analytic simulations, the role of residual off-diagonal terms neglected in the first transformation were considered in the discussion presented below. As depicted in Table. S.1 (refer supplementary information), the off-diagonal contributions comprises of the double-quantum (DQ) and single-quantum (SQ) satellite transitions operators. To fold the above off-diagonal contributions, a second transformation function `$S_{2}$' was employed. A brief description of the procedure employed in the derivation of effective Floquet Hamiltonians from the second transformation `$S_{2}$'  is outlined below. The diagonal corrections from the first transformation are included along `$H_{0}$' and the off-diagonal operators $\left( \left(\hat{ST} \right)_{\pm1} and \left(\hat{D} \right)_{\pm1} \right)$ form the perturbation. \\
\begin{align}
\label{eq:effham1a}  
H_{0}& =\omega_{Q}I_{F}
            +G_{CT}^{(1)}  \left( \hat{CT}_{S} \right)_{0} 
            +i\ G_{TQ}^{(1)} \left( \hat{T}_{AS} \right)_{0} 
            +G_{ZQ}^{(1)}  \left(T^{(2)0}\right)_{0}              
\end{align}
\begin{align}
\label{eq:43}
H_{1}&=G_{ST}^{(1)} \left\lbrace  \left( \hat{ST}_{S}^{(r)} \right)_{+1} + \left( \hat{ST}_{S}^{(cr)} \right)_{-1}  \right\rbrace 
           +G_{DQ}^{(1)} \left\lbrace  \left( \hat{D}_{S}^{(r)} \right)_{+1} + \left( \hat{D}_{S}^{(cr)} \right)_{-1}  \right\rbrace    
\end{align}
In the above equation, the coefficients `$G_{ST}^{(1)} =\sum_{i=0}^{N_{1}} G_{ST,i}^{(1)} $' and `$G_{DQ}^{(1)} =\sum_{i=0}^{N_{1}} G_{DQ,i}^{(1)} $' denote the off-diagonal coefficients resulting from the first-transformation and are described in Table. S.4 (in the supplementary information). Depending on the desired level of accuracy, the off-diagonal contributions (the order is denoted by value of $N_{1}$) from the first-transformation are incorporated accordingly. 
Employing the transformation function `$S_{2}$',
\begin{align} 
\label{s2transcase3}
S_{2}&=C_{ST}^{(2)} \left\lbrace  \left( \hat{ST}_{S}^{(r)} \right)_{+1} - \left( \hat{ST}_{S}^{(cr)} \right)_{-1}  \right\rbrace +
             C_{DQ}^{(2)}  \left\lbrace  \left( \hat{D}_{S}^{(r)} \right)_{+1} - \left( \hat{D}_{S}^{(cr)} \right)_{-1}  \right\rbrace
\end{align}
the off-diagonal contributions to `$H_{1}$' (Eq:~\ref{eq:43}) are completely folded. In contrast to the previous description involving single transformation, the higher order contributions in the present case are evaluated using the commutator between $S_{2}$ and $H_{1}$. Analogous to the previous description, a general expression illustrating the various contributions could be derived using the general expression presented below.
\begin{align}
H_{n}^{(2)}=\sum_{n=2}^{\infty} \dfrac{(i)^{n-1}}{n \times (n-2)!} \left[ \underbrace{\left[S_{2},...............\left[S_{2} \right. \right.}_{n-1} ,H_{1} {\left. \left. \right].............. .\right]}  \right]
\end{align}
The diagonal and off-diagonal contributions resulting from the second transformation are tabulated in Table. S.5 (refer supplementary information).

As illustrated in Table. S.5 (refer supplementary information), the even order terms comprise of diagonal contributions $\left(\text{arising from ZQ} (T^{(2)0})_{0}, \right. \\ \left. \text{CT} \left( \hat{CT}_{S} \right)_{0}  \text{and TQ} \left( \hat{T}_{AS} \right)_{0} \right)$ , while the odd-order terms, represent the off-diagonal contributions
 $\left( \text{from SQ $\left( \hat{ST}_{S}^{(r,cr)} \right)_{\pm1}$ and DQ $\left( \hat{D}_{S}^{(r,cr)} \right)_{\pm1}$} \right) $ .\\
Following the standard procedure, the effective Hamiltonian after second transformation is derived systematically to the desired level of accuracy.
\begin{align}
H_{eff}&= e^{i \lambda^{2} S_{2}}\ e^{i \lambda S_{1}} \ H_{F} \  e^{-i \lambda S_{1}}\  e^{-i \lambda^{2} S_{2}}\ \nonumber \\ 
&=\omega_{Q}I_{F}
            +G_{CT}^{(2)}  \left( \hat{CT}_{S} \right)_{0} 
            +i\ G_{TQ}^{(2)}  \left( \hat{T}_{AS} \right)_{0}  
            +G_{ZQ}^{(2)} \left(T^{(2)0}\right)_{0} 
            \label{eq:effham2} 
\end{align}
In Eq.~\ref{eq:effham2}, the coefficients `$G_{CT}^{(2)}$', `$G_{TQ}^{(2)}$', `$G_{ZQ}^{(2)}$' represent diagonal contributions resulting from both the first and second transformation.
\begin{align}
G_{CT}^{(2)}=\sum_{i=0}^{N_{1}} G_{CT,i}^{(1)}+\sum_{j=0}^{N_{2}} G_{CT,j}^{(2)} \nonumber \\
G_{TQ}^{(2)}=\sum_{i=0}^{N_{1}} G_{TQ,i}^{(1)}+\sum_{j=0}^{N_{2}} G_{TQ,j}^{(2)} \nonumber \\
G_{ZQ}^{(2)}=\sum_{i=0}^{N_{1}} G_{ZQ,i}^{(1)}+\sum_{j=0}^{N_{2}} G_{ZQ,j}^{(2)}
\end{align}
A detailed description of the above coefficients is given in Table. S.6 of the supplementary information.

\begin{center}
\begin{figure}[H]
    \centering
     \includegraphics[width=0.5\textwidth, angle=0]{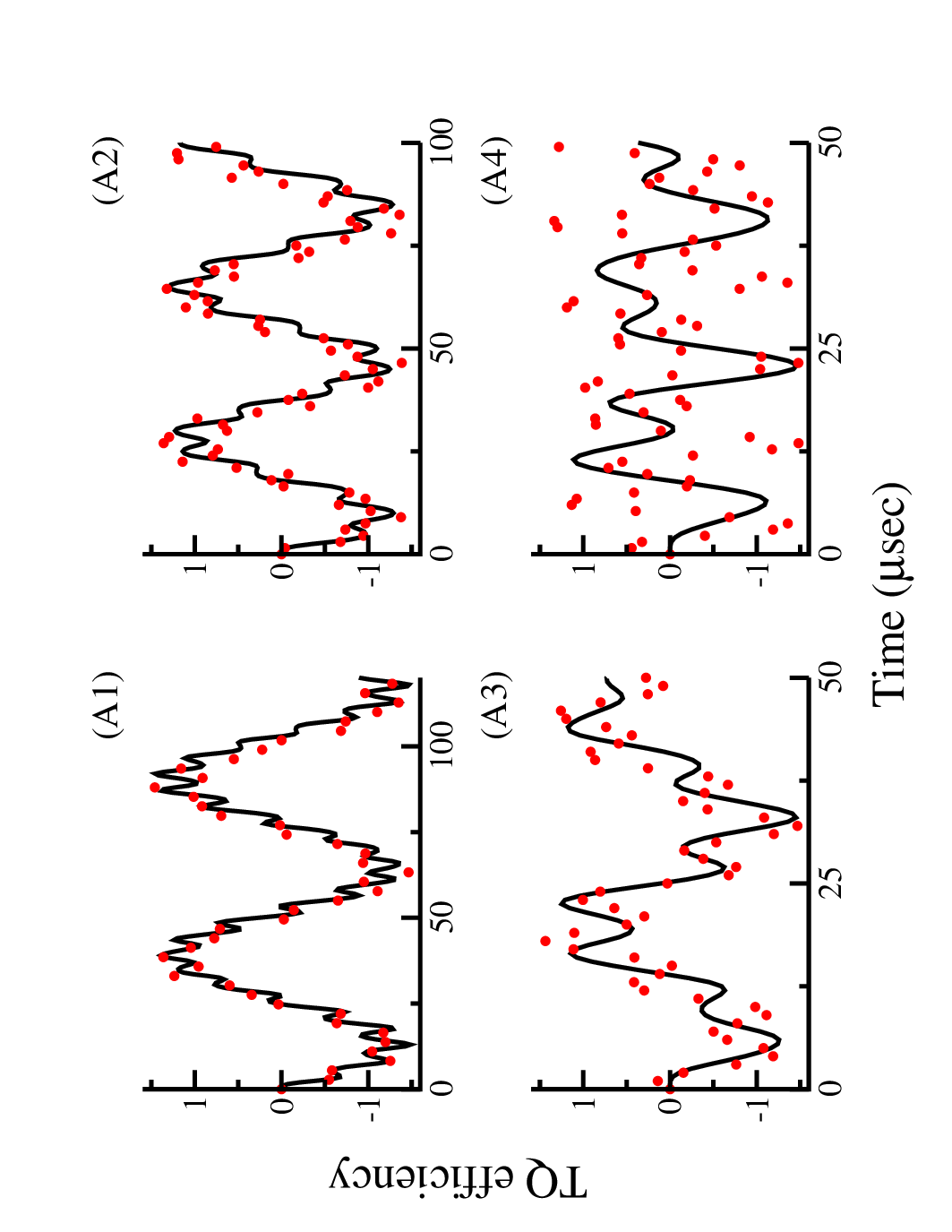}    
\caption{Case-III (b): Comparison of numerical (black thick line) and analytic simulations (red dots) based on effective Hamiltonians derived from the second  transformation. The off-diagonal contributions to order $\lambda^{3}$ from the first transformation $(N_{1}=3)$ and diagonal corrections to order $\lambda^{n}$ from the second transformation $(N_{2}>2)$ were included in the derivation of the effective Hamiltonians. In the simulations depicted, the quadrupole coupling constant $\left(C_{Q}=\rfrac{\omega_{Q}}{\pi}\right)$ is varied A1) $C_{Q}=500$ kHz, A2) $C_{Q}=400$ kHz, A3) $C_{Q}=300$ kHz, A4) $C_{Q}=200$ kHz, employing an excitation pulse of constant RF amplitude, $(\rfrac{\omega_{1}}{2\pi})=100$ kHz. The simulations correspond to a single crystal.}
\label{fig:lambda3_n} 
    \end{figure} 
\end{center}

\begin{center}
\begin{figure}[H]
    \centering
     \includegraphics[width=0.5\textwidth, angle=0]{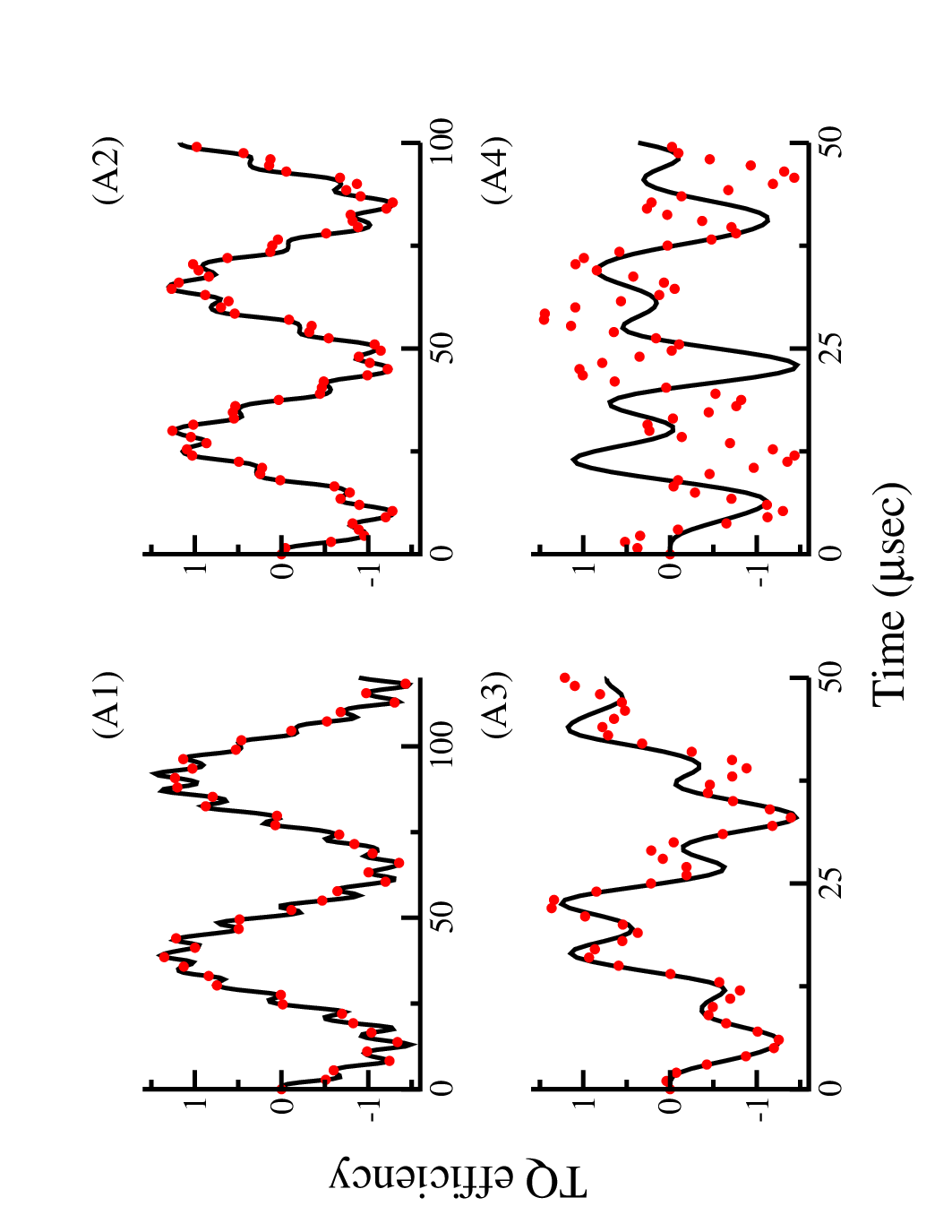}    
\caption{Case-IV (a): Comparison of numerical (black thick line) and analytic simulations (red dots) based on effective Hamiltonians derived from the second  transformation. The off-diagonal contributions to order $\lambda^{n}$ from the first transformation $(N_{1}>3)$ and diagonal corrections to order $\lambda^{2}$ from the second transformation $(N_{2}=2)$ were included in the derivation of the effective Hamiltonians. In the simulations depicted, the quadrupole coupling constant $\left(C_{Q}=\rfrac{\omega_{Q}}{\pi}\right)$ is varied A1) $C_{Q}=500$ kHz, A2) $C_{Q}=400$ kHz, A3) $C_{Q}=300$ kHz, A4) $C_{Q}=200$ kHz, employing an excitation pulse of constant RF amplitude, $(\rfrac{\omega_{1}}{2\pi})=100$ kHz. The simulations correspond to a single crystal.}
\label{fig:lambdan_3}
    \end{figure} 
\end{center}

\begin{center}
\begin{figure}[H]
    \centering
     \includegraphics[width=0.5\textwidth, angle=0]{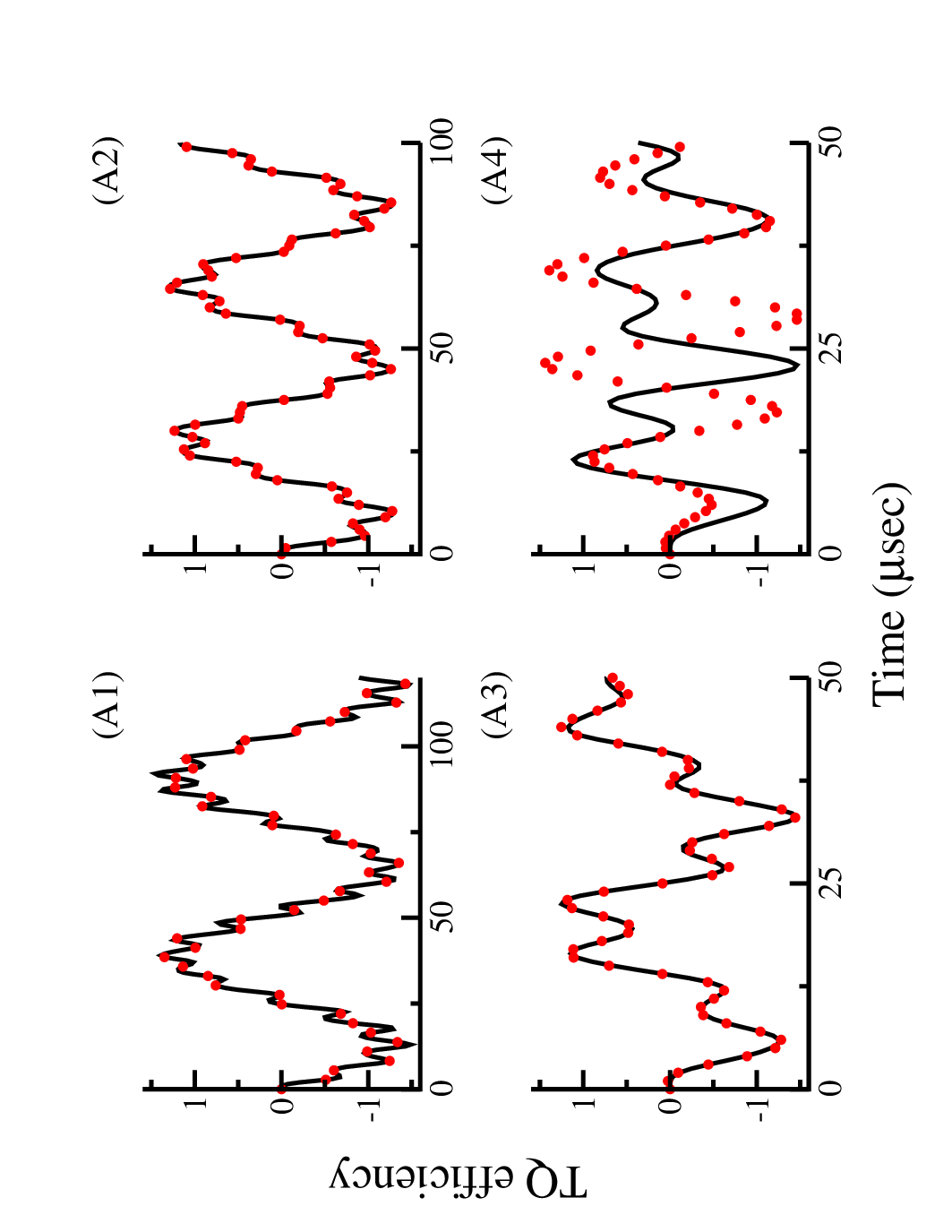}    
\caption{Case-IV (b): Comparison of numerical (black thick line) and analytic simulations  (red dots) based on effective Hamiltonians derived from the second  transformation. The off-diagonal contributions to order $\lambda^{n}$ from the first transformation $(N_{1}>3)$ and diagonal corrections to order $\lambda^{n}$ from the second transformation $(N_{2}>2)$ were included in the derivation of the effective Hamiltonians. In the simulations depicted, the quadrupole coupling constant $\left(C_{Q}=\rfrac{\omega_{Q}}{\pi}\right)$ is varied A1) $C_{Q}=500$ kHz, A2) $C_{Q}=400$ kHz, A3) $C_{Q}=300$ kHz, A4) $C_{Q}=200$ kHz, employing an excitation pulse of constant RF amplitude, $(\rfrac{\omega_{1}}{2\pi})=100$ kHz. The simulations correspond to a single crystal.}
\label{fig:lambdan_n}
    \end{figure} 
\end{center}

\begin{center}
\begin{figure}[H]
    \centering
     \includegraphics[width=0.5\textwidth, angle=0]{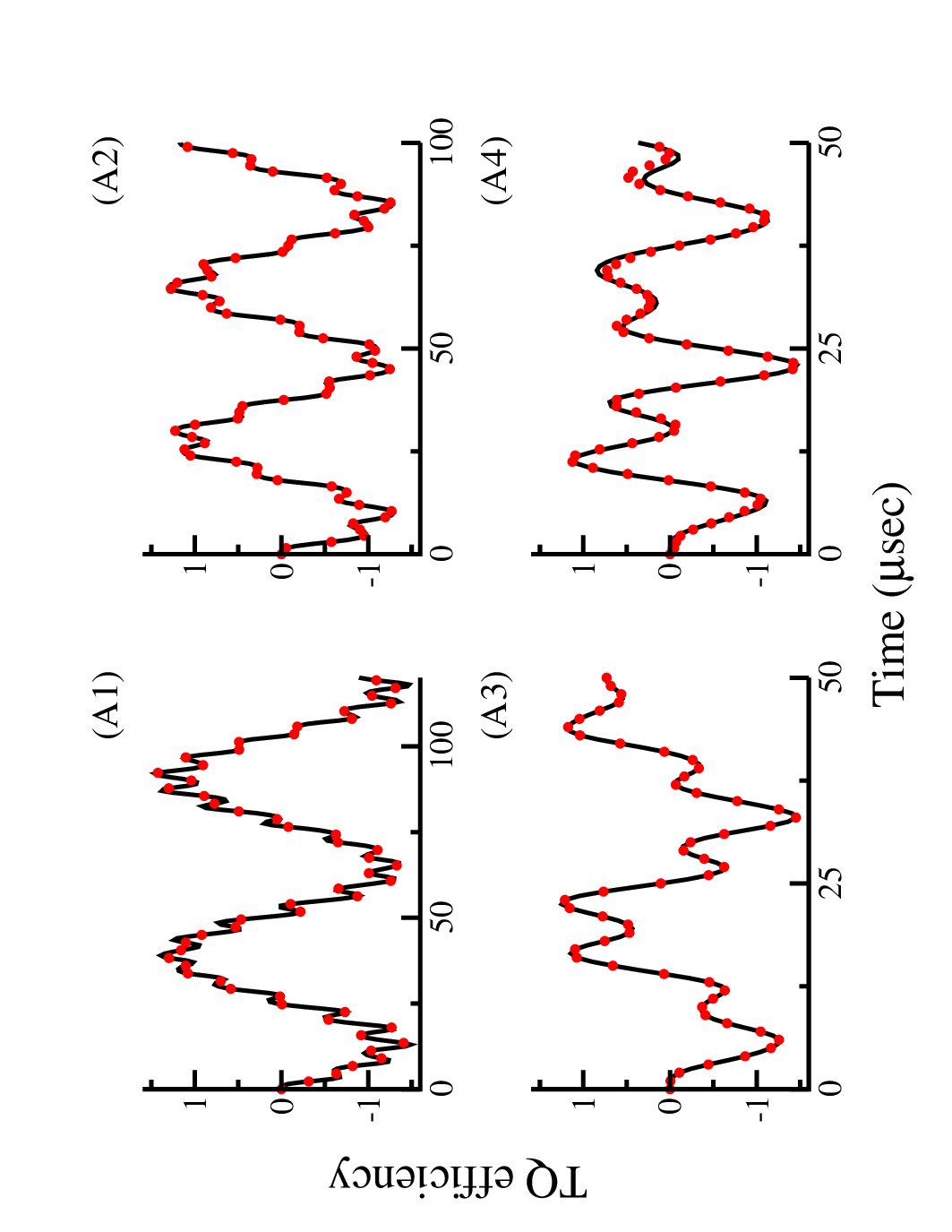}    
\caption{Comparison of numerical (black thick line) and analytic simulations  (red dots) based on effective Hamiltonians derived from the third transformation. The off-diagonal contributions to order $\lambda^{n}$ from the second transformation $(N_{2}>3)$ and diagonal corrections to order $\lambda^{n}$ from the third transformation $(N_{3}>2)$ were included in the derivation of the effective Hamiltonians. In the simulations depicted, the quadrupole coupling constant $\left(C_{Q}=\rfrac{\omega_{Q}}{\pi}\right)$ is varied A1) $C_{Q}=500$ kHz, A2) $C_{Q}=400$ kHz, A3) $C_{Q}=300$ kHz, A4) $C_{Q}=200$ kHz, employing an excitation pulse of constant RF amplitude, $(\rfrac{\omega_{1}}{2\pi})=100$ kHz. The simulations correspond to a single crystal.}
\label{fig:hr_s3}
    \end{figure} 
\end{center}

Analogous to the description in the previous section, both the initial density operator and the detection operators are transformed by the second transformation function, `$S_{2}$'. A detailed  description of this procedure is outlined in the supplementary information. Subsequently after simplification, the TQ efficiency is calculated using the expression given below.
\begin{mdframed}[innertopmargin=0pt,userdefinedwidth=14.3cm,align=center,middlelinewidth=1pt,roundcorner=10pt] 
\begin{align}
  \left\langle T^{(3)-3} \right\rangle=  \left( \Phi_{1}^{3} \ \right) &
\left\lbrace -\dfrac{1}{4} \ K_{+1}^{\left( \theta_{1},\theta_{2} \right)}\  K_{+2}^{\left( \theta_{1},\theta_{2} \right)}\ S_{(\theta_{TQ})}   
 +\dfrac{1}{4} \ K_{-1}^{\left( \theta_{1},\theta_{2} \right)}\  K_{-2}^{\left( \theta_{1},\theta_{2} \right)}\ S_{(\theta_{CT})} \right.  \nonumber \\
&\ \ \left. -\dfrac{3}{2}\ S_{(\theta_{1})}\ S_{(\theta_{2})}\ C_{(\theta_{RF})}\  S_{(\theta_{ZQ})} \right.  \nonumber \\
&\ \ \left. -\left\lbrace \dfrac{1}{2}\ \left(S_{(\theta_{1})} \right)^{2}+\left(S_{(\theta_{2})} \right)^{2} \right\rbrace\ S_{(\theta_{RF})}\ C_{(\theta_{ZQ})}  \right\rbrace 
\label{lambdas2}
\end{align}
\end{mdframed}
Based on the coefficients described in Table. S.7 (refer supplementary information) and considering the leading terms, a simplified form of the above equation is derived.$\left(i.e.,\  \theta_{TQ}=\dfrac{3 \omega_{1}^{3} t_{p1}}{2 \Omega_{Q}^{2}}  \quad ; \quad  
\theta_{CT}=-2 \omega_{1} t_{p1}+\dfrac{3 \omega_{1}^{3} t_{p1}}{2 \Omega_{Q}^{2}} \quad ; \quad
\theta_{RF}=\omega_{1} t_{p1} \quad ; \quad \right. \\
\left. \theta_{ZQ}=(\Delta-\omega_{Q}) t_{p1} -\dfrac{3 \omega_{1}^{2} t_{p1}}{2\ \Omega_{Q}} \right)$  
\begin{align}
\left\langle T^{(3)-3} (t_{p1})\right\rangle \propto \left\lbrace -\dfrac{1}{4}\ S_{\left( \theta_{TQ} \right)} + \dfrac{1}{4}\ S_{\left( \theta_{CT} \right)}-\dfrac{3}{2}\ C_{(\theta_{RF})}\  S_{(\theta_{ZQ})}+ S_{(\theta_{RF})} \ C_{(\theta_{ZQ})}   \right\rbrace
\end{align}
\begin{table}[H]
\caption{Definition of coefficients employed in the perturbing Hamiltonians for Case-III (a,b) and Case-IV (a,b)}
\centering
\resizebox{12cm}{!} {
\begin{tabular}{||l|c|c|c|}
\hline 
 \multicolumn{2}{||c|}{\textbf{}} &$G_{ST}^{(1)}$ &$G_{DQ}^{(1)}$  \\ \hline \hline
 \multicolumn{2}{||c|}{\textbf{}} & &  \\
\multicolumn{2}{||c|}{\shortstack{\textbf{Case-III (a) and Case-IV(a)} \\[0.2cm] $\left(N_{1} =3\right)$}}  &$+ \dfrac{1}{3} \left(\dfrac{\omega_{1}}{2} \right)  \left(\theta\right)^{2} $ & $-\dfrac{\omega_{1}}{2\sqrt{2}}  \left(\theta \right)$ \\  [.5cm]
\multicolumn{2}{||c|}{\shortstack{\textbf{Case-III (b) and Case-IV(b)} \\[0.2cm] $\left(N_{1} >3\right)$}}   & $ +\left( \dfrac{\omega_{1} }{2}\right) \left\lbrace \dfrac{1}{3\times 1!} \left(\theta \right)^{2} \right. $ &$ -\dfrac{\omega_{1}}{2\sqrt{2}} S_{\left(\theta \right)} $  \\ [.5cm]
 \multicolumn{2}{||c|}{\textbf{}} & $ \left. -\dfrac{1}{5\times 3!} \left(\theta \right)^{4} +\dfrac{1}{7\times 5!} \left(\theta \right)^{6} -....\right\rbrace  $ & \\ [.5cm]
 \multicolumn{2}{||c|}{\textbf{}} & &  \\ \hline \hline 
 &$G_{CT}^{(2)}$ &$G_{TQ}^{(2)}$ &$G_{ZQ}^{(2)}$ \\ \hline
 & & & \\
Case-III (a) &$\sum\limits_{i=0}^{3} G_{CT,i}^{(1)}+\sum\limits_{j=1}^{2} G_{CT,j}^{(2)} $ &$ \sum\limits_{i=0}^{3} G_{TQ,i}^{(1)}+\sum\limits_{j=1}^{2} G_{TQ,j}^{(2)}$ & $\sum\limits_{i=0}^{3} G_{ZQ,i}^{(1)}+\sum\limits_{j=1}^{2} G_{ZQ,j}^{(2)} $ \\ [.5cm]
Case-III (b) &$\sum\limits_{i=0}^{3} G_{CT,i}^{(1)}+\sum\limits_{j=1}^{N_{2}} G_{CT,j}^{(2)} $ &$ \sum\limits_{i=0}^{3} G_{TQ,i}^{(1)}+\sum\limits_{j=1}^{N_{2}} G_{TQ,j}^{(2)}$ & $\sum\limits_{i=0}^{3} G_{ZQ,i}^{(1)}+\sum\limits_{j=1}^{N_{2}} G_{ZQ,j}^{(2)} $ \\ [.5cm]
Case-IV (a) &$\sum\limits_{i=0}^{N_{1}} G_{CT,i}^{(1)}+\sum\limits_{j=1}^{2} G_{CT,j}^{(2)} $ &$ \sum\limits_{i=0}^{N_{1}} G_{TQ,i}^{(1)}+\sum\limits_{j=1}^{2} G_{TQ,j}^{(2)}$ & $\sum\limits_{i=0}^{N_{1}} G_{ZQ,i}^{(1)}+\sum\limits_{j=1}^{2} G_{ZQ,j}^{(2)} $ \\ [.5cm]
Case-IV (b) &$\sum\limits_{i=0}^{N_{1}} G_{CT,i}^{(1)}+\sum\limits_{j=1}^{N_{2}} G_{CT,j}^{(2)} $ &$ \sum\limits_{i=0}^{N_{1}} G_{TQ,i}^{(1)}+\sum\limits_{j=1}^{N_{2}} G_{TQ,j}^{(2)}$ & $\sum\limits_{i=0}^{N_{1}} G_{ZQ,i}^{(1)}+\sum\limits_{j=1}^{N_{2}} G_{ZQ,j}^{(2)} $ \\ [.5cm]
& & & \\ \hline
\multicolumn{4}{||c|}{\textbf{}} \\
\multicolumn{4}{||c|}{\textbf{$\theta = \left(\dfrac{\sqrt{3}\omega_{1}}{\Omega_{Q}}\right) \quad ; \quad C_{(\theta)}=\cos\left(\theta \right) \quad ; \quad S_{(\theta)}=\sin\left(\theta \right)$}} \\ 
\multicolumn{4}{||c|}{\textbf{}} \\     \hline
\end{tabular}
}
\label{tab:corrham1} 
\end{table} 
To illustrate the role of the off-diagonal contributions from the first transformation, a systematic analysis analogous to the one in the previous section was performed. To begin with, off-diagonal contributions to order `$\lambda^{3}$' were included and diagonal corrections to the resulting effective Hamiltonian from the second transformation were evaluated to order `$\lambda^{2}$' (henceforth referred as Case-III(a) )  and order `$\lambda^{n}$' (referred as Case-III(b)). The diagonal and off-diagonal terms employed in the simulations are summarized in Table.~\ref{tab:corrham1}
\begin{center}
\begin{figure}[H]
    \centering
     \includegraphics[width=0.5\textwidth, angle=0]{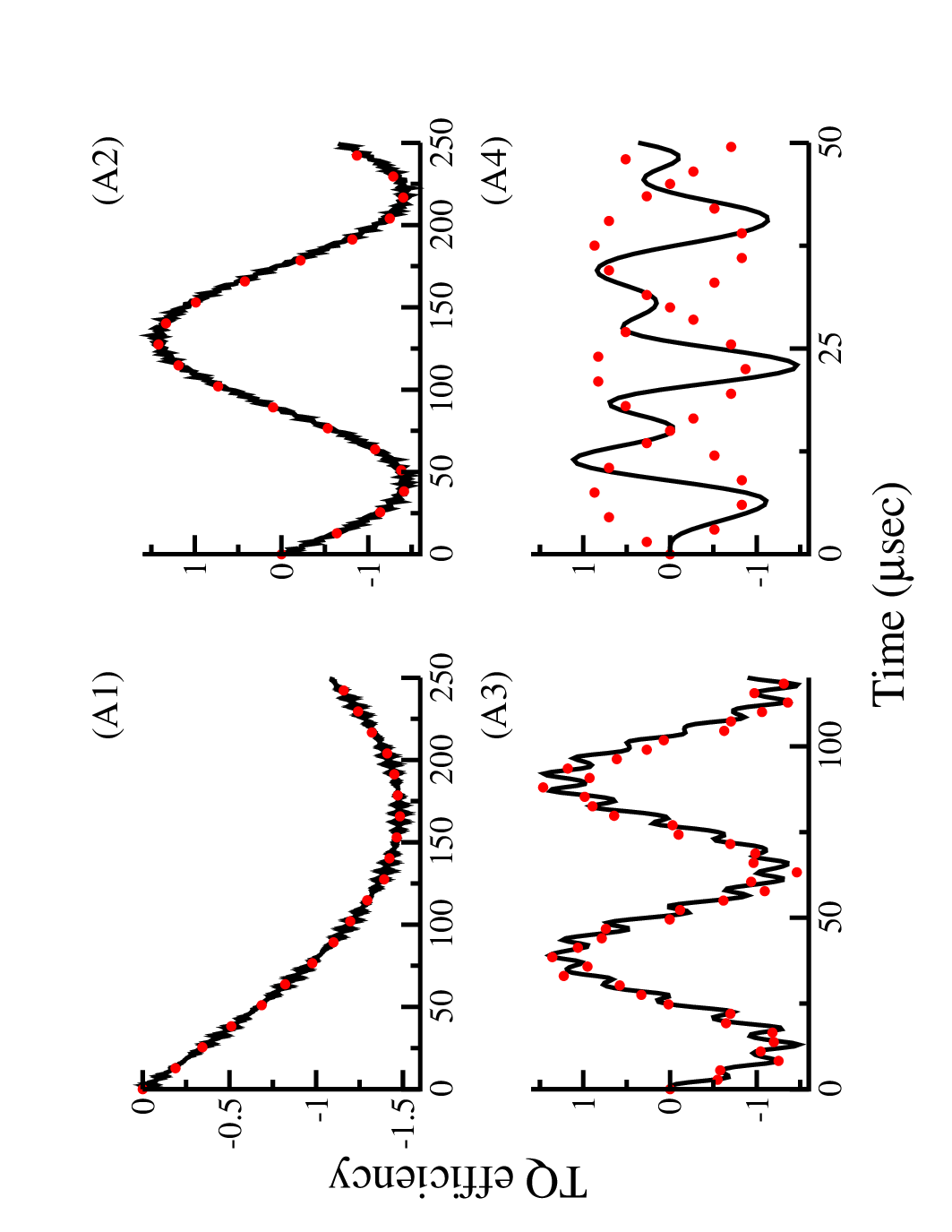}    
\caption{Case-III (a): Comparison of numerical (black thick line) and analytic simulations (red dots) based on effective Hamiltonians derived from the second  transformation. The off-diagonal contributions to order $\lambda^{3}$ from the first transformation $(N_{1}=3)$ and diagonal corrections to order $\lambda^{2}$ from the second transformation $(N_{2}=2)$ were included in the derivation of the effective Hamiltonians. In the simulations depicted, the quadrupole coupling constant $\left(C_{Q}=\rfrac{\omega_{Q}}{\pi}\right)$ is varied A1) $C_{Q}=2$ MHz, A2) $C_{Q}=1$ MHz, A3) $C_{Q}=500$ kHz, A4) $C_{Q}=200$ kHz, employing an excitation pulse of constant RF amplitude, $(\rfrac{\omega_{1}}{2\pi})=100$ kHz. The simulations correspond to a single crystal.}
\label{fig:lambda3_3}
    \end{figure} 
\end{center} 
As depicted in Figure.~\ref{fig:lambda3_3}, the analytic simulations from the effective Hamiltonian are in excellent agreement with SIMPSON simulations for the case corresponding to $C_{Q}=500$ kHz. Hence, the discrepancy observed in Figure.~\ref{fig:lambdan} is mainly due to the neglect of the off-diagonal corrections from the first transformation. Nevertheless, the analytic simulations do not match with the numerical simulations when the quadrupolar frequency `$\omega_{Q}$' is equal to the amplitude ($\omega_{1}$) of the RF pulse (see panel A4 in Figure.~\ref{fig:lambda3_3}).\\
To explain the discrepancy (for quadrupolar coupling constants less than 500 kHz) additional set of simulations incorporating higher order diagonal corrections (Case-III (b)) were performed. In Figure.~\ref{fig:lambda3_n}, TQ excitation efficiency corresponding to Case-III (b) is depicted by lowering the quadrupolar coupling constant in steps of 100 kHz. 
As depicted in Figure.~\ref{fig:lambda3_n}, the discrepancy observed in the analytic simulations increases when the quadrupolar frequency is lowered. To further improve the accuracy of the analytic simulations, off-diagonal contributions to `$10^{th} $' order resulting from the first transformation were included in the perturbing Hamiltonian ($N_{1}=10$, in Eq.~\ref{eq:effham1a}). Subsequently, the role of the diagonal corrections from the second transformation upto II order (Case-IV (a) $N_{1}>3$;$N_{2}=2$) and the $n^{th}$ order (Case-IV (b) $N_{1}>3$;$N_{2}>2$) were investigated. A detailed description illustrating the form of the coefficients employed in the perturbing and effective Hamiltonians is listed in Table.~\ref{tab:corrham1}. \\
The simulations corresponding to Case-IV (a) and Case-IV (b) are depicted through Figures.~\ref{fig:lambdan_3} and ~\ref{fig:lambdan_n}, respectively. 
As illustrated, the analytic simulations corresponding to Case-IV (b) Figure.~\ref{fig:lambdan_n} are in better agreement when compared to the simulations presented in Figures.~\ref{fig:lambdan} -~\ref{fig:lambdan_3}. The minor deviations observed in Figure.~\ref{fig:hr_s3} (panel A4, $\rfrac{\omega_{Q}}{\pi}=100$ kHz, $(\rfrac{\omega_{1}}{2\pi})=100$ kHz) could be further improved by incorporating the off-diagonal corrections to order `$\lambda^{3}$' resulting from the second transformation and is demonstrated in Figure.~\ref{fig:hr_s3}.
As depicted, the analytic simulations based on effective Floquet Hamiltonians is in excellent agreement with the SIMPSON simulations. To further validate the approach, TQ excitation in systems with lower quadrupolar frequency (lower than the RF amplitude) were further investigated.
\begin{center}
\begin{figure}[H]
    \centering
     \includegraphics[width=0.5\textwidth, angle=0]{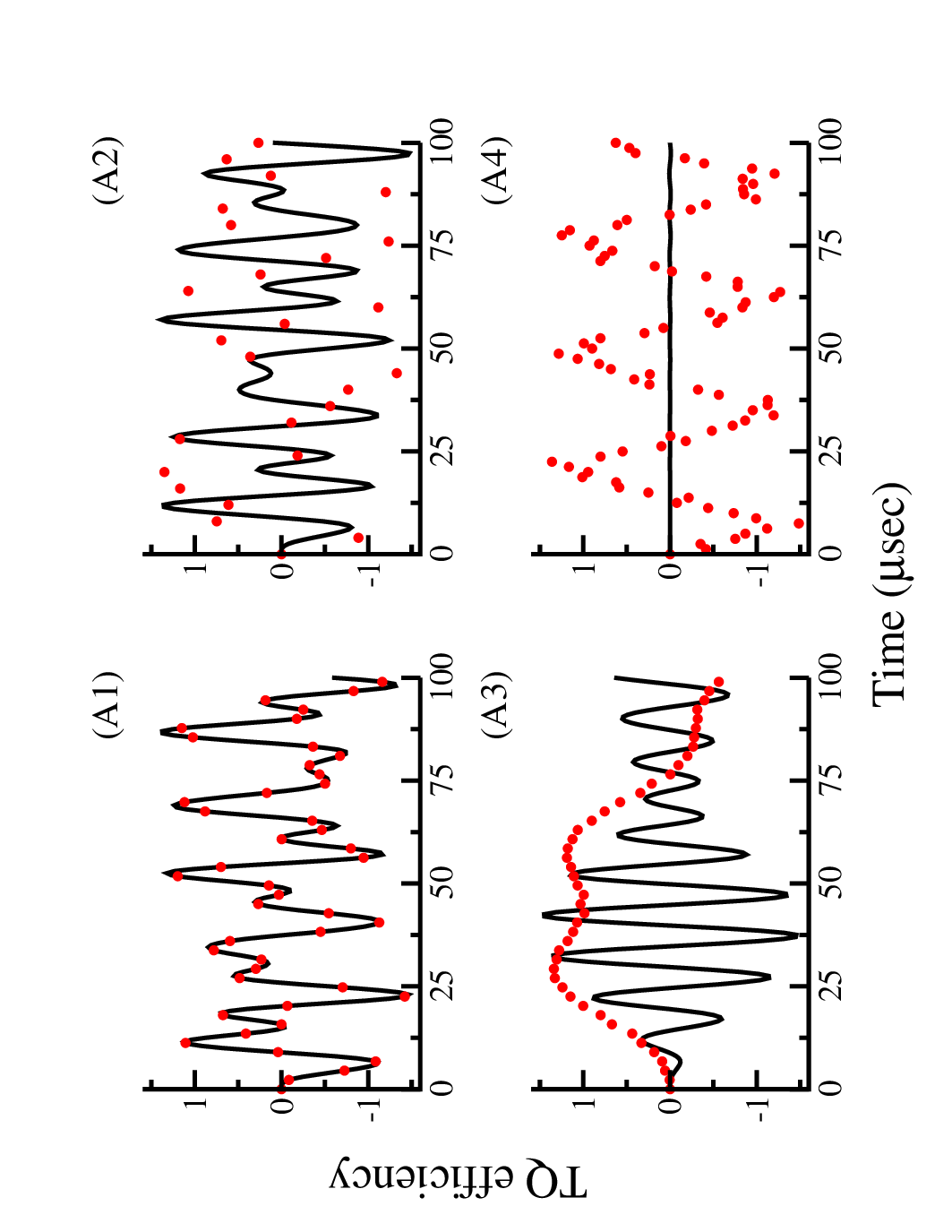}    
\caption{Comparison of numerical (black thick line) and analytic simulations  (red dots) based on effective Hamiltonians derived from the tenth transformation. In the simulations depicted, the quadrupole coupling constant$\left(C_{Q}=\rfrac{\omega_{Q}}{\pi}\right)$ is varied A1) $C_{Q}=200$ kHz, A2) $C_{Q}=150$ kHz, A3) $C_{Q}=50$ kHz, A4) $C_{Q}=1$ kHz, employing an excitation pulse of constant RF amplitude, $(\rfrac{\omega_{1}}{2\pi})=100$ kHz. The simulations correspond to a single crystal.}
\label{fig:hr_fail}
    \end{figure} 
\end{center}  
Interestingly, the analytic simulations (as depicted in Figure.~\ref{fig:hr_fail}, panels A3, A4), are in complete disagreement when the quadrupolar frequency is lower than the RF amplitude. The observed discrepancy could be attributed to the lack of convergence of perturbations with decreasing magnitudes of the quadrupolar coupling constant. Hence, the calculations in the quadrupolar interaction frame are valid only for systems wherein, the quadrupolar frequency is greater than or equal to the amplitude of the pulse.
\clearpage
To explain the observed discrepancy in the analytic simulations, an alternate method is proposed in the following section and forms the basis for the description in the weak coupling regime.
\subsubsection{Weak coupling ($\omega_{Q}<< \omega_{1}$)}
When the frequency, `$\omega$' of the oscillating field (or pulse) is adjusted to the Larmor frequency `$\omega_{0}$' ($\omega=\omega_{0}$), Eq.~\ref{eq:geq} reduces to the familiar form,
\begin{align}
\tilde{H}(t)=-\hbar \omega_{1} I_{x}-\hbar \Omega_{Q} T^{(2)0}
\label{eq:ham1}
\end{align}
In cases, where the amplitude of the RF pulse exceeds the magnitude of the quadrupolar frequency, the RF interaction Hamiltonian (Eq.~\ref{eq:ham1}) is quantized along the z-axis using the transformation function `$U_{2}$' ($U_{2}=e^{i\rfrac{\pi}{2}I_{y}}$)
\begin{align}
\tilde{\tilde{H}}=&U_{2}\ \tilde{H}(t) \ U_{2}^{-1} \nonumber \\
&=-\hbar  \omega_{1} I_{z}+\left( \dfrac{\hbar \Omega_{Q}}{2} \right) T^{(2)0}- \sqrt{\dfrac{3}{2}} \left( \dfrac{\hbar \Omega_{Q}}{2} \right) \left(\Phi_{1}^{2} T^{(2)2}+\Phi_{1}^{-2} T^{(2)-2} \right)
\label{eq:r2grot}
\end{align}
To further simplify the description, the above Hamiltonian is transformed into the RF interaction frame defined by the transformation function, `$U_{3}$' ($U_{3}=exp \left(-i \omega_{1}t I_{z}\right) $)
\begin{align}
\tilde{\tilde{\tilde{H}}}(t)=&U_{3}\ \tilde{\tilde{H}} \ U_{3}^{-1} \nonumber \\
&=\left( \dfrac{\hbar \Omega_{Q}}{2} \right) T^{(2)0}- \sqrt{\dfrac{3}{2}} \left( \dfrac{\hbar \Omega_{Q}}{2} \right) \left(\Phi_{1}^{2} T^{(2)2} e^{-2i\omega_{1}t}+\Phi_{1}^{-2} T^{(2)-2}e^{2i\omega_{1}t} \right)
\end{align}
As illustrated above, the Hamiltonian in the tilted RF interaction frame is time-dependent. In accord with the description presented in the previous section, the above time-dependent Hamiltonian is transformed into a time-independent Floquet Hamiltonian.\\
\begin{align}
H_{F}=\omega_{1}I_{F}+\left( \dfrac{\hbar \Omega_{Q}}{2} \right) \left(T^{(2)0}\right)_{0} - \sqrt{\dfrac{3}{2}} \left( \dfrac{\hbar \Omega_{Q}}{2} \right) \left\lbrace \left(\Phi_{1}^{2} T^{(2)2}\right)_{2} + \left( \Phi_{1}^{-2} T^{(2)-2} \right)_{-2} \right\rbrace
\end{align}
To facilitate analytic description, the above untransformed Floquet Hamiltonian is  re-expressed as a sum of zero-order and perturbing Hamiltonian. The perturbing Hamiltonian ($H_{1}$) comprises of both diagonal ($H_{1,d}$) and off-diagonal terms ($H_{1,od}$).
\begin{gather}
H_{0}=\omega_{1}I_{F}
H_{1}=H_{1,d}+H_{1,od} \nonumber\\
H_{1,d}=+\left( \dfrac{ \Omega_{Q}}{2} \right) \left(T^{(2)0}\right)_{0} ;\quad H_{1,od}= - \sqrt{\dfrac{3}{2}} \left( \dfrac{ \Omega_{Q}}{2} \right) \left\lbrace \left(\Phi_{1}^{2} T^{(2)2}\right)_{2} + \left(\Phi_{1}^{-2} T^{(2)-2} \right)_{-2} \right\rbrace
\end{gather}
In contrast to the analytic description present in Regime-I, the quadrupolar interaction acts like a perturbation and plays an important role in the excitation of MQ transitions in Regime-II. Employing the transformation function, `$S_{1}$'.
\begin{gather}
H_{eff}= e^{i \lambda S_{1}} \ H_{F}\  e^{-i \lambda S_{1}} \nonumber \\
S_{1}=C_{DR}^{(1)}   \left\lbrace \left(\Phi_{1}^{2} T^{(2)2}\right)_{2} - \left(\Phi_{1}^{-2} T^{(2)-2} \right)_{-2} \right\rbrace
\label{eq:s1lr}
\intertext{where,}
C_{DR}^{(1)}=-i  \sqrt{\dfrac{3}{2}} \left(\dfrac{\Omega_{Q}} {4\omega_{1}}\right)
\end{gather}
the off-diagonal contributions in $H_{1}$ (i.e. $H_{1,od}$) is folded and the higher order corrections to the effective Floquet Hamiltonian are derived using the relations described in Table.~\ref{tab:coneffham}. To first-order, the effective Hamiltonian comprises of `$H_{1,d}$'. 
\begin{align}
H_{1}^{(1)}&=\left( \dfrac{\hbar \Omega_{Q}}{2} \right) \left(T^{(2)0}\right)_{0}
\end{align}
A detailed derivation of the commutator relations involving the transformation `$S_{1}$' and `$H_{1}$' to various orders of `$\lambda$' are tabulated in Table. S.8 (refer supplementary information). As illustrated in Table. S.8 (refer supplementary information), the diagonal corrections mainly comprise of ZQ operators $\left( \left(T^{(k)0} \right)_{0}; k=1,2,3 \right)$, while the off-diagonal contributions are represented through the DQ operators $\left( \left(T^{(k)\pm 2} \right)_{\pm 2}\right)$. Below, a pedagogical description analogous to Regime-I is attempted to explicate the role of the higher-order contributions in the excitation process.\\
\textbf{I. Effective Hamiltonians from single transformation, `$S_{1}$'}\\
To begin with, let the effective Hamiltonian (comprising of diagonal corrections only) describing the excitation process in Regime-II be represented by, 
\begin{align}
H_{eff}&= e^{i \lambda S_{1}} \ H_{F} \  e^{-i \lambda S_{1}} \nonumber \\ 
&=\omega_{1}I_{F}
            +\dfrac{i}{\sqrt{5}}\  G_{1R}^{(1)}  \left(T^{(1)0}\right)_{0}
            +\ G_{2R}^{(1)} \left(T^{(2)0}\right)_{0}
            +\dfrac{i}{\sqrt{5}}\ G_{3R}^{(1)} \left(T^{(3)0}\right)_{0} \nonumber \\
            \label{eq:effham1lr} 
\intertext{where,}
G_{1R}^{(1)} &=\sum_{i=1}^{N_{1}} G_{1R,i}^{(1)}   \quad ; \quad 
G_{2R}^{(1)} =\sum_{i=1}^{N_{1}} G_{2R,i}^{(1)}    \quad ; \quad    
G_{3R}^{(1)} =\sum_{i=1}^{N_{1}} G_{3R,i}^{(1)}           
\end{align}
and $N_{1}$ represents the order of the correction from the first transformation. A detailed description of the coefficients is tabulated in Table.~\ref{tab:coeffhamdia2lr}.
\begin{table}[H]
\caption{Definition of the coefficients employed in the derivation of effective Hamiltonian (Eq.~\ref{eq:effham1lr}) based on first transformation}
\centering
\resizebox{13cm}{!} {
\begin{tabular}{||c|c|c|}
\hline 
 $G_{1R}^{(1)}$ &$G_{2R}^{(1)}$ &$G_{3R}^{(1)}$ \\ \hline \hline
 & & \\
 $G_{1R,1}^{(1)}=0$&$G_{2R,1}^{(1)}=\dfrac{\Omega_{Q}}{2}  $ &$G_{3R,1}^{(1)}=0$  \\ [.5cm]
 $G_{1R,2}^{(1)}=\dfrac{1}{2 \times 0!} \left(\sqrt{3} \Omega_{Q}\right) \left( \xi \right)  $ &  $G_{2R,2}^{(1)}=0 $ & $G_{3R,2}^{(1)}=\dfrac{1}{2 \times 0!} \left(\frac{\sqrt{3} \Omega_{Q}}{2}\right) \left( \xi \right)  $ \\ [.5cm]
 $G_{1R,3}^{(1)}= 0 $ &  $G_{2R,3}^{(1)}=-\dfrac{1}{2!} \dfrac{\Omega_{Q}}{2} \left( \xi \right)^{2} $& $G_{3R,3}^{(1)}=0  $ \\ [.5cm]
$G_{1R,4}^{(1)}=-\dfrac{1}{4 \times 2!} \left(\sqrt{3} \Omega_{Q}\right) \left( \xi \right)^{3}$&$G_{2R,4}^{(1)}=0$ & $G_{3R,4}^{(1)}=-\dfrac{1}{4 \times 2!} \left(\frac{\sqrt{3} \Omega_{Q}}{2}\right) \left( \xi \right)^{3}$ \\ [.5cm]
$G_{1R,5}^{(1)}=0$&$G_{2R,5}^{(1)}=\dfrac{1}{4!} \dfrac{\Omega_{Q}}{2} \left( \xi \right)^{4}$ & $G_{3R,5}^{(1)}=0$ \\ [.5cm]
$G_{1R,6}^{(1)}=\dfrac{1}{6 \times 4!} \left(\sqrt{3} \Omega_{Q}\right) \left( \xi \right)^{5}$&$G_{2R,6}^{(1)}=0$ & $G_{3R,6}^{(1)}=\dfrac{1}{6 \times 4!} \left(\frac{\sqrt{3} \Omega_{Q}}{2}\right) \left( \xi \right)^{5}$ \\ [.5cm]
$G_{1R,7}^{(1)}=0$&$G_{2R,7}^{(1)}=-\dfrac{1}{6!} \dfrac{\Omega_{Q}}{2} \left( \xi \right)^{6}$ & $G_{3R,7}^{(1)}=0$ \\ [.5cm]
 .& .&. \\[.5cm]
. &. &. \\ \hline
 $G_{1R}^{(1)}=\left(\sqrt{3} \Omega_{Q} \right) \left\lbrace +\dfrac{1}{2\times 0!} \left( \xi \right) \right. $ & $G_{2R}^{(1)}=\left(\dfrac{\Omega_{Q}}{2} \right)  C_{ \left( \xi \right)} $ &$G_{3R}^{(1)}=\left(\dfrac{\sqrt{3}\Omega_{Q}} {2} \right) \left\lbrace +\dfrac{1}{2\times 0!} \left( \xi \right) \right.$  \\ [.5cm]
$\left. -\dfrac{1}{4\times 2!}\left( \xi \right)^{3} +\dfrac{1}{6\times 4!} \left( \xi \right)^{5} +....\right\rbrace$ & &$\left. -\dfrac{1}{4\times 2!}\left(\xi \right)^{3} +\dfrac{1}{6\times 4!} \left( \xi \right)^{5} +....\right\rbrace$ \\\hline
 \multicolumn{3}{||c|}{\textbf{}} \\
\multicolumn{3}{||c|}{\textbf{$\xi = \left(\dfrac{\sqrt{3}\Omega_{Q}} {4\omega_{1}}\right) \quad ; \quad C_{(\xi)}=\cos\left(\xi \right) $}} \\ 
\multicolumn{3}{||c|}{\textbf{}} \\     \hline
\end{tabular}
}
\label{tab:coeffhamdia2lr} 
\end{table}
To have a consistent description, the initial density operator ($\rho_{F}(0)=(I_{z})_{0}$) along with the detection operator `$T^{(3)-3}$', is transformed by the transformation function `$S_{1}$' (please refer to the supplementary information for more details).

Subsequently, employing the effective Hamiltonian (Eq.~\ref{eq:effham1lr}), the evolution of the initial density operator in Regime-II is calculated. In contrast to the description in Regime-I, only SQ and TQ coherences are created by the pulse in Regime-II.
\begin{align}
\tilde{\rho}_{F}(t_{p1})=\tilde{\rho}_{F}(t_{p1})_{SQ}+\tilde{\rho}_{F}(t_{p1})_{TQ}
\label{eq:denaex1lr}
\end{align} 
A detailed description of the calculations is illustrated in the supplementary information.\\
Subsequently, the TQ signal in Regime-II is calculated by the expression given below.
\begin{align}
\left\langle T^{(3)-3} (t_{p1})\right\rangle \propto \left\lbrace -\dfrac{1}{8}\ S_{\left( \theta_{TQ} \right)} +\dfrac{1}{8}\  S_{\left( \theta_{CT} \right)}-\dfrac{3}{8}\  S_{\left( \theta_{STA} \right)}-\dfrac{3}{8}\  S_{\left( \theta_{STB} \right)}\right\rbrace
\end{align}
where,\\
$\theta_{TQ}=3 \omega_{1}t_{p1}+\dfrac{3  \Omega_{Q}^{2}\ t_{p1}}{16\ \omega_{1}} \quad ; \quad
\theta_{CT}=\omega_{1}t_{p1}+\dfrac{3  \Omega_{Q}^{2}\ t_{p1}}{16\ \omega_{1}} \quad ; \quad \\
\theta_{STA}=\omega_{1}t_{p1}+\dfrac{\Omega_{Q}}{2} \left(1-\dfrac{3 \Omega_{Q}^{2} }{32\ \omega_{1}^{2}}\right)t_{p1} \quad ; \quad  
\theta_{STB}=\omega_{1}t_{p1}-\dfrac{\Omega_{Q}}{2} \left(1-\dfrac{3 \Omega_{Q}^{2} }{32\ \omega_{1}^{2}}\right)t_{p1}  $ \\
A detailed description of the coefficients is presented in the supplementary information.

\begin{center}
\begin{figure}[H]
    \centering
     \includegraphics[width=0.5\textwidth, angle=0]{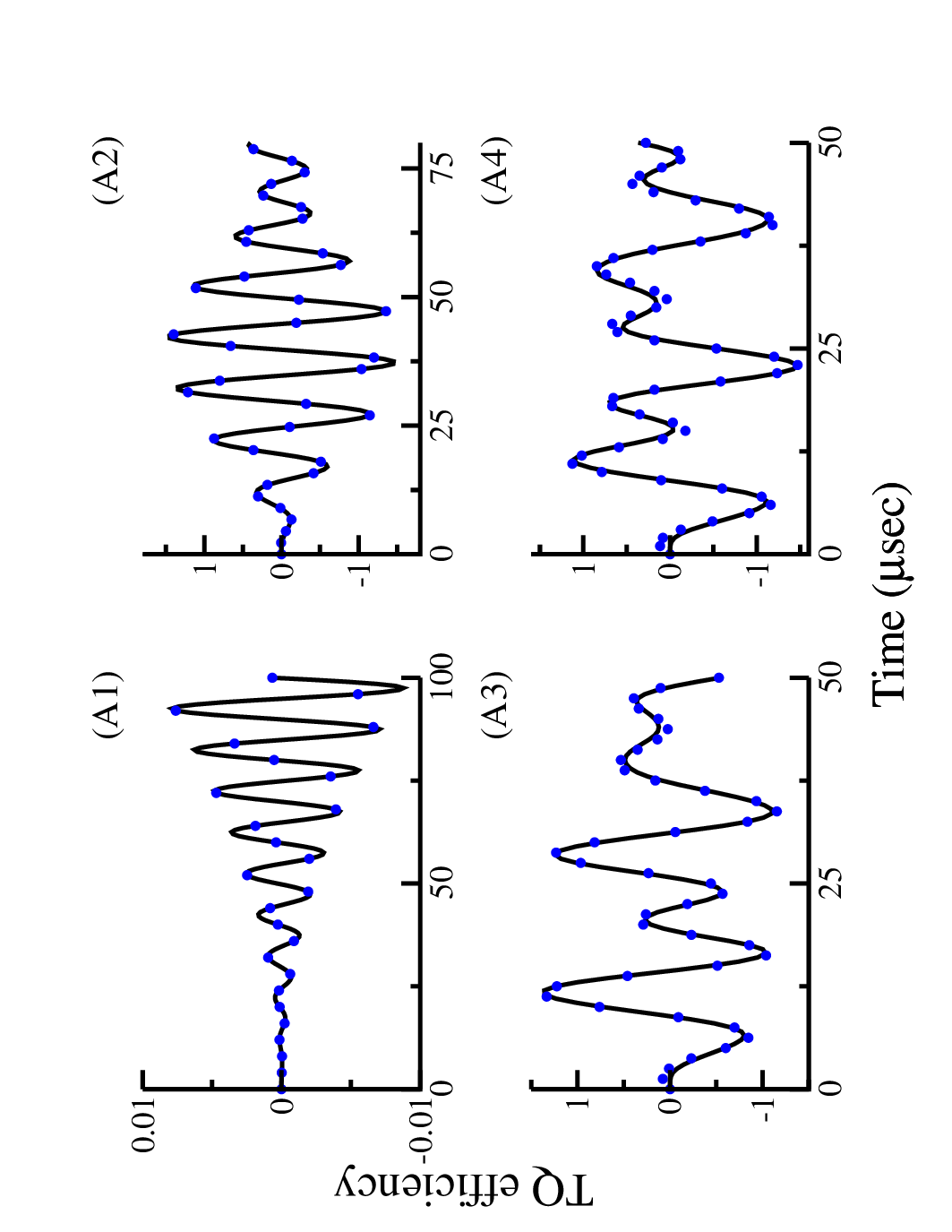}    
\caption{Case-I: Comparison of numerical (black thick line) and analytic simulations (blue dots) based on effective Hamiltonians derived from a single transformation (in Regime-II) comprising of diagonal corrections to order $\lambda^{2}$ $(N_{1}=3)$.  In the simulations depicted, the quadrupole coupling constant $\left(C_{Q}=\rfrac{\omega_{Q}}{\pi}\right)$ is varied A1) $C_{Q}=1$ kHz, A2) $C_{Q}=50$ kHz, A3) $C_{Q}=150$ kHz, A4) $C_{Q}=200$ kHz, employing an excitation pulse of constant RF amplitude, $(\rfrac{\omega_{1}}{2\pi})=100$ kHz. The simulations correspond to a single crystal.}
\label{fig:lambda3lr}
    \end{figure} 
\end{center}

\begin{center}
\begin{figure}[H]
    \centering
     \includegraphics[width=0.5\textwidth, angle=0]{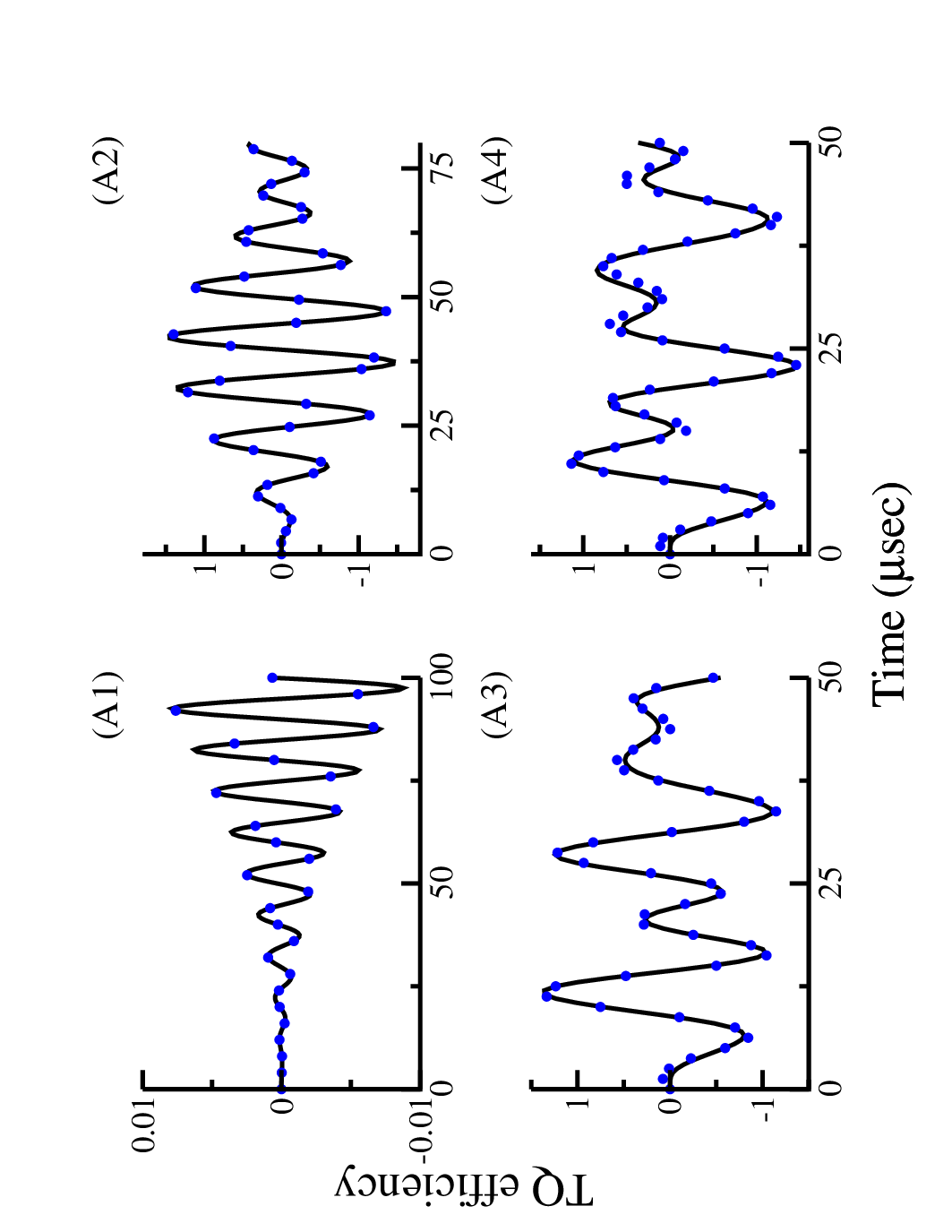}    
\caption{Case-II: Comparison of numerical (black thick line) and analytic simulations (blue dots) based on effective Hamiltonians derived from a single transformation (in Regime-II) comprising of diagonal corrections to order $\lambda^{n}$ $(N_{1}>3)$.  In the simulations depicted, the quadrupole coupling constant $\left(C_{Q}=\rfrac{\omega_{Q}}{\pi}\right)$ is varied A1) $C_{Q}=1$ kHz, A2) $C_{Q}=50$ kHz, A3) $C_{Q}=150$ kHz, A4) $C_{Q}=200$ kHz, employing an excitation pulse of constant RF amplitude, $(\rfrac{\omega_{1}}{2\pi})=100$ kHz. The simulations correspond to a single crystal.}
\label{fig:lambdanlr}
    \end{figure} 
\end{center}

Analogous to the description in Regime-I (refer Eq.~\ref{eq:lambda3}), the TQ efficiency comprises of four terms. To illustrate the role of higher order corrections, simulations based on effective Hamiltonians incorporating diagonal corrections (refer Table.~\ref{tab:coeffham1lr}) to order `$\lambda^{2}$' (Case-I) and `$\lambda^{n}$' (Case-II) are illustrated in Figures.~\ref{fig:lambda3lr} and ~\ref{fig:lambdanlr} respectively. In the simulations depicted, the quadrupolar coupling constant was varied from 1 kHz to 200 kHz with a constant RF amplitude of 100 kHz. As depicted in Figures.~\ref{fig:lambda3lr} and ~\ref{fig:lambdanlr}, the analytic simulations are in excellent agreement with SIMPSON simulations. In contrast to Regime-I, the efficiency of excitation increases with the quadrupolar coupling constant (for a constant RF amplitude that is larger than or equal to the quadrupolar frequency).
\begin{table}[H]
\caption{Coefficients employed in the derivation of Effective Hamiltonians for Case-I and Case-II in Regime-II based on first transformation}
\centering
\resizebox{13cm}{!} {
\begin{tabular}{||l|c|c|c|}
\hline 
 &$G_{1R}^{(1)}$ &$G_{2R}^{(1)}$ &$G_{3R}^{(1)}$ \\ \hline \hline
 & & & \\
Case-I&$\left\lbrace \left(\dfrac{ \sqrt{3} \Omega_{Q}} {2} \right) \left( \xi \right) \right\rbrace $ & $+\left( \dfrac{ \Omega_{Q}}{2} \right) -\dfrac{1}{2!} \left( \dfrac{ \Omega_{Q}}{2} \right)  \left( \xi \right)^{2}$ & $\dfrac{1}{2} \left\lbrace \left(\dfrac{ \sqrt{3} \Omega_{Q}} {2} \right) \left( \xi \right) \right\rbrace $ \\  [.5cm]
Case-II&$  \left(\sqrt{3} \Omega_{Q} \right) \left\lbrace +\dfrac{1}{2\times 0!} \left( \xi \right) \right. $ &$\left(\dfrac{\Omega_{Q}}{2} \right)  C_{ \left( \xi \right)} $ &$  \left(\dfrac{\sqrt{3}\Omega_{Q}} {2} \right) \left\lbrace +\dfrac{1}{2\times 0!} \left( \xi \right) \right. $  \\ [.5cm]
 &$\left. -\dfrac{1}{4\times 2!}\left( \xi \right)^{3} +\dfrac{1}{6\times 4!} \left( \xi \right)^{5} +....\right\rbrace$&  &$\left. -\dfrac{1}{4\times 2!}\left( \xi \right)^{3} +\dfrac{1}{6\times 4!} \left( \xi \right)^{5} +....\right\rbrace  $ \\ [.5cm]
 & & & \\ \hline
\multicolumn{4}{||c|}{\textbf{}} \\
\multicolumn{4}{||c|}{\textbf{$\xi = \left(\dfrac{\sqrt{3}\Omega_{Q}} {4\omega_{1}}\right) \quad ; \quad C_{(\xi)}=\cos\left(\xi \right) $}} \\ 
\multicolumn{4}{||c|}{\textbf{}} \\     \hline 
\end{tabular}
}
\label{tab:coeffham1lr} 
\end{table}

\begin{center}
\begin{figure}[H]
    \centering
     \includegraphics[width=0.5\textwidth, angle=0]{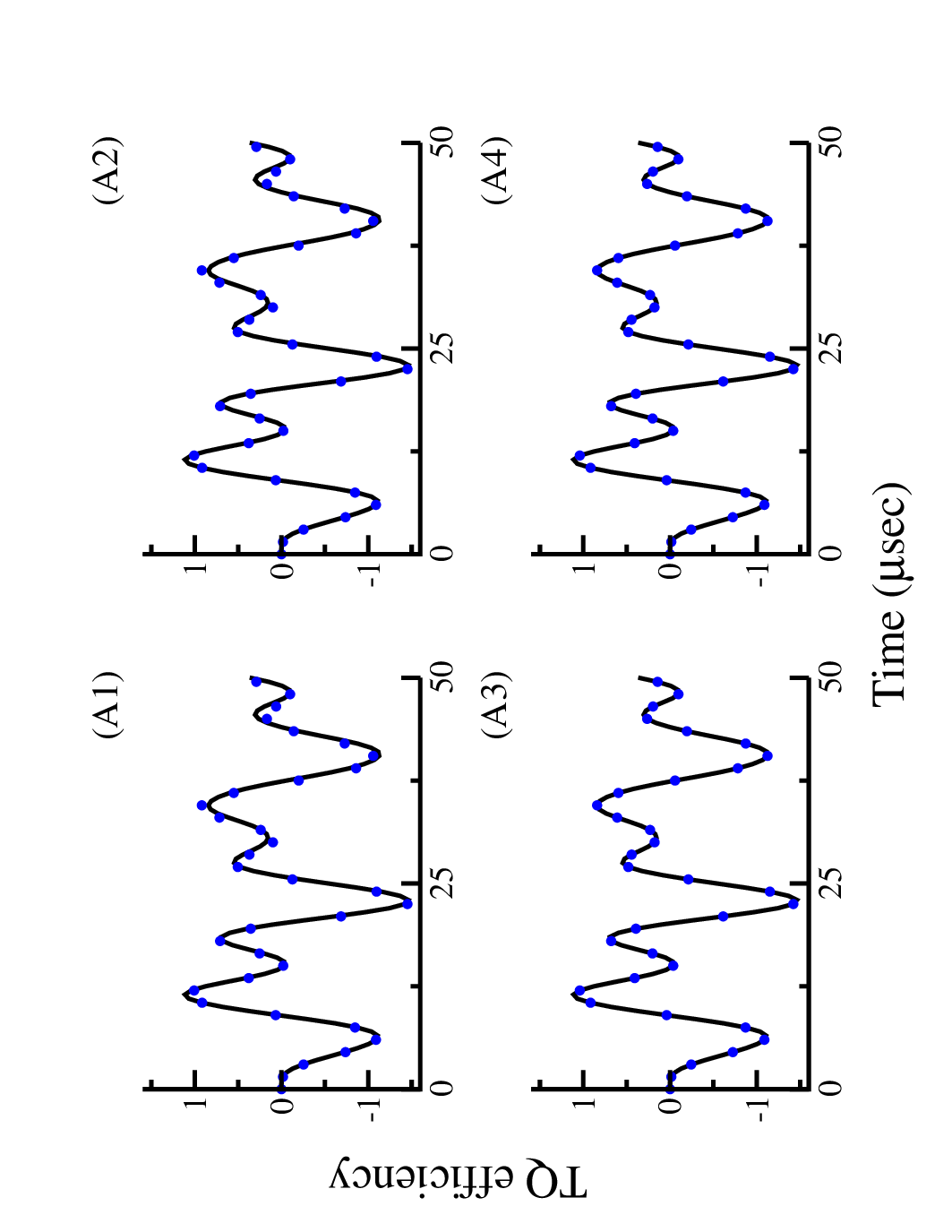}    
\caption{Comparison of numerical (black thick line) and analytic simulations (blue dots) based on effective Hamiltonians derived from Regime-II corresponding to the quadrupole coupling constant $\left(C_{Q}=\rfrac{\omega_{Q}}{\pi}\right)$, $C_{Q}=200$ kHz and RF amplitude, $(\rfrac{\omega_{1}}{2\pi})=100$ kHz.The simulation results from Case-III (a) (panel A1), Case-III (b) (panel A2), Case-IV (a) (panel A3) and Case-IV (b) (panel A4) are illustrated. The simulations correspond to a single crystal.}
\label{fig:lambda3_3lr}
    \end{figure} 
\end{center}

When the amplitude of the RF pulse is exactly equal to the quadrupolar frequency ($\omega_{Q}$), the minor deviations that are observed (in Panel A4 of Figure.~\ref{fig:lambdanlr}) could further be improved  by a second transformation. A detailed description of this procedure is outlined in the supplementary information.\\
\begin{table}[H]
\caption{Definitions of coefficients employed in the perturbing and effective Hamiltonians for Case-III(a,b) and Case-IV(a,b) in Regime-II} 
\centering
\resizebox{13cm}{!} {
\begin{tabular}{||l|c|c|c|}
\hline 
 \multicolumn{2}{||c|}{\textbf{}}&$G_{DR}^{(1)}$ &$G_{TR}^{(1)}$  \\ \hline \hline
\multicolumn{2}{||c|}{\textbf{}} & &  \\
\multicolumn{2}{||c|}{\shortstack{\textbf{Case-III (a) and Case-IV(a)} \\[0.2cm] $\left(N_{1} =3\right)$}}&$+\dfrac{1}{3\times 1!} \left(\dfrac{\sqrt{3}\Omega_{Q}} {2\sqrt{2}} \right) \left( \xi\right)^{2} $ & $-\left(\dfrac{\Omega_{Q}}{2\sqrt{2}}\right) \left( \xi \right) $ \\  [.5cm]
\multicolumn{2}{||c|}{\shortstack{\textbf{Case-III (b) and Case-IV(b)} \\[0.2cm] $\left(N_{1} >3\right)$}}  &$+\left(\dfrac{\sqrt{3}\Omega_{Q}} {2\sqrt{2}} \right) \left\lbrace \dfrac{1}{3\times 1!}  \left( \xi \right)^{2} \right.$ & $-\left(\dfrac{\Omega_{Q}}{2\sqrt{2}}\right)\ S_{\left( \xi \right)}$ \\ [.5cm]
 \multicolumn{2}{||c|}{\textbf{}}& $\left. -\dfrac{1}{5\times 3!}  \left( \xi \right)^{4} +\dfrac{1}{7\times 5!}  \left( \xi\right)^{6} -....\right\rbrace   $ & \\ [.5cm]
\multicolumn{2}{||c|}{\textbf{}} & &  \\ \hline \hline 
 &$G_{1R}^{(2)}$ &$G_{2R}^{(2)}$ &$G_{3R}^{(2)}$ \\ \hline
 & & & \\
Case-III (a) &$\sum\limits_{i=1}^{3} G_{1R,i}^{(1)}+\sum\limits_{j=1}^{2} G_{1R,j}^{(2)} $ &$ \sum\limits_{i=1}^{3} G_{2R,i}^{(1)}+\sum\limits_{j=1}^{2} G_{2R,j}^{(2)}$ & $\sum\limits_{i=1}^{3} G_{3R,i}^{(1)}+\sum\limits_{j=1}^{2} G_{3R,j}^{(2)} $ \\ [.5cm]
Case-III (b) &$\sum\limits_{i=1}^{3} G_{1R,i}^{(1)}+\sum\limits_{j=1}^{N_{2}} G_{1R,j}^{(2)} $ &$ \sum\limits_{i=1}^{3} G_{2R,i}^{(1)}+\sum\limits_{j=1}^{N_{2}} G_{2R,j}^{(2)}$ & $\sum\limits_{i=1}^{3} G_{3R,i}^{(1)}+\sum\limits_{j=1}^{N_{2}} G_{3R,j}^{(2)} $ \\ [.5cm]
Case-IV (a) &$\sum\limits_{i=1}^{N_{1}} G_{1R,i}^{(1)}+\sum\limits_{j=1}^{2} G_{1R,j}^{(2)} $ &$ \sum\limits_{i=1}^{N_{1}} G_{2R,i}^{(1)}+\sum\limits_{j=1}^{2} G_{2R,j}^{(2)}$ & $\sum\limits_{i=1}^{N_{1}} G_{3R,i}^{(1)}+\sum\limits_{j=1}^{2} G_{3R,j}^{(2)} $ \\ [.5cm]
Case-IV (b) &$\sum\limits_{i=1}^{N_{1}} G_{1R,i}^{(1)}+\sum\limits_{j=1}^{N_{2}} G_{1R,j}^{(2)} $ &$ \sum\limits_{i=1}^{N_{1}} G_{2R,i}^{(1)}+\sum\limits_{j=1}^{N_{2}} G_{2R,j}^{(2)}$ & $\sum\limits_{i=1}^{N_{1}} G_{3R,i}^{(1)}+\sum\limits_{j=1}^{N_{2}} G_{3R,j}^{(2)} $ \\ [.5cm]
& & & \\ \hline
 \multicolumn{4}{||c|}{\textbf{}} \\
\multicolumn{4}{||c|}{\textbf{$\xi = \left(\dfrac{\sqrt{3}\Omega_{Q}} {4\omega_{1}}\right) \quad ; \quad C_{(\xi)}=\cos\left(\xi \right) \quad ; \quad S_{\left( \xi \right)}=\sin\left(\xi \right)$}} \\ 
\multicolumn{4}{||c|}{\textbf{}} \\     \hline
\end{tabular}
}
\label{tab:corrham1lr} 
\end{table}
To illustrate the role of the higher order off-diagonal contributions, simulations depicting the excitation profile corresponding to Case-III(a) (Panel A1, $N_{1}=3$ and $N_{2}=2$),Case-III(b) (Panel A2, $N_{1}=3$ and $N_{2}=10$), Case-IV(a) (Panel A3, $N_{1}=10$ and $N_{2}=2$), Case-IV(b) (Panel A4, $N_{1}=10$ and $N_{2}=10$) are depicted in Figure.~\ref{fig:lambda3_3lr}. As depicted, the minor discrepancies observed in Figure.~\ref{fig:lambdanlr} (Panel A4) are completely addressed with the inclusion of higher order off-diagonal contributions (refer Table.~\ref{tab:corrham1lr}). Hence, in the case of Regime-II,the analytic simulations based on effective Hamiltonians derived from a single transformation yield results in excellent agreement to those obtained from exact numerical methods.
\clearpage
\subsection{Powder Sample}
To further validate the effective Hamiltonian approach, the calculations described in the previous sections were extended to describe the excitation in a powder sample. In the past, analytic description of the excitation process have remained hindered due to the distribution of the quadrupolar coupling constant (spatial anisotropy) present in a powder sample. Additionally, the incorporation of the orientation dependence during the excitation process was unclear within the existing frameworks. To this end, the formalism presented in this article is tailor made to describe the excitation process both in isotropic and anisotropic solids. As described in the theory section, the quadrupolar interaction represented through `$\Omega_{Q}$' becomes equal to `$\omega_{Q}^{(\alpha \beta \gamma)}$' for a powder sample$\left(\Omega_{Q}=\omega_{Q}^{(\alpha \beta \gamma)}\right)$ . Consequently, an anisotropic offset term `$\Delta$',($\Delta=\Omega_{Q}-\omega_{Q}^{(\alpha \beta \gamma)}$) (corresponding to the $T^{(2)0}$ operator) is present along the zero-order Hamiltonian. In the case of a single crystal, the offset term tends to zero and the excitation profile is orientation independent.\\
To investigate the exactness of the proposed effective Floquet Hamiltonians, TQ excitation in a powder sample was investigated systematically by the inclusion of higher order corrections. In the simulations presented below, the theoretical framework presented in Regime-I was employed to calculate the excitation profile in the presence of higher order diagonal and off-diagonal contributions.\\
\begin{center}
\begin{figure}[H]
    \centering
     \includegraphics[width=0.45\textwidth, angle=0]{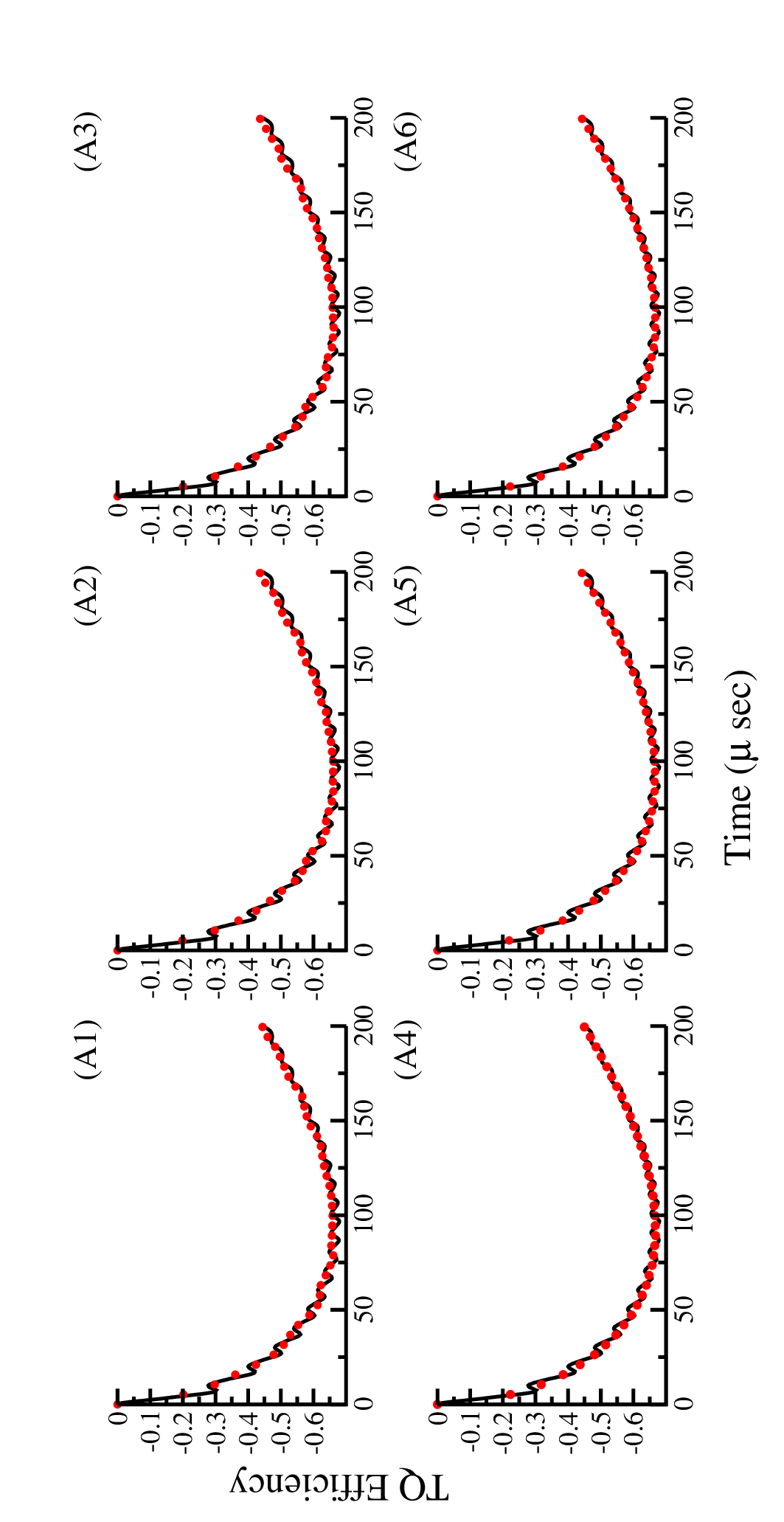}    
\caption{Comparison of numerical (black thick line) and analytic simulations (red dots) based on effective Hamiltonians derived from Regime-I corresponding to the quadrupole coupling constant  $\left(C_{Q}=\rfrac{\omega_{Q}}{\pi}\right)$, $C_{Q}=4$ MHz and RF amplitude, $(\rfrac{\omega_{1}}{2\pi})=100$ kHz. The simulation results from the first transformation, Case-I (panel A1), Case-II (panel A4) and second transformation, Case-III (a) (panel A2), Case-III (b) (panel A3), Case-IV (a) (panel A5) and Case-IV (b) (panel A6) are illustrated. The powder simulations were performed using a crystal file having 28656 orientations ($\alpha, \beta$).}
\label{fig:pow4mhz}
    \end{figure} 
\end{center}

\begin{center}
\begin{figure}[H]
    \centering
     \includegraphics[width=0.45\textwidth, angle=0]{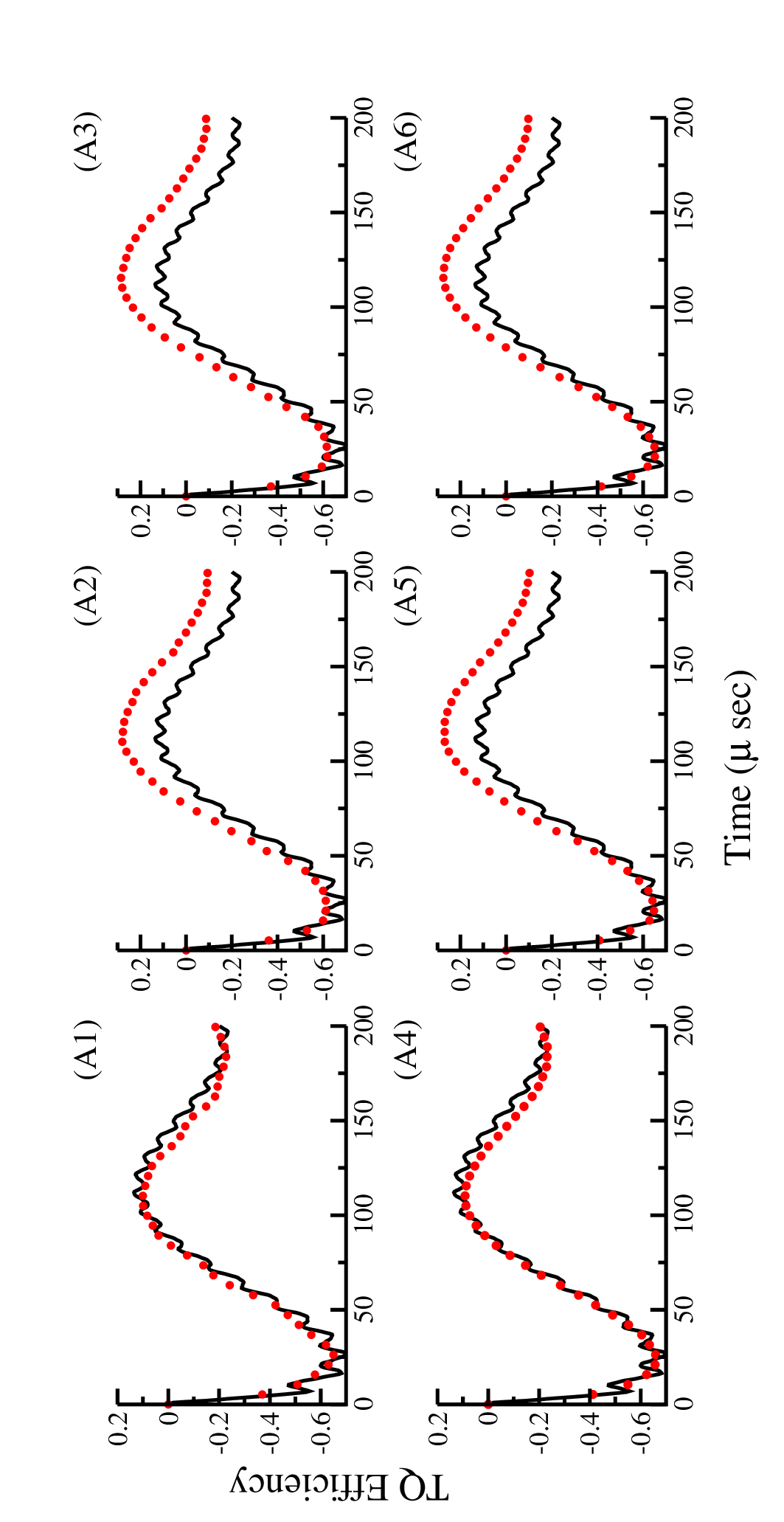}    
\caption{Comparison of numerical (black thick line) and analytic simulations (red dots) based on effective Hamiltonians derived from Regime-I corresponding to the quadrupole coupling constant $\left(C_{Q}=\rfrac{\omega_{Q}}{\pi}\right)$, $C_{Q}=2$ MHz and RF amplitude, $(\rfrac{\omega_{1}}{2\pi})=100$ kHz. The simulation results from the first transformation, Case-I (panel A1), Case-II (panel A4) and second transformation, Case-III (a) (panel A2), Case-III (b) (panel A3), Case-IV (a) (panel A5) and Case-IV (b) (panel A6) are illustrated. The powder simulations were performed using a crystal file having 28656 orientations ($\alpha, \beta$).}
\label{fig:pow2mhz}
    \end{figure} 
\end{center}

\begin{center}
\begin{figure}[H] 
    \centering
     \includegraphics[width=0.5\textwidth, angle=0]{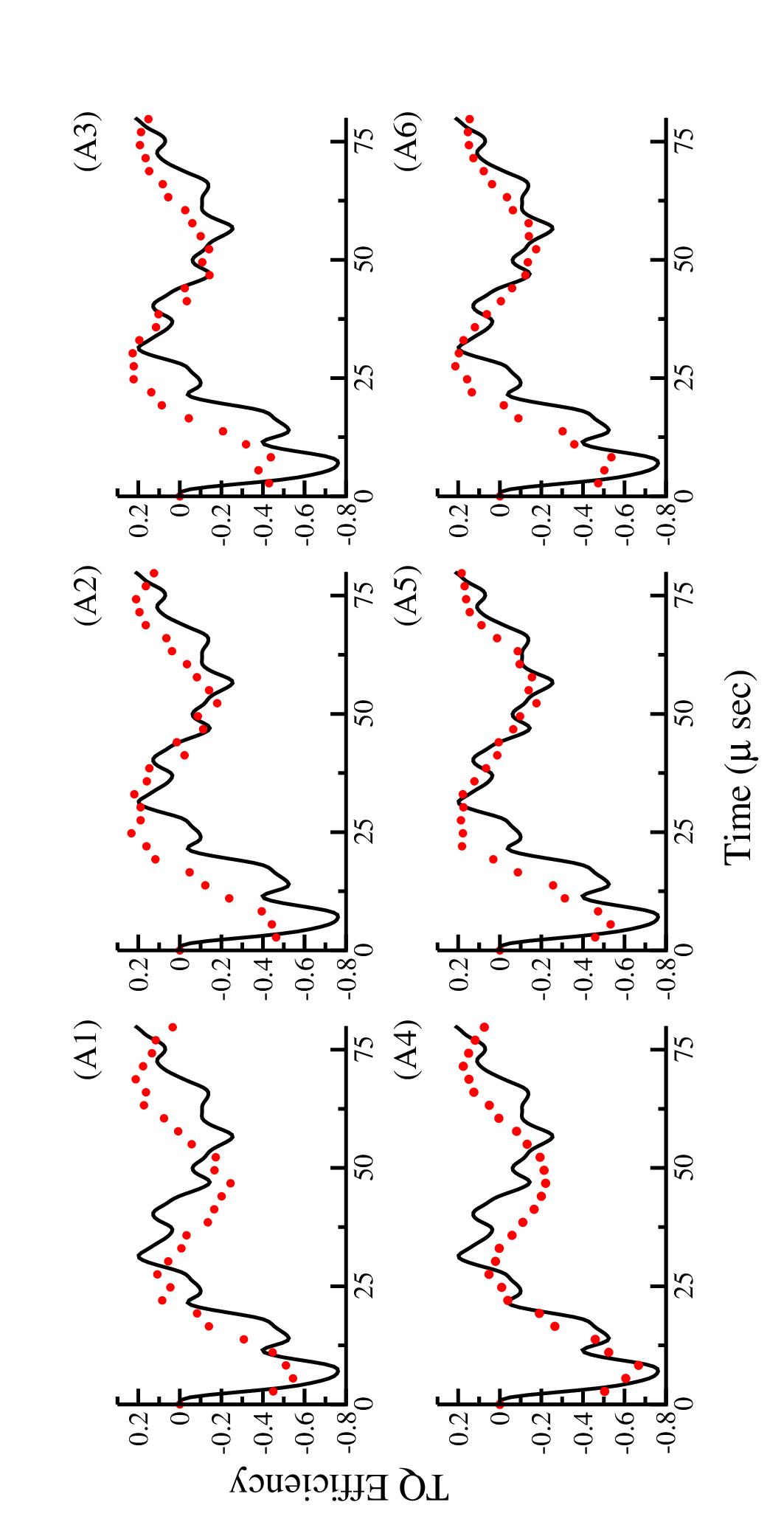}    
\caption{Comparison of numerical (black thick line) and analytic simulations (red dots) based on effective Hamiltonians derived from Regime-I corresponding to the quadrupole coupling constant $\left(C_{Q}=\rfrac{\omega_{Q}}{\pi}\right)$, $C_{Q}=1$ MHz and RF amplitude, $(\rfrac{\omega_{1}}{2\pi})=100$ kHz. The simulation results from the first transformation, Case-I (panel A1), Case-II (panel A4) and second transformation, Case-III (a) (panel A2), Case-III (b) (panel A3), Case-IV (a) (panel A5) and Case-IV (b) (panel A6) are illustrated. The powder simulations were performed using a crystal file having 28656 orientations ($\alpha, \beta$).}
\label{fig:pow1mhz}
    \end{figure} 
\end{center}
\begin{center}
\begin{figure}[H]
    \centering
     \includegraphics[width=0.5\textwidth, angle=0]{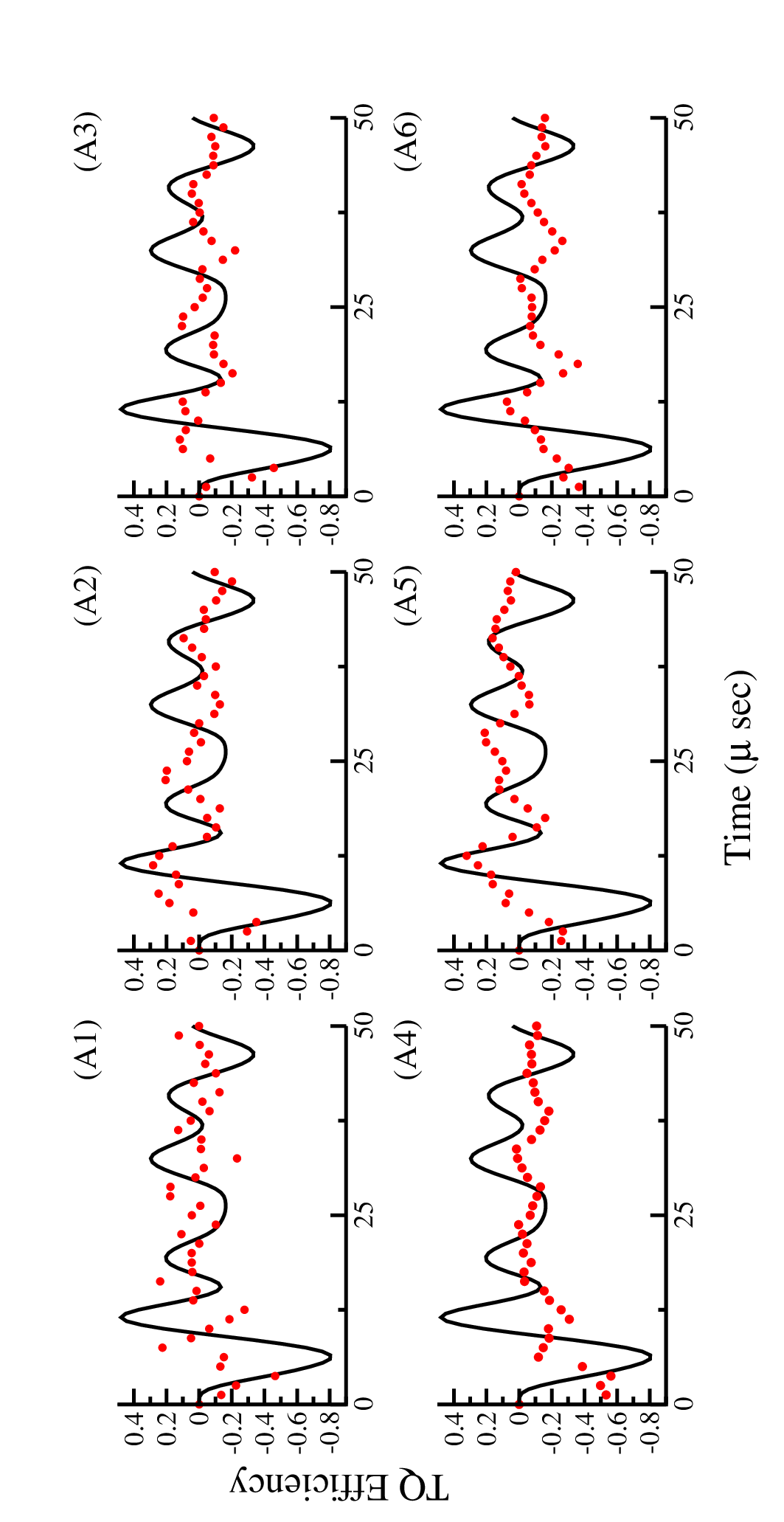}    
\caption{Comparison of numerical (black thick line) and analytic simulations (red dots) based on effective Hamiltonians derived from Regime-I corresponding to the quadrupole coupling constant $\left(C_{Q}=\rfrac{\omega_{Q}}{\pi}\right)$, $C_{Q}=500$ kHz and RF amplitude, $(\rfrac{\omega_{1}}{2\pi})=100$ kHz. The simulation results from the first transformation, Case-I (panel A1), Case-II (panel A4) and second transformation, Case-III (a) (panel A2), Case-III (b) (panel A3), Case-IV (a) (panel A5) and Case-IV (b) (panel A6) are illustrated. The powder simulations were performed using a crystal file having 28656 orientations ($\alpha, \beta$).}
\label{fig:pow500khz}
    \end{figure} 
\end{center}
\begin{center}
\begin{figure}[H]
    \centering
     \includegraphics[width=0.5\textwidth, angle=0]{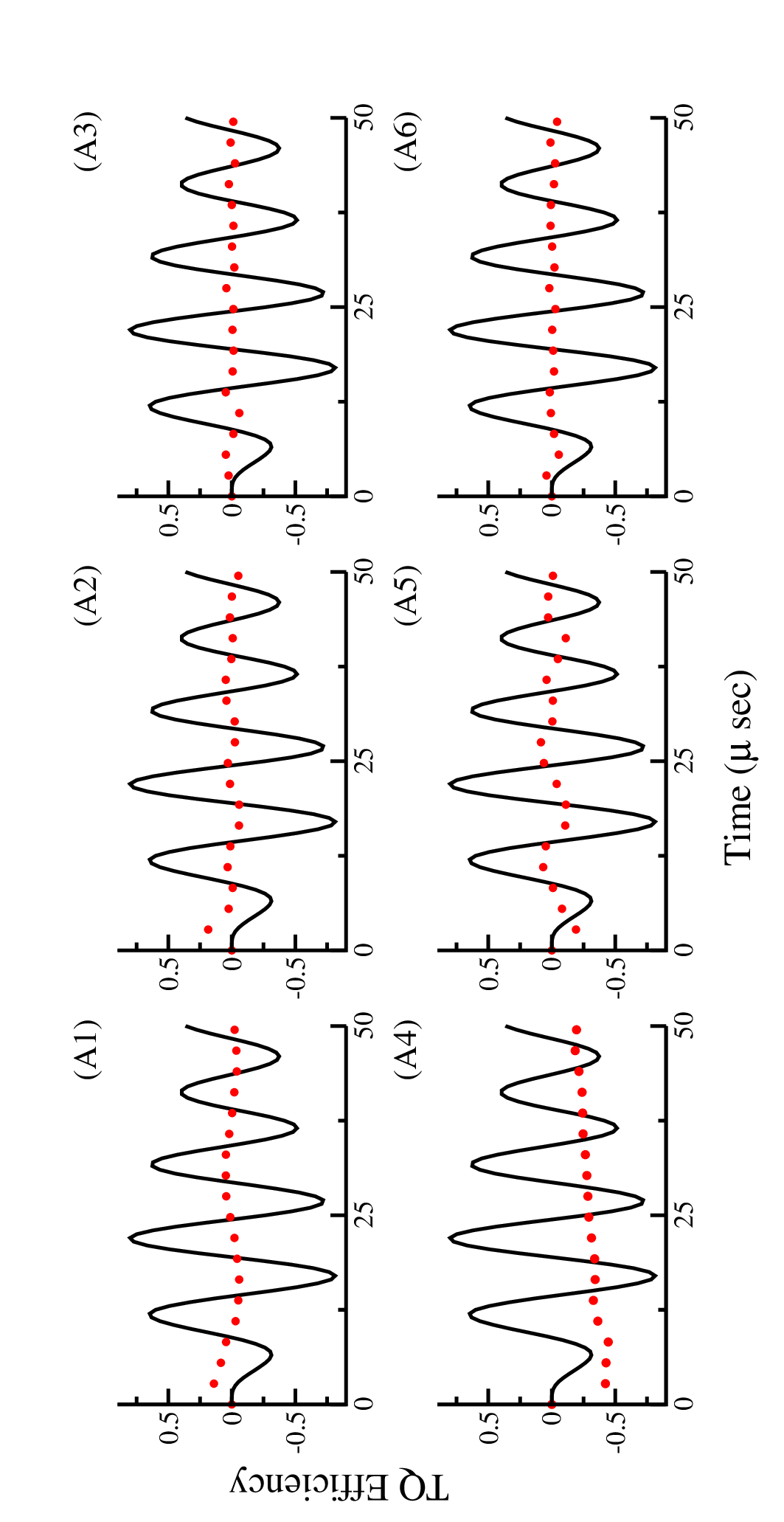}    
\caption{Comparison of numerical (black thick line) and analytic simulations (red dots) based on effective Hamiltonians derived from Regime-I corresponding to the quadrupole coupling constant $\left(C_{Q}=\rfrac{\omega_{Q}}{\pi}\right)$, $C_{Q}=200$ kHz and RF amplitude, $(\rfrac{\omega_{1}}{2\pi})=100$ kHz. The simulation results from the first transformation, Case-I (panel A1), Case-II (panel A4) and second transformation, Case-III (a) (panel A2), Case-III (b) (panel A3), Case-IV (a) (panel A5) and Case-IV (b) (panel A6) are illustrated. The powder simulations were performed using a crystal file having 28656 orientations ($\alpha, \beta$).}
\label{fig:pow200khz}
    \end{figure} 
\end{center}

As depicted in Figure.~\ref{fig:pow4mhz}, in the strong coupling regime ($C_{Q}=4$ MHz), the analytic simulations based on Regime-I corresponding to Case-I is in excellent agreement with SIMPSON simulations. In the extreme strong coupling regime, the analytical expression depicting the TQ excitation reduces to a much simpler form (analogous to Eq.~\ref{eq:t3vn} for the single crystal). 
\begin{align}
 \left\langle T^{(3)-3} (t_{p1}) \right\rangle&=-\dfrac{3}{2} \sin\left(\dfrac{3 \omega_{1}^{3} t_{p1}}{2\Omega_{Q}^{2}} \right)
\end{align}
where, `$\Omega_{Q}=\omega_{Q}$' for single crystal and `$\Omega_{Q}=\omega_{Q}^{(\alpha \beta \gamma)} $' for a powder sample.\\
However, with decreasing quadrupolar coupling constants, significant derivations are observed in the simulations irrespective of the inclusion of higher order (both diagonal and off-diagonal contributions). In all the simulations (Figures.~\ref{fig:pow4mhz},~\ref{fig:pow2mhz},~\ref{fig:pow1mhz}),  effective Hamiltonian derived from Regime-I were employed in the analytic simulations.
Nevertheless, such a simplified description ceases with a decrease in the quadrupolar coupling constant. In contrast to the calculations in the single crystal, (corresponding the quadrupolar coupling constants, $C_{Q}=1$ MHz; $C_{Q}=500$ kHz; and $C_{Q}=200$ kHz), the analytic simulations for the same in a powder sample are in complete disagreement with SIMPSON simulations. Additionally, as described in Figures.~\ref{fig:pow500khz} and ~\ref{fig:pow200khz}, the analytical simulations based on Regime-I are in complete disagreement when the magnitude of the quadrupolar frequency is equal to the amplitude of the RF pulse.

\begin{center}
\begin{figure}[H]
    \centering
     \includegraphics[width=0.45\textwidth, angle=0]{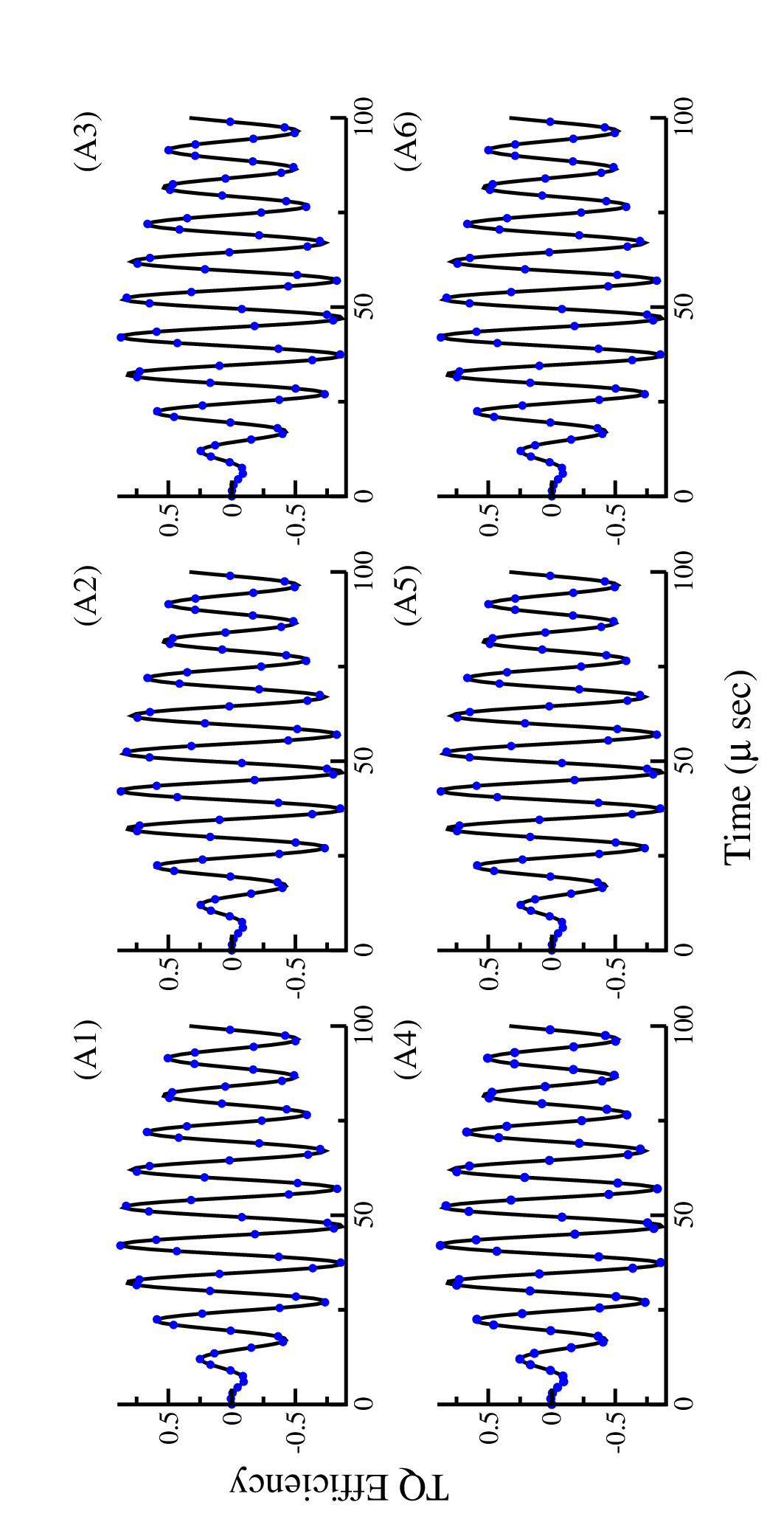}    
\caption{Comparison of numerical (black thick line) and analytic simulations (blue dots) based on effective Hamiltonians derived from Regime-II corresponding to the quadrupole coupling constant $\left(C_{Q}=\rfrac{\omega_{Q}}{\pi}\right)$, $C_{Q}=100$ kHz and RF amplitude, $(\rfrac{\omega_{1}}{2\pi})=100$ kHz. The simulation results from the first transformation, Case-I (panel A1), Case-II (panel A4) and second transformation, Case-III (a) (panel A2), Case-III (b) (panel A3), Case-IV (a) (panel A5) and Case-IV (b) (panel A6) are illustrated. The powder simulations were performed using a crystal file having 28656 orientations ($\alpha, \beta$).}
\label{fig:pow100khzlr}
    \end{figure} 
\end{center}

\begin{center}
\begin{figure}[H]
    \centering
     \includegraphics[width=0.45\textwidth, angle=0]{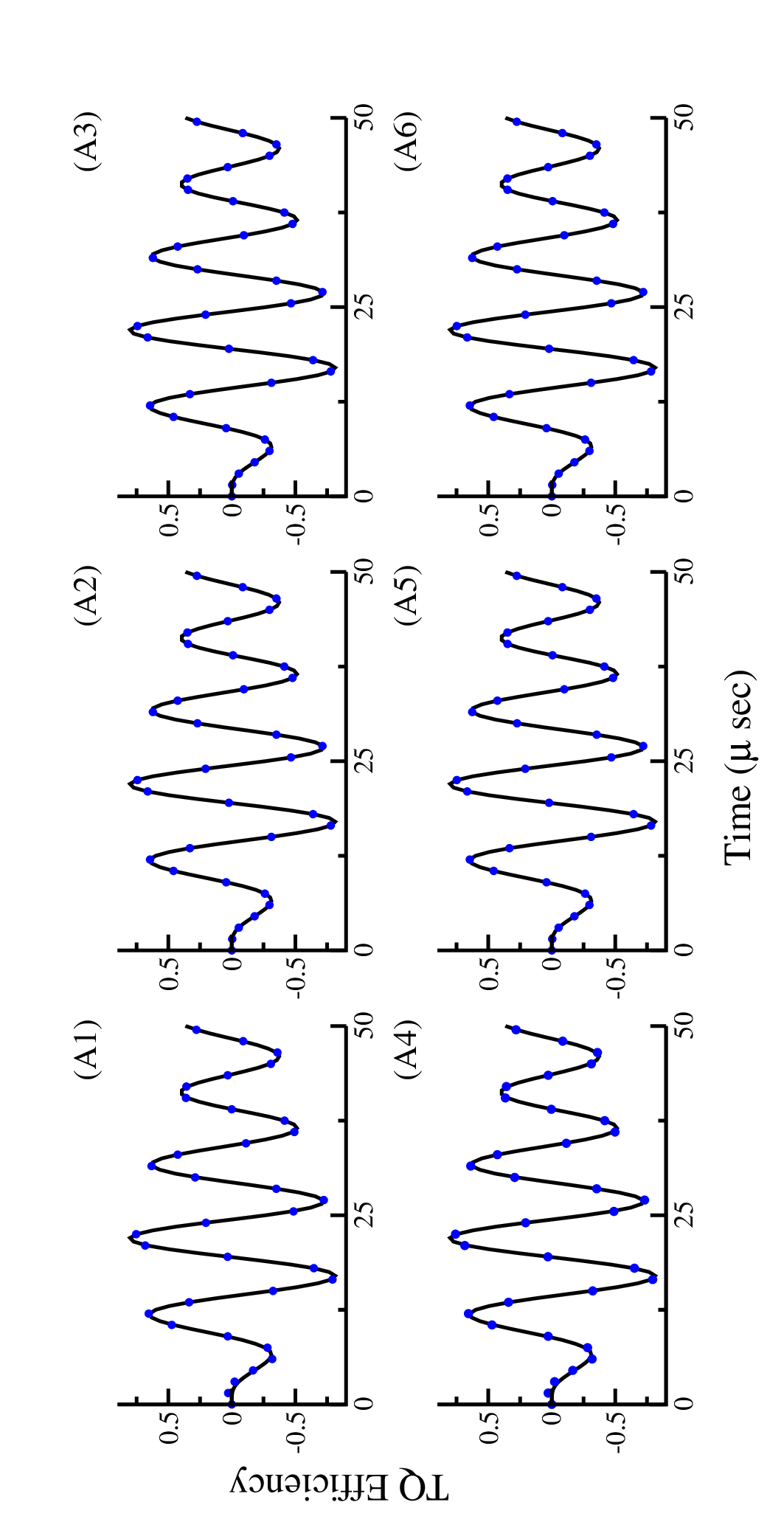}    
\caption{Comparison of numerical (black thick line) and analytic simulations (blue dots) based on effective Hamiltonians derived from Regime-II corresponding to the quadrupole coupling constant $\left(C_{Q}=\rfrac{\omega_{Q}}{\pi}\right)$, $C_{Q}=200$ kHz and RF amplitude, $(\rfrac{\omega_{1}}{2\pi})=100$ kHz. The simulation results from the first transformation, Case-I (panel A1), Case-II (panel A4) and second transformation, Case-III (a) (panel A2), Case-III (b) (panel A3), Case-IV (a) (panel A5) and Case-IV (b) (panel A6) are illustrated. The powder simulations were performed using a crystal file having 28656 orientations ($\alpha, \beta$).}
\label{fig:pow200khzlr}
    \end{figure}  
\end{center}

In an alternate approach, powder simulations based on the theoretical description presented in Regime-II were employed to describe the excitation profile for systems with quadrupolar frequency comparable to or lower than RF amplitude of the pulse. For the sake of illustration, analytic powder simulations employing quadrupolar coupling constants ($C_{Q}=100$ kHz; $C_{Q}=200$ kHz) are depicted in Figures.~\ref{fig:pow100khzlr},~\ref{fig:pow200khzlr}, respectively. As depicted, the powder simulations based as the effective Hamiltonian derived from Regime-II yield results in excellent agreement with SIMPSON simulations.\\

Hence, the methodology presented in the strong and weak coupling regimes is suitable for describing the excitation process in a powder sample. In the intermediate regime ($1<\rfrac{\omega_{Q}}{\omega_{1}}<20$), the effective Hamiltonians based on either regimes (Regime-I and Regime-II) yield results in complete disagreement with those obtained from exact numerical methods. \\

\begin{center}
\begin{figure}[H]
    \centering
     \includegraphics[width=0.5\textwidth, angle=0]{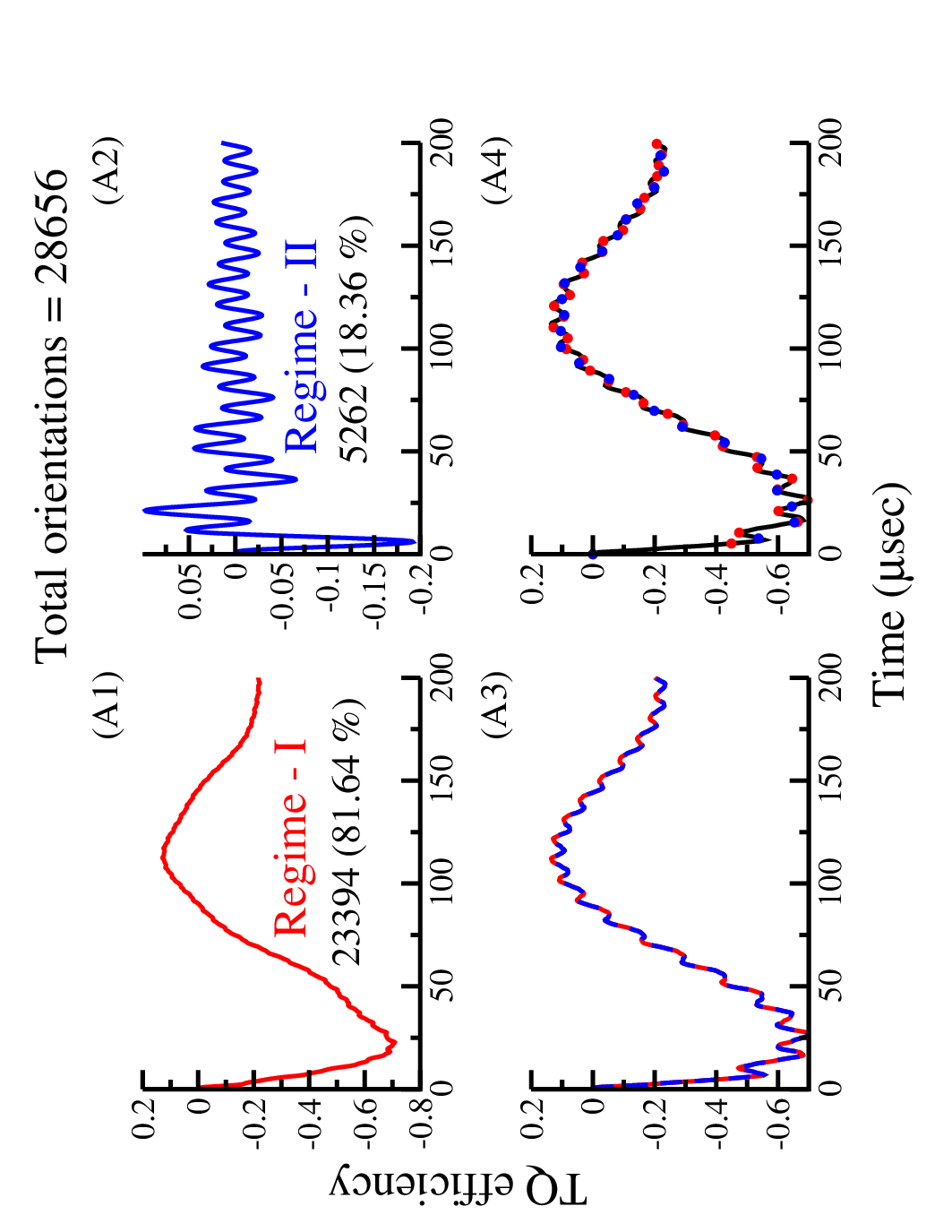}    
\caption{Comparison of numerical (black thick line) and analytic simulations (blue dots) based on effective Hamiltonians derived from both regimes corresponding to the quadrupole coupling constant $\left(C_{Q}=\rfrac{\omega_{Q}}{\pi}\right)$, $C_{Q}=2$ MHz and RF amplitude, $(\rfrac{\omega_{1}}{2\pi})=100$ kHz. The analytic simulations emerging from the effective Hamiltonians derived from Regime-I only (see panel A1), Regime-II only (see panel A2), hybrid method (combination of Regime-I and Regime-II) ( in panel A3) are depicted. In panel A4, the analytic simulations from the hybrid method (combination of Regime-I (red) and Regime-II (blue)) are compared with exact numerical simulations (black line). The choice of Regime-I and Regime-II is purely dependent on the magnitude of $\omega_{Q}^{(\alpha \beta \gamma)}$ relative to the RF amplitude. When $\omega_{Q}^{(\alpha \beta \gamma)} < \omega_{1}$, Regime-II is employed, $\omega_{Q}^{(\alpha \beta \gamma)} > \omega_{1}$, Regime-I is employed. The powder simulations were performed using a crystal file having 28656 orientations ($\alpha, \beta$).}
\label{fig:pow2r2mhz}
    \end{figure} 
\end{center}

\begin{center}
\begin{figure}[H]
    \centering
     \includegraphics[width=0.5\textwidth, angle=0]{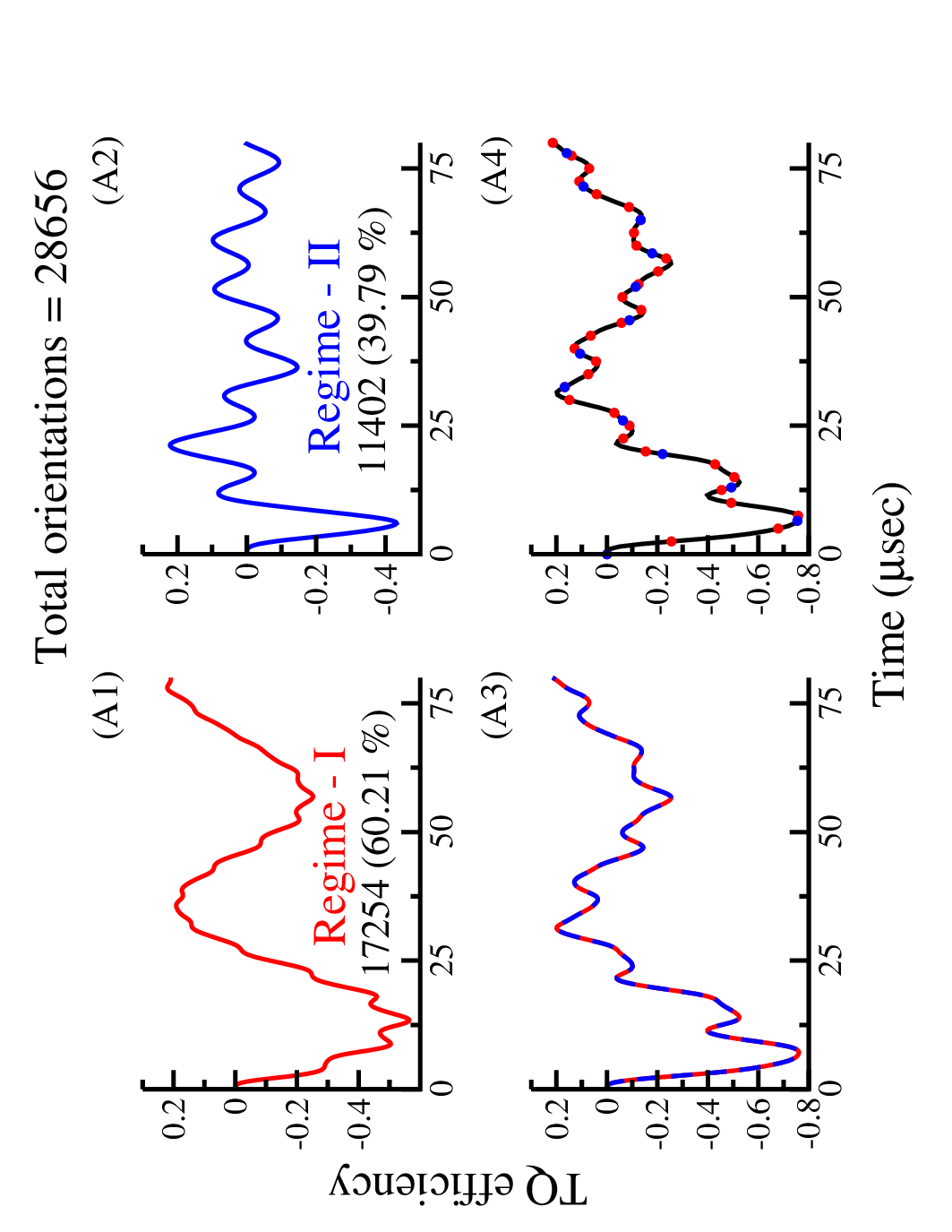}    
\caption{Comparison of numerical (black thick line) and analytic simulations (blue dots) based on effective Hamiltonians derived from both regimes corresponding to the quadrupole coupling constant $\left(C_{Q}=\rfrac{\omega_{Q}}{\pi}\right)$, $C_{Q}=1$ MHz and RF amplitude, $(\rfrac{\omega_{1}}{2\pi})=100$ kHz. The analytic simulations emerging from the effective Hamiltonians derived from Regime-I only (see panel A1), Regime-II only (see panel A2), hybrid method (combination of Regime-I and Regime-II) ( in panel A3) are depicted. In panel A4, the analytic simulations from the hybrid method (combination of Regime-I (red) and Regime-II (blue)) are compared with exact numerical simulations (black line). The choice of Regime-I and Regime-II is purely dependent on the magnitude of $\omega_{Q}^{(\alpha \beta \gamma)}$ relative to the RF amplitude. When $\omega_{Q}^{(\alpha \beta \gamma)} < \omega_{1}$, Regime-II is employed, $\omega_{Q}^{(\alpha \beta \gamma)} > \omega_{1}$, Regime-I is employed. The powder simulations were performed using a crystal file having 28656 orientations ($\alpha, \beta$).}
\label{fig:pow2r1mhz}
    \end{figure} 
\end{center}

\begin{center}
\begin{figure}[H]
    \centering
     \includegraphics[width=0.5\textwidth, angle=0]{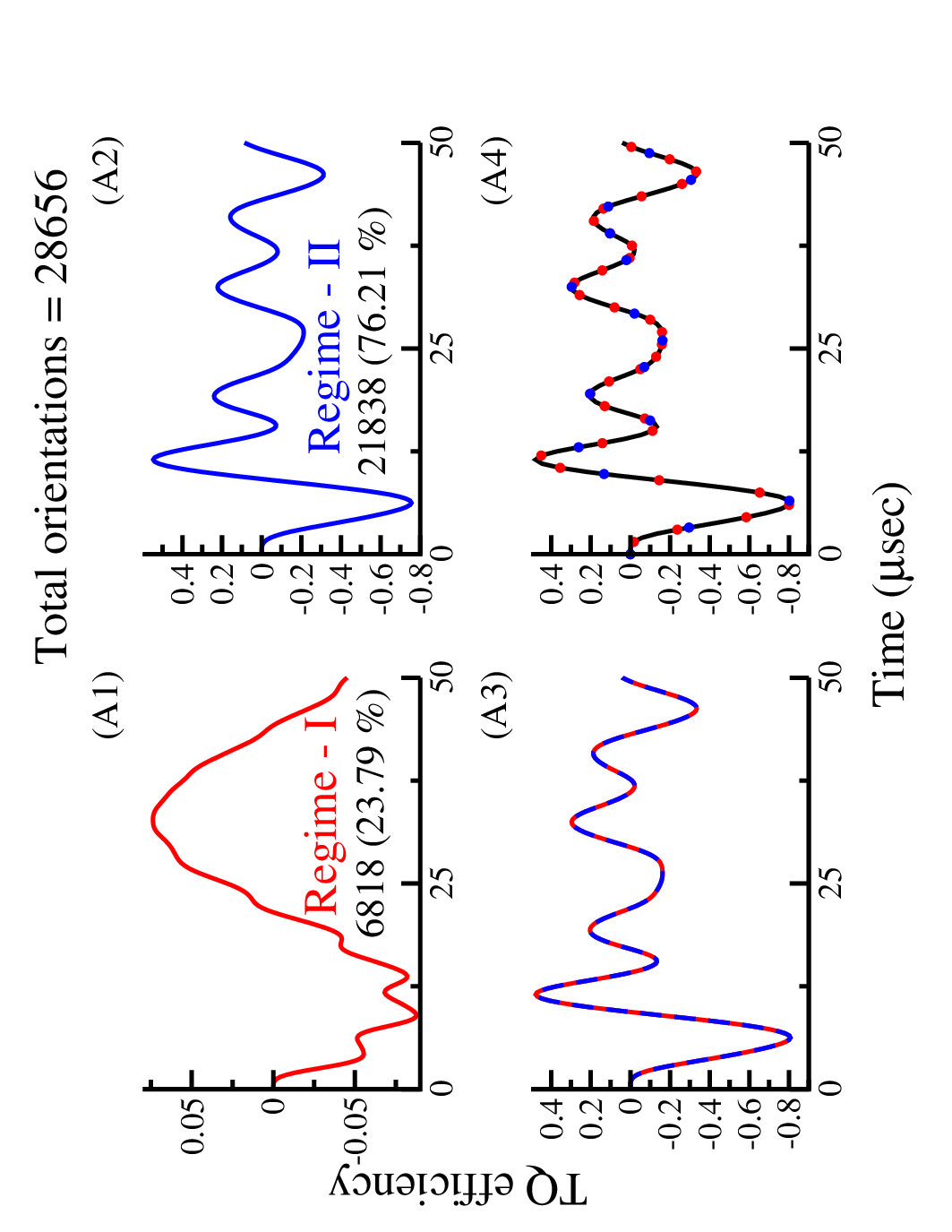}    
\caption{Comparison of numerical (black thick line) and analytic simulations (blue dots) based on effective Hamiltonians derived from both regimes corresponding to the quadrupole coupling constant $\left(C_{Q}=\rfrac{\omega_{Q}}{\pi}\right)$, $C_{Q}=500$ kHz and RF amplitude, $(\rfrac{\omega_{1}}{2\pi})=100$ kHz. The analytic simulations emerging from the effective Hamiltonians derived from Regime-I only (see panel A1), Regime-II only (see panel A2), hybrid method (combination of Regime-I and Regime-II) ( in panel A3) are depicted. In panel A4, the analytic simulations from the hybrid method (combination of Regime-I (red) and Regime-II (blue)) are compared with exact numerical simulations (black line). The choice of Regime-I and Regime-II is purely dependent on the magnitude of $\omega_{Q}^{(\alpha \beta \gamma)}$ relative to the RF amplitude. When $\omega_{Q}^{(\alpha \beta \gamma)} < \omega_{1}$, Regime-II is employed, $\omega_{Q}^{(\alpha \beta \gamma)} > \omega_{1}$, Regime-I is employed. The powder simulations were performed using a crystal file having 28656 orientations ($\alpha, \beta$).}
\label{fig:pow2r500khz}
    \end{figure} 
\end{center}

To address this issue, an alternate approach defined as the "hybrid method"(purely hypothetical) that employs effective Hamiltonians derived from Regime-I and Regime-II is proposed to describe the excitation process. Since a  powder sample comprises of a distribution of quadrupolar coupling constants, the choice of the interaction frames (quadrupolar interaction frame for Regime-I and tilted RF interaction frame for Regime-II) play an important role in the convergence of the perturbation corrections. For crystallite orientations with `$\omega_{Q}^{(\alpha \beta \gamma)}<\omega_{1}$', the effective Hamiltonians based on Regime-II were employed (Eqs. S.38 and S.41 in the supplementary information), while for `$\omega_{Q}^{(\alpha \beta \gamma)}>\omega_{1}$', effective Hamiltonians based in Regime-I were employed (Eq.~\ref{eq:effham2},~\ref{lambdas2}) in simulating the excitation profile in the intermediate regime. A schematic depiction of the excitation profiles in the intermediate regimes are illustrated in Figures.~\ref{fig:pow2r2mhz} -~\ref{fig:pow2r500khz}.

\begin{center}
\begin{figure}[H]
    \centering
     \includegraphics[width=0.5\textwidth, angle=0]{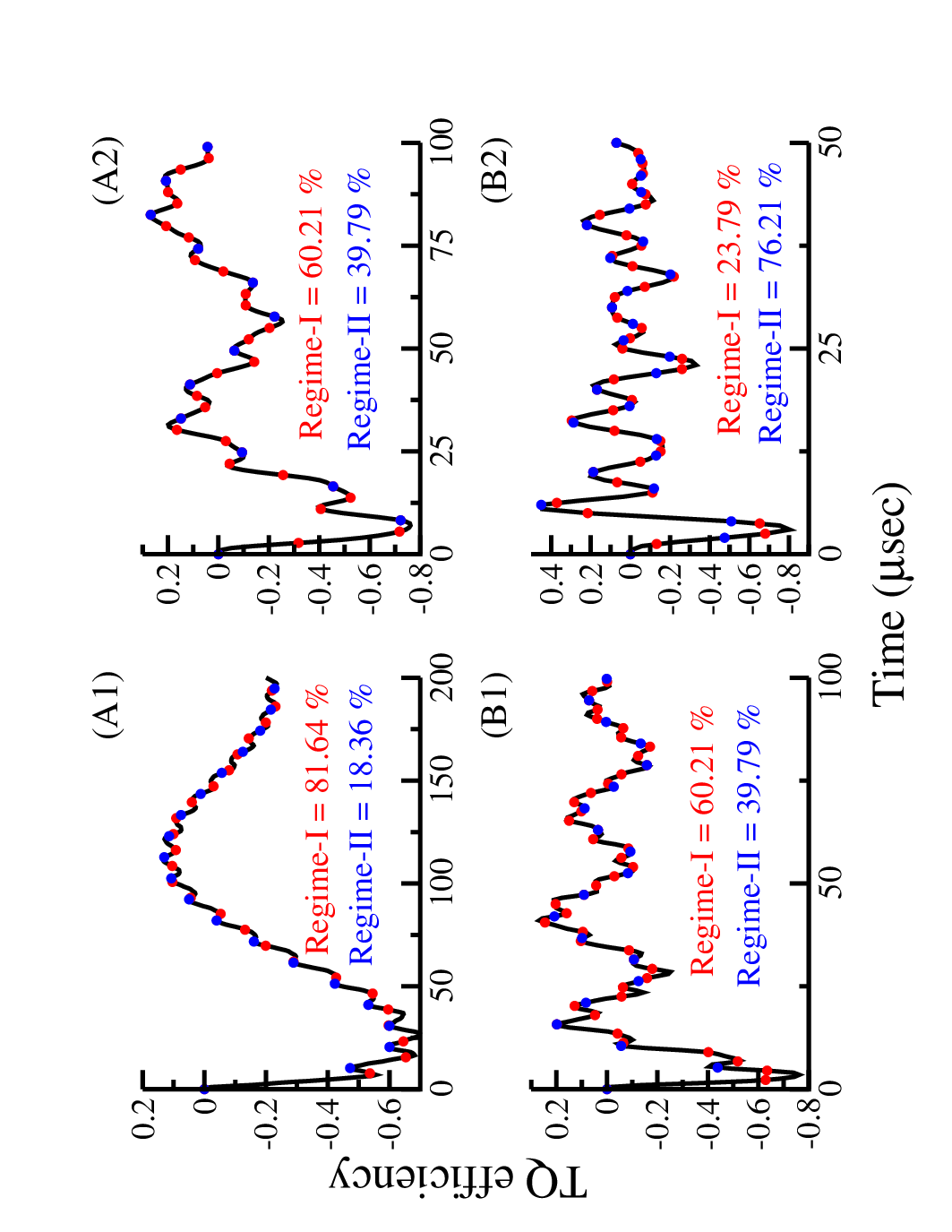}    
\caption{Simulations illustrating the role of the RF amplitude in the hybrid simulations in a powder sample. The magnitude of the quadrupole coupling constant $\left(C_{Q}=\rfrac{\omega_{Q}}{\pi}\right)$ is varied along the row, while the amplitude of the pulse is varied along the column. The following parameters were employed in the simulations: A1) $C_{Q}=2$ MHz, $(\rfrac{\omega_{1}}{2\pi})=100$ kHz, A2) $C_{Q}=1$ MHz, $(\rfrac{\omega_{1}}{2\pi})=100$ kHz, B1) $C_{Q}=2$ MHz, $(\rfrac{\omega_{1}}{2\pi})=200$ kHz, B2) $C_{Q}=1$ MHz, $(\rfrac{\omega_{1}}{2\pi})=200$ kHz. The red and blue dots represent the analytic simulations from Regime-I and II respectively, while the black line denotes the results from numerical simulations.The powder simulations were performed using a crystal file having 28656 orientations ($\alpha, \beta$).}
\label{fig:pow100_200khz}
    \end{figure} 
\end{center}

In Figure.~\ref{fig:pow2r2mhz}, the excitation profile in the intermediate regime corresponding to the analytic simulations emerging from the effective Hamiltonians derived from the procedure described in Regime-I (panel A1) and Regime-II (panel A2) are depicted, respectively. In panel A3, analytic simulations emerging from the proposed hybrid method is depicted. In the proposed hybrid method, depending on the magnitude of the quadrupolar frequency relative to the RF amplitude, effective Hamiltonians derived from Regime-I and Regime-II were employed to simulate the TQ excitation profiles. In panel A4, the analytic simulations from the proposed hybrid method are compared with SIMPSON simulations. The powder simulations were performed using a crystal file having 28656 orientations ($\alpha, \beta$). Of the 28656 orientations that were employed, $82\%$ of the crystallites had their `$\omega_{Q}^{(\alpha \beta \gamma)}>\omega_{1}$' and $18\%$ had `$\omega_{Q}^{(\alpha \beta \gamma)}<\omega_{1}$'. As depicted, the analytic simulations emerging from the proposed `hybrid method' are in excellent agreement with the numerical simulations. To further substantiate this approach, additional set of simulations corresponding to ($ C_{Q}=1$ MHz; $ C_{Q}=500$ kHz) are depicted in Figures.~\ref{fig:pow2r1mhz},~\ref{fig:pow2r500khz}, respectively. As the magnitude of the quadrupolar coupling constant decreases, the percentage of crystallites that adhere to the dynamics governed by Regime-II should in-principle increase owing to the scaling introduced by the powder averaging. This argument of ours is justified in the analytic simulations depicted in panel-A4 of Figures.~\ref{fig:pow2r1mhz} and~\ref{fig:pow2r500khz}. 
To further substantiate this reasoning, the excitation profile for a given quadrupolar coupling constant was calculated by varying the amplitude of the excitation pulse. As depicted in Figure.~\ref{fig:pow100_200khz}, the number of crystallite orientations within `$\omega_{Q}^{(\alpha \beta \gamma)}<\omega_{1}$' increases with the amplitude of the pulse (panels A1,B1 and panels A2, B2).
Hence, the choice of the appropriate interaction frame (whether it is quadrupolar interaction frame or tilted RF interaction frame) plays an important role in the exactness of the derived effective Hamiltonians. The hybrid method proposed in this article is extremely beneficial in the analytic description of powder samples and could be employed to build theoretical models for quantifying experimental data involving powder samples. From a theoretical standpoint, the role of MAS in this demarcation between the two regimes deserves a formal description that requires a multi-modal Floquet analyses and will be addressed in future publications. 

\clearpage
\section{Conclusions and Perspectives}
\begin{center}
\begin{figure}[H]
    \centering
     \includegraphics[width=0.6\textwidth, angle=0]{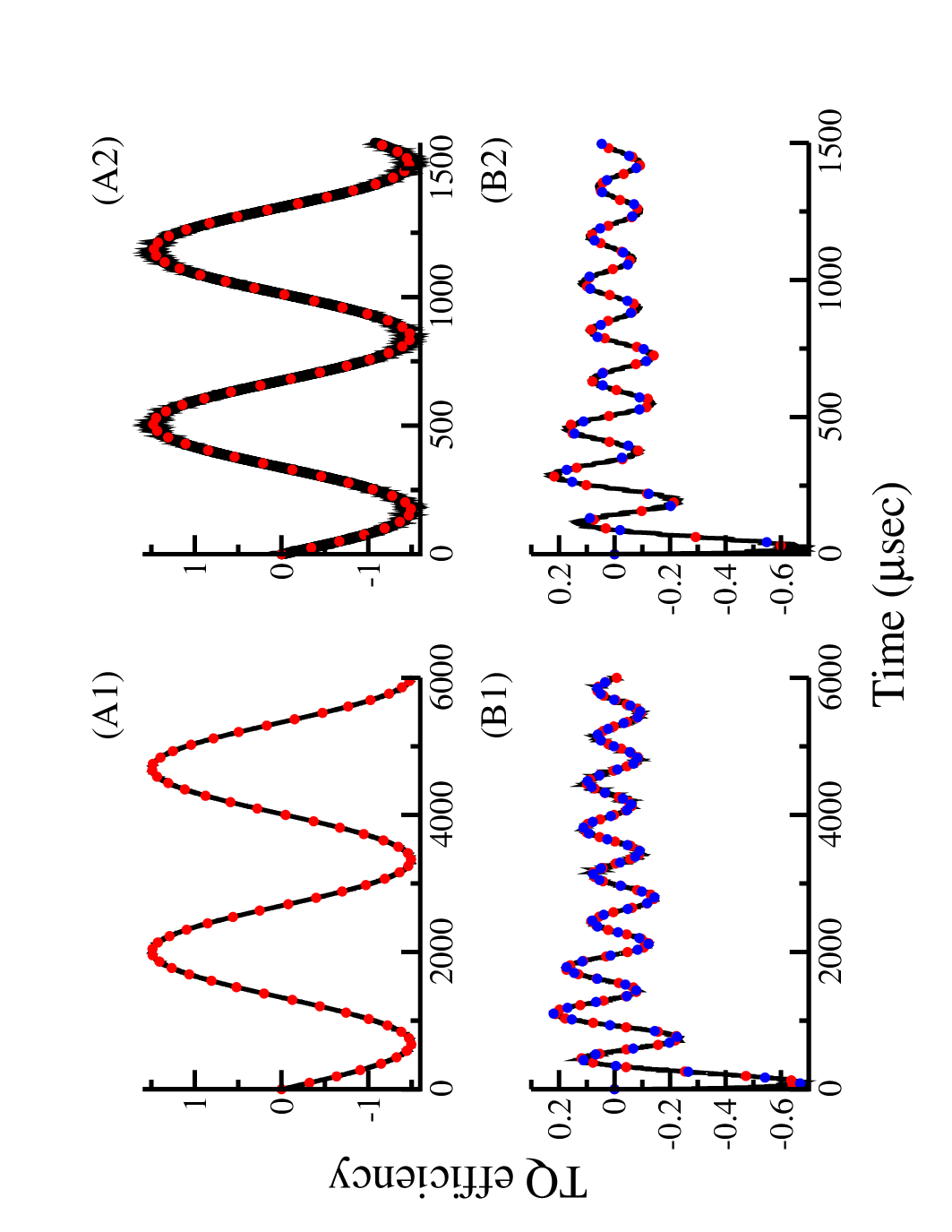}    
\caption{Comparison of TQ excitation in a single crystal (panel A) and powder sample (panel B) corresponding to the strong coupling regime. The simulations (both in single crystal and powder sample) were derived based on the theoretical framework presented in Regime-I. In contrast to the single crystal, the TQ signal in powder sample (panel B) decays with time, clearly illustrating the interference effects between the different crystallite orientations. The following parameters were employed in the simulations: A1) $C_{Q}=4$ MHz, $(\rfrac{\omega_{1}}{2\pi})=100$ kHz, A2) $C_{Q}=2$ MHz, $(\rfrac{\omega_{1}}{2\pi})=100$ kHz, B1) $C_{Q}=4$ MHz, $(\rfrac{\omega_{1}}{2\pi})=100$ kHz and B2) $C_{Q}=2$ MHz, $(\rfrac{\omega_{1}}{2\pi})=100$ kHz.}
\label{fig:con1}
    \end{figure} 
\end{center}

\begin{center}
\begin{figure}[H]
    \centering
     \includegraphics[width=0.6\textwidth, angle=0]{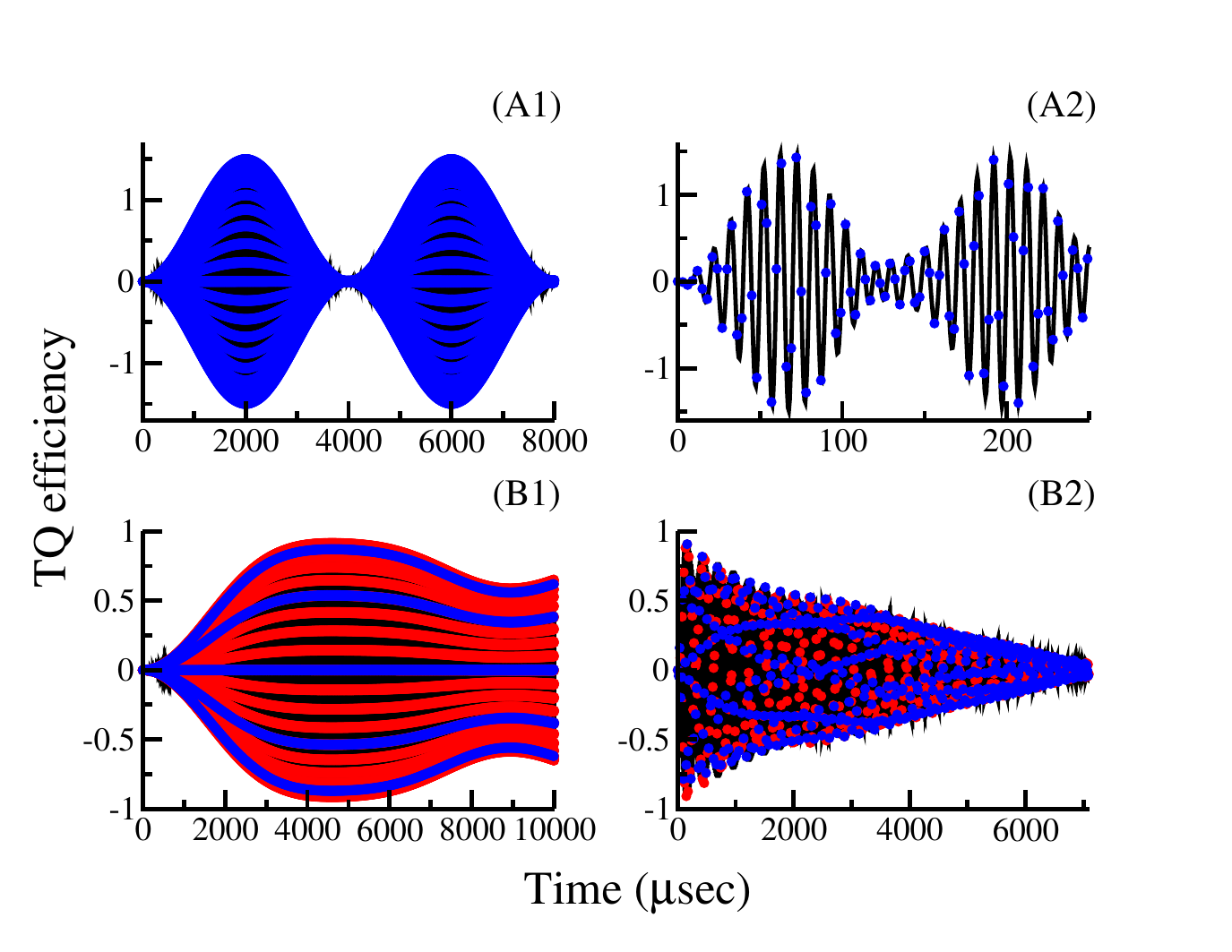}    
\caption{Comparison of TQ excitation in a single crystal (panel A) and powder sample (panel B) corresponding to the weak coupling regime. The simulations (both in single crystal and powder sample) were derived based on the theoretical framework presented in Regime-II. In contrast to the single crystal, the TQ signal in powder sample (panel B) decays with time, clearly illustrating the interference effects between the different crystallite orientations. The following parameters were employed in the simulations: A1) $C_{Q}=1$ kHz, $(\rfrac{\omega_{1}}{2\pi})=100$ kHz, A2) $C_{Q}=30$ kHz, $(\rfrac{\omega_{1}}{2\pi})=100$ kHz, B1) $C_{Q}=1$ kHz, $(\rfrac{\omega_{1}}{2\pi})=100$ kHz, and B2) $C_{Q}=30$ kHz, $(\rfrac{\omega_{1}}{2\pi})=100$ kHz.}
\label{fig:con2}
    \end{figure} 
\end{center}

In summary, the present study highlights the role of the interaction frames in improving the exactness of the effective Floquet Hamiltonians employed in the description of MQ excitation of quadrupolar nuclei. Since the contact transformation method employed in the derivation of effective Floquet Hamiltonian is based on perturbation theory, the convergence of the perturbation corrections play an important role in the accuracy of the analytic methods based on effective Hamiltonians. From an operational point of view, the choice of a particular interaction frame depends delicately on both the nature of the sample and the extrinsic parameters (pulse parameters such as RF amplitude) employed in the experiments. For example, in the case of a single crystal, the classification into strong, intermediate and weak coupling regimes depends primarily on the magnitude of the quadrupolar frequency relative to the amplitude of the RF pulse. In cases where the magnitude of the quadrupolar frequency largely exceeds the RF amplitude, the effective Hamiltonians derived from the quadrupolar interaction frame (Regime- I) provide an accurate description of the excitation process. Alternatively, when the quadrupolar frequency is lower than the amplitude of the excitation pulse, the description in the RF interaction frame (Regime-II) is necessary. However, in both these cases, the convergence of the perturbation corrections/series is faster and the number of transformations required in the derivation of the effective Floquet Hamiltonian is limited to a single unitary transformation. Consequently, in the strong coupling regime (Regime-I, Eq.~\ref{eq:con1}), the TQ signal expression reduces to a much simpler form.
\begin{align}
\label{eq:con1}
  \left\langle T^{(3)-3} (t_{p1}) \right\rangle&=-  \left( \Phi_{1}^{3} \ \right) 
\left\lbrace \dfrac{3}{2}   \sin\left( \dfrac{3\  \omega_{1}^{3}\ t_{p1}}{2\ \Omega_{Q}^{2}}\right)     \right\rbrace 
\end{align}
In a similar vein, the TQ signal in the weak-coupling regime (Regime-II, Eq.~\ref{eq:con3}  ) reduces to a much simpler form.
\begin{align}
\label{eq:con3}
\left\langle T^{(3)-3} (t_{p1}) \right\rangle&=   \left( \Phi_{1}^{3} \  \right) 
\left\lbrace \dfrac{3}{2}\ \sin\left( \omega_{1}\ t_{p1}\right) \sin^{2}\left( \dfrac{\Omega_{Q}\ t_{p1}}{4}\right)\right\rbrace
\end{align}
As discussed in the previous section, the above expressions are equally valid for describing the excitation in both isotropic (single crystal, $\Omega_{Q}=\omega_{Q}$ ) and anisotropic (powder samples, $\Omega_{Q}=\omega_{Q}^{(\alpha \beta \gamma)}$ ) solids. Below, the excitation profile emerging from the above analytic expressions are depicted in Figures.~\ref{fig:con1} and ~\ref{fig:con2} corresponding to the strong and weak-coupling regimes, respectively.  For comparative purpose, the simulations corresponding to the single crystal (panels A1, A2) are compared with those from powder sample (panels B1, B2).

As depicted, in the case of a single crystal, the excitation profiles both in the strong and weak-coupling regimes are oscillatory (periodic) and resemble to the Rabi oscillations\cite{rabi}. Interestingly, in the case of a powder sample (Figure.\ref{fig:con1} , panels B1, B2), Figure.~\ref{fig:con2} , panels B1, B2) the oscillations decrease in intensity with time and are no longer periodic. The dissipation of the signal in the time-domain could be explained through the analytic expressions described above. As described above, in the case of a single crystal (both in strong and weak-coupling regimes), the excitation profile is described by a single trigonometric function. On the contrary, in a powder sample, the signal in the time domain (at each time point) is an ensemble average over all possible orientations. Hence, the time-domain signal in a powder sample has contributions from a distribution of quadrupolar frequency associated with individual crystallites leading to interference between the different trigonometic terms. Consequently, the signal intensity decreases with time (or gets damped) in a powder sample and was also reported recently in a theoretical study involving spin I=$\rfrac{1}{2}$ nuclei\cite{ranjan}.

\begin{center}
\begin{figure}[H]
    \centering
     \includegraphics[width=0.6\textwidth, angle=0]{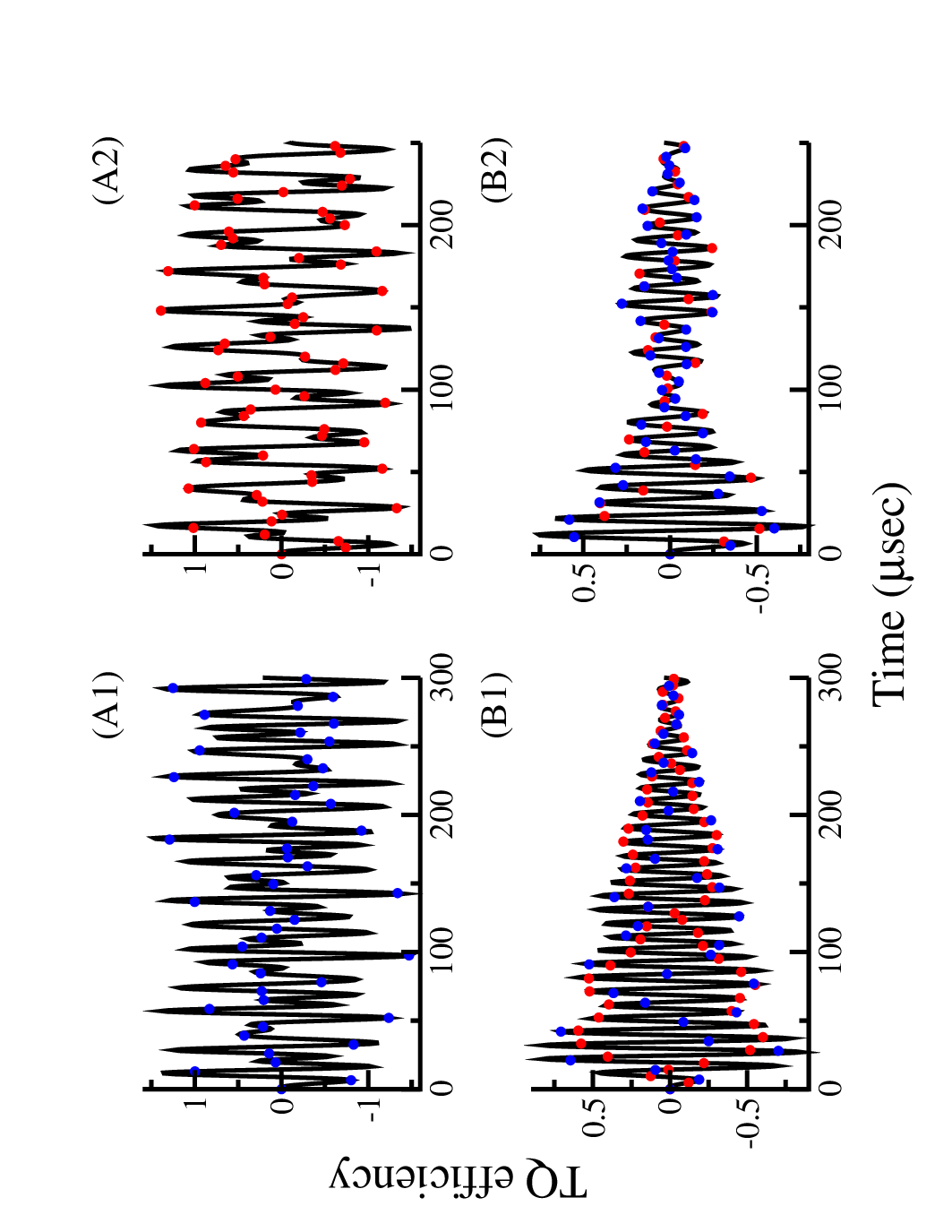}    
\caption{Comparison of TQ excitation in a single crystal (panel A) and powder sample (panel B) corresponding to the intermediate coupling regime. The simulations (both in single crystal and powder sample) were derived based on the theoretical framework presented in Regime-II. In contrast to the single crystal, the TQ signal in powder sample (panel B) decays with time, clearly illustrating the interference effects between the different crystallite orientations. The following parameters were employed in the simulations: A1) $C_{Q}=150$ kHz, $(\rfrac{\omega_{1}}{2\pi})=100$ kHz, A2) $C_{Q}=250$ kHz, $(\rfrac{\omega_{1}}{2\pi})=100$ kHz, B1) $C_{Q}=150$ kHz, $(\rfrac{\omega_{1}}{2\pi})=100$ kHz, and B2) $C_{Q}=250$ kHz, $(\rfrac{\omega_{1}}{2\pi})=100$ kHz.}
\label{fig:con3}
    \end{figure} 
\end{center}

To further explore this aspect, the above simulations were extended to study the excitation process in the intermediate regime. As discussed in the previous section, depending on the nature of the sample, the following criterion is employed in the classification of the intermediate regime. In the case of single crystal, when the magnitude of the quadrupolar frequency is greater than the RF amplitude of the pulse, the spin physics in the intermediate regime (approximately defined by the condition $1<\rfrac{\omega_{Q}}{\omega_{1}}<10$) is governed by the calculations presented in Regime-I. Consequently, a series of unitary transformations are necessary to improve the exactness of the effective Hamiltonians and the excitation profile has contributions from all the four trigonometric terms present in Eq.~\ref{lambdas2}. In a similar vein, when the amplitude of the excitation pulse exceeds the quadrupolar frequency, the framework presented in Regime-II is employed in the derivation of effective Floquet Hamiltonians in the intermediate regime (approximately defined by the condition $1<\rfrac{\omega_{1}}{\omega_{Q}}<5$) and the TQ signal is evaluated using Eq. S.41 (given in the supplementary information). To illustrate the difference between the excitation process in a single crystal and powder sample, analytic simulations in the intermediate regime (from both cases) are illustrated in Figure.~\ref{fig:con3}.
In contrast to the simulations presented in Figures.~\ref{fig:con1}  and ~\ref{fig:con2}, the oscillations in the single crystal get damped and are non-periodic. This unexpected result could be explained through the interference effects resulting from the presence of the four trigonometric terms present in the TQ signal expression (Eqs.~\ref{lambdas2}  and S.41). Due to scaling of the quadrupolar frequency in a powder sample, the definition of the intermediate regime differs from those described for a single crystal. Accordingly, depending on the relative magnitude of the quadrupolar frequency to that of the RF amplitude, the following conditions are proposed to classify the intermediate regime in a powder sample: (a) $\omega_{Q} > \omega_{1}$ (intermediate regime $1<\rfrac{\omega_{Q}}{\omega_{1}}<20$), (b)$\omega_{Q} < \omega_{1}$ (intermediate regime $1<\rfrac{\omega_{1}}{\omega_{Q}}<3$). Hence, the damping effects observed in a powder sample should be severe due to the presence of the trigonometric terms in addition to the ensemble effect originating from different crystallite orientations present in a powder sample. The simulations presented in panels (B1, B2 of Figure.~\ref{fig:con3}), justify our explanation.
In contrast to a single crystal, the demarcation of the various regimes in a powder sample are less straightforward. Depending on the relative magnitude of the quadrupolar coupling constant to the amplitude of the RF pulse, the extent of contributions from the two regimes (Regime-I and Regime-II) varies. As illustrated in Table.~\ref{tab:con1}, the number of crystallites in a powder sample with `$\omega_{Q}^{(\alpha \beta \gamma)}< \omega_{1}$' increases when the quadrupolar frequency is lowered. Accordingly, the theoretical framework presented in the Regime-II should be employed in the derivation of the effective Floquet Hamiltonians. In a similar vein, for the remaining crystallites with `$\omega_{Q}^{(\alpha \beta \gamma)}> \omega_{1}$' , the theoretical framework  presented in Regime-I is suitable for analytic description. Hence, analytical description of the spin dynamics for a powder sample in the intermediate regime entails the need for the “hybrid method”. 
\begin{center}
\begin{table}[H]
\caption{Classification of crystallite orientations into Regime-I and Regime-II based on the relative magnitude of the anisotropic quadruple coupling constant `$\rfrac{\omega_{Q}}{2\pi}$' to the amplitude of the exciting pulse `$\rfrac{\omega_{1}}{2\pi}$' when RF amplitude is always at $\rfrac{\omega_{1}}{2\pi}=100$ kHz}
\centering
\resizebox{13cm}{!} {
\begin{tabular}{||c|c|c|c|}
\hline 
 \textbf{$\omega_{1}:\omega_{Q}$}&\textbf{No. of crystallite }  &\textbf{No. of crystallite }&\textbf{Total No. of crystallite } \\
 &\textbf{orientations in Regime-I} & \textbf{orientations in Regime-II} & \textbf{orientations}\\ \hline \hline
 & & & \\
 $1:20$& $26064\ (90.95 \%)$ & $2592\ (9.05 \%)$   & $28656$ \\ [0.5cm]
 $1:10$& $23394\ (81.64 \%)$ & $5262\ (18.36 \%)$   & $28656$ \\ [0.5cm]
 $1:5$  & $17254\ (60.21 \%)$ & $11402\ (39.79 \%)$ & $28656$ \\ [0.5cm]
 $1:3$  & $ 8954\ (31.25 \%)$  & $19702\ (68.75 \%)$ & $28656$ \\ [0.5cm]
 $1:2$  & $0$&$28656\ (100.00 \%)$ & $28656$ \\  [0.5cm]
  & & & \\\hline 
\end{tabular}
}
\label{tab:con1} 
\end{table}
\end{center}
In the past, theoretical descriptions have often remained confined to the strong coupling regime. Since majority of the quadrupolar nuclei have large quadrupolar coupling constants, descriptions often remained confined to strong couplings regimes only. The analytic theory proposed in this article presents a more general framework highlighting the importance of the intermediate regime in the description of the excitation process in both isotropic and anisotropic solids. The higher order corrections and the transformations required are systematically derivable from the proposed recursive relations and are computationally efficient. Since the transition from one regime to the other regime depends on the amplitude of the RF pulse, the proposed hybrid method is well suited for quantifying the experimental data involving quadrupolar nuclei in solids. The combined effects of sample spinning in addition to the modulation frequency of a typical multiple pulse sequence entails a multimodal Floquet analysis of the MQ experiment and would be discussed in future publications. \\
\\
\\
\textbf{Acknowledgments:}
This research work was supported by a research grant to R.R by the Department of Science and Technology (No.DST,SR/S1/PC-07/2008), Government of India and V.G. would like to thank IISER Mohali for the graduate assistantship.

\clearpage
\bibliography{sref} 
\bibliographystyle{apsrev4-1}

\end{document}